\documentclass[screen=true, table, nonacm]{acmart}
\usepackage[utf8]{inputenc}

\usepackage{booktabs}
\usepackage{units}
\usepackage{amsfonts}
\usepackage{mathtools}
\usepackage{amsmath,marvosym}
\usepackage{xifthen}
\usepackage{listings}
\usepackage{xcolor}
\usepackage{todonotes}
\usepackage{graphicx}
\usepackage{lineno}
\usepackage{pgf}
\usepackage{collcell}
\usepackage[normalem]{ulem}

\usepackage{wrapfig}
\usepackage{lscape}

\usepackage[figuresright]{rotating}
\newenvironment{amssidewaysfigure}
  {\begin{sidewaysfigure}\vspace*{.7\textwidth}\begin{minipage}{\textheight}\centering}
  {\end{minipage}\end{sidewaysfigure}}

\acmBooktitle{}
\begin{document}

\newcommand*{\MinNumber}{0}%
\newcommand*{\MaxNumber}{7}%

\newcommand{\ApplyGradient}[3]{%
  \pgfmathsetmacro{\PercentColor}{100.0*(#1-#2)/(#3-#2)}%
  \edef\x{\noexpand\cellcolor{black!\PercentColor}}\x\textcolor{white}{#1}%
}
\newcommand{\ApplyGradientTrick}[4]{%
  \pgfmathsetmacro{\PercentColor}{100.0*(#1-#3)/(#4-#3)}%
  \edef\x{\noexpand\cellcolor{black!\PercentColor}}\x\textcolor{black!\PercentColor}{#1}\x\textcolor{white}{#2}%
}

\title[A Review of Intrusion Detection Systems and Their Evaluation in the IoT]{A Review of Intrusion Detection Systems and Their Evaluation in the IoT}

\author{Luca Arnaboldi}
 \affiliation{%
   \institution{University of Edinburgh}
  \streetaddress{Holyrood}
   \city{Edinburgh}
   \postcode{EH8 8AQ}
   \country{UK}}
 \email{luca.arnaboldi@ed.ac.uk}
 \authornote{Work done whilst at Newcastle University}
 \author{Charles Morisset}
 \affiliation{%
   \institution{Newcastle University}
  \streetaddress{1 Urban Science Square}
   \city{Newcastle Upon Tyne}
   \postcode{NE4 5TG}
   \country{UK}}
 \email{charles.morisset@ncl.ac.uk}

\begin{abstract}
Intrusion Detection Systems (IDS) are key components for securing critical infrastructures, capable of detecting malicious activities on networks or hosts.
The procedure of implementing a IDS for Internet of Things (IoT) networks is not without challenges due to the variability of these systems and specifically the difficulty in accessing data. 
The specifics of these very constrained devices render the design of an IDS capable of dealing with the varied attacks a very challenging problem and a very active research subject. 
In the current state of literature, a number of approaches have been proposed to improve the efficiency of intrusion detection, catering to some of these limitations, such as resource constraints and mobility.
In this article, we review works on IDS specifically for these kinds of devices from 2008 to 2018, collecting a total of 51 different IDS papers. 
We summarise the current themes of the field, summarise the techniques employed to train and deploy the IDSs and provide a qualitative evaluations of these approaches.
While these works provide valuable insights and solutions for sub-parts of these constraints, we discuss the limitations of these solutions as a whole, in particular what kinds of attacks these approaches struggle to detect and the setup limitations that are unique to this kind of system. 
We find that although several paper claim novelty of their approach little inter paper comparisons have been made, that there is a dire need for sharing of datasets and almost no shared code repositories, consequently raising the need for a thorough comparative evaluation.
\end{abstract}


\maketitle

\section{Introduction}

An IDS is in essence a monitor placed on a device and/or network that analyses incoming messages, to detect attacks and/or unwanted traffic, and when paired with an intrusion prevention system can be used to stop attacks before they affect the system. 
They are trained using system behaviour data or from existing attack databases, and use these patterns to make the detection.
They are widely deployed in a variety of systems and can often be considered as a first line of defence against intruders.
Deployment strategies differ; however, they can broadly be classified into Host Based (HIDS) and Network Based (NIDS).
A HIDS monitors activities on the device itself such as system calls or shell commands to discover unauthorised behaviours or accesses. 
They are normally very fine grained traces of behaviour on a single device.
A NIDS on the other hand looks at the network data to determine the likelihood of intrusion. 
This approach is more flexible when looking at large systems of devices and will have a smaller overhead on the devices themselves; however, it has less granularity.
To optimise detection it is sometime desirable to mix the approaches and have both implemented on a system.
This technique is often referred to as collaborative intrusion detection (CID). 
Whilst this approach theoretically provides a wider coverage it also needs a very structured architecture, and is therefore much more complex to implement.

Standard approaches used to train IDSs include using a database of known attacks ({\em misuse detection}) and testing systems to create a ``benchmark" behaviour and flag any anomaly as a potential attack ({\em behaviour based detection})~\cite{mell2003overview}. 
However it has become more and more common to mix up the approaches to make up for their respective detriments, namely the inability for misuse detection to detect unknown attacks and the high false positive rate of anomaly detection. 
This is generally referred to as an Hybrid IDS, which can be either a combination of these two techniques or a combination of sub categories such as rule based together with anomaly. 
In combination with the different kinds of deployment options described it is possible to create various different IDS frameworks to suit different scenarios.

One such scenario that this paper focuses on is the Internet of Things (IoT). 
The IoT is an umbrella term to describe the embedding of sensors, actuators, computing and connectivity to physical objects like buildings, vehicles, wearable, and many more.
This might describe anything from a smart television to a building or factory to potentially a whole city of interconnected devices monitoring various aspects; traffic levels, temperature, number of people. 
These systems aim to improve various aspects of our lives~\cite{IOTSurv2010}; however, with these improvements, several issues arise in terms of security and privacy, as made evident by various surveys~\cite{IOTSecSurv2014,IOTSecSurv2013,IOTSecChallenges2015}.

A lot of challenges that are cited in the literature arise as an outcome of the restraints of these devices. The devices themselves are often very simple and are built to perform a specific task, with little room for variation, consequently they are unable to be adapted to apply security solutions and rely externally for security. 
In this scenarios it is beneficial to implement an IDS on such systems as it can perform the much needed security without needing to change the IoT setup. 
It is of particular concern to ensure that an IoT system is secure as this emergent paradigm relies heavily on data collection and as recent attacks show may lead to serious privacy concerns~\cite{IOTPrivacySurv2014}. 
Whilst data is a key attack objective, the devices themselves make for desirable targets as their limited capabilities make it easy to take down whole networks e.g. through battery drainage or through gain of illicit access to the devices themselves, in turn allowing to employ them as a botnets.
In these very heterogeneous and dynamic systems it becomes very difficult to find a single IDS solution and several approaches will be discussed.

There are various different approaches proposed to secure IoT systems by means of IDSs, and several other surveys have discussed them extensively~\cite{butun2014survey,zarpelao2017survey,elrawy2018intrusion}; however, whilst a summation of work is done, no work has been done towards evaluating the different approaches. 
We conducted a systematic review selecting papers from the year of 2008 to 2018. 
The papers were selected if they fit the criteria we identified in Sec.~\ref{sec:idsIoT}.
The criteria was selected as the core metrics that made designing an IDS for the IoT different than that for a more traditional system.
Our results have found that although several papers are proposing new approaches to deal with the new attacks pertinent to the IoT, very few cross evaluate with previously proposed approaches.
This lack of cross evaluation is quite worrying as there are several approaches all solving the same problem, without clear guidance on why one cannot use the existing techniques.
In this survey we attempt to understand why this is the case an systematically evaluate the pros and cons of the different approaches.

A unified evaluation criteria allows for informed decision making, and gives users a clearer sense of what the abilities of an IDS are. 
IDS Solutions for the IoT take many forms and are often bespoke to a specific scenario, so it is very difficult for a potential implementer to make a decision of whether it would be suitable for their needs. 
Furthermore, due to the highly dynamic nature of IoT environments the adaptability of these systems becomes an important effectiveness metric and the results need to fit their scenario.
To compare the effectiveness of an IDS several metrics can be used e.g. F Score, false positive rate, mean error etc; however, results can be scenario dependent or be reliant on setup specifics. 
It is desirable, on the other hand, to evaluate each IDS under the same circumstances of deployment and against the same attack range to get an accurate comparison.

Out of the 51 surveyed works only three provided source code artefacts in their papers. 
Although we attempted to contact all authors of the works to compare the various approaches, we were only able to obtain one further artefact. 
This discovery is a very worrying trend, as new techniques keep getting proposed it becomes unclear as to the strengths and weaknesses of each approach. 
Without open access to previous work it becomes impossible to build on top of previous efforts making it harder to innovate the field and leading to repetition of efforts and lack of advancement.

The contributions of this chapter are the following; 
\begin{enumerate}
    \item A comprehensive survey of the current state of the art in IoT intrusion detection consisting of 51 papers;
    \item A break down  of different approaches and designs currently proposed; 
    \item  A summary of the current literature assessing pros and cons of the different approaches.
\end{enumerate}

\textit{The remainder of the paper is broken down as follows:}
\begin{itemize}
    \item Section~\ref{sec:ids-relwork} provides a list of related works surveying IDSs in the IoT;
    \item Section~\ref{sec:idsIoT} provides a discussion on adjustments and specific requirements to deploy and IDS in the context of IoT;
    \item Section~\ref{sec:idsArch} provides a breakdown of deployment architectures for IoT systems;
    \item Section~\ref{sec:idstech} provides a breakdown of techniques and algorithms used to train and deploy IDSs as well as their limitations and advantages;
    \item Section~\ref{sec:eval-survey} discusses current evaluation criteria and methods, the way IDS are currently evaluated and a discussion on their effectiveness.
    \item Section~\ref{sec:tool-collection} discusses our approach to collecting tools for unified evaluation - we note this was unsuccessful.
    \item Section~\ref{sec:idssum} summarises our literature review of 51 proposed IDS tools for IoT, categorises them and breaks down each approach;
    \item Section~\ref{sec:chaptConc} concludes and discusses current trends, challenges and the current gaps in the literature;
\end{itemize}

\section{Related Work}
\label{sec:ids-relwork}

Despite the maturity of the field of IDS research, the current IDS solutions are inadequate for wide usage in IoT deployments. 
To address this difficulty researchers have proposed new means to do intrusion detection that can adapt to the IoT constraints and circumvent them.
In this survey we investigate the current state of intrusion detection, present the difficulties in adapting different IDS training/detection techniques to the IoT and comprehensively summarise what the state of the art in IDS in the IoT is.

The discussion of how to protect IoT systems by means of IDSs is a very hot topic for research and several other works have reviewed the literature of IDSs in IoT or similar paradigms~\cite{butun2014survey,zarpelao2017survey,elrawy2018intrusion}. 
Whilst a summation of the work in the area exists, little to no work has attempted to answer the questions of how to quantify an IDSs effectiveness in the scope of an IoT System. 

Prior to the IoT, another form of constrained network environment was wireless sensor networks (WSN).
A WSN would be considered as a sub category of a IoT system, with very specific configurations and several of the same key security concepts and limitations.
Butun et al. (2013)~\cite{butun2014survey}, provide one of the early works in surveying IDSs for Wireless Sensor Networks.
The authors also progress to discuss MANETS or Mobile Ad hoc networks, another constrained network configuration that is more dynamic than WSNs.
The survey talks about some of the most common approaches used in the detection of attacks on these systems.
Unlike a lot of IoT configurations, WSNs are often homogeneous, this allows for a lot of statistical approaches to be used which are often not able to capture attacks in the IoT.
The authors provide an example of this using forwarding percentage of packets by a node, and discuss the effectiveness of these approaches.
The authors discuss a lot of the techniques used to detect attacks in WSNs putting specific focus on the advantages of using each of the approaches.
Unlike the IoT these systems don't have downsides of multi protocols, and interconnect multi network systems and therefore there is much less focus on how the IDS is deployed and how the different detection agents may interact.
The evaluation of the approaches revolves around their applicability to scenarios and no formal comparison of the approaches is provided.
This nonetheless is an extensive and very good analysis around the techniques that can be used in MANETS and WSN and how they work.

Departing from WSNs, Zarpel\~ao et al.(2017)~\cite{zarpelao2017survey}, propose one of the first efforts classifying IDS research for the IoT. 
Their aim is to identify leading trends, open issues, and future research possibilities. 
Their proposed classification is structured around the: detection method, IDS placement strategy,security threat and validation strategy.
The authors discuss the IDS from the point of view of how the IDS is deployed, how the detection is done and also what attacks are detected.
The survey takes a form of a literature summary where for each paper a summary of these three concepts is provided.
The authors also discuss the issue of how these approaches are validated.
The authors find that many different approaches are used, and no single strategy is used.
This is mentioned as a core issues as it makes the comparison of the different approaches difficult, not to mention some approaches didn't provide any validation at all.

The most recent survey for IDSs in the IoT, to the best of our knowledge, is the work of Elrawy et al. (2018)~\cite{elrawy2018intrusion}. 
In their work the authors extensively discuss the structure and architecture of the IoT, potential threats, and different applications.
A summary is provided of techniques used in different papers, with a summary of their advantages and disadvantages.
The structure provided is informative and does a good job of introducing the different concepts prior to discussing the IDSs.
The authors then provide a short summary of each paper and summarise the techniques they use and deployment strategy in a table.
the survey also discusses the authors evaluation results, the intuition of this is very good however, as mentioned by the authors, due to the different ways of evaluation is is impossible to draw comparisons from these results.
Across these surveys there are many shared themes.
As different advances are done in network architectures several new approaches are developed to defend them showing the need for new analysis and surveys.
They recurring trend is also that there is a lack of evaluation of these approaches, so none of the surveys are able to compare the effectiveness of the different IDS.

Each of these reviews have chosen around 20 different IDS papers, and our more extensive review has collected over 50, and yet we have no easy way to compare the different approaches.
This makes it difficult to establish open problems, choose the best approach for a scenario and to assess the validity of the proposed literature.
This raises the need for ways to compare different IDSs  in a uniform and comparable manner.

\section{An overview of Intrusion Detection for the IoT}
\label{sec:idsIoT}

As the IoT takes over several key aspects of our lives, including homes, factories and even healthcare, so it becomes very important to keep them secure.
However, these systems vary greatly and function differently from traditional internet systems, so new solutions need to be devised to ensure their safety. 
From a security perspective it is important to understand the implications of the IoT on how solutions are designed.
We know for certain that some specifics of IoT make it impossible to implement certain solutions, e.g. resource constrains such as limited battery life, low bandwidth, small processing capabilities, and memory constraints make computationally intensive security protocols impossible to implement. 
However it is less certain how these impact the ability for an IDS to be implemented successfully.
We identify four main concerns to consider:

\begin{enumerate}
    \item \textit{New kinds of attacks} which are often not pertinent to standard systems.
    Consequently arising the need for the IDS detection to be expanded.
    Taking these observations in mind current solutions need to be adjusted to be able to detect these new avenues of intrusion.
    In traditional networks, the system administrator deploys IDS agents in nodes with higher computing capacity. 
    Whilst in the context of IoT networks which are usually composed of nodes with resource constraints, this may not be an option.
    We summarise a list of new vulnerability vectors in the IoT that would impact the ability for an IDS to detect intrusions as:
    \begin{enumerate}
        \item[i.] \textit{Node compromise attacks}, devices in IoT systems are often vulnerable to physical takeover, allowing for malicious behaviour from previously benign members of the network; 
        \item[ii.] \textit{Communication errors}, due to deployment in disadvantaged environments~\cite{secheverria-disadvantaged}, IoT is very prone to network errors which can be a source of irregular traffic; 
        \item[iii.] \textit{Battery drainage attacks}, the constrained resources may lead to selfish device behaviours, due to device self preservation constant polling may lead to battery drainage so devices will stay offline for large periods of time, leading to difficulty in behavioural patterning and harder network diagnostics; \item[iv.] \textit{Routing attacks}, the IoT is often deployed as an open network through WiFi or similar technologies, this opens up the network to external connections as well as new attacks to do with routing, such as sinkhole attacks;
        \item[v.] \textit{ Compromised communication}, as a consequence of low bandwidth it is difficult for devices to reach far away destinations,so the network configuration becomes very important, therefore, if a node is identified as a key connector between two of the devices and is consequently compromised, communication in the network is broken; 
    \end{enumerate}
    \item \textit{Placement} of the IDS itself represents a unique challenge.
    In traditional networks end systems are directly connected to specific nodes (e.g., wireless access points, switches, and routers) that are responsible for forwarding the packets to the destination~\cite{zarpelao2017survey}. 
    In the IoT on the other hand the network may have
    \item \textit{Multi hop routing} and even be partially or fully \textit{disconnected}, so the ability to survey the full traffic may be infeasible.
    \item \textit{New communication protocols}, most modern IDSs are built to work with traditional web protocols (HTTP, TCP, REST architecture etc.), due to the constraints of these devices the IoT often operates on completely new protocols (6LowPAN, CoAP, ZigBee etc.), making traditional IDSs simple unable to comprehend the protocols.
    To compound this even further there is an issue of  \textit{multi protocol systems} that are very common and require even more adaptation.
\end{enumerate}

These core issues, are the main principles in consideration when creating an IDS for IoT systems.

\section{Types of Intrusion Detection Systems in IoT Context}
\label{sec:idsArch}

The IoT does not reside in a bubble, it is very much integrated with the various internet infrastructures such as cloud, fog and edge computing; therefore, these need to be considered as key components of an IDS design strategy. 
When talking about IoT it becomes difficult to generalise in the same way traditional IDS classifications have done; one cannot, for example, easily say that the IDS is placed centrally or distributed, as where it's placed as well as how it is placed, is a core factor of importance i.e. it could be placed centrally in the fog, or centrally in the cloud or even on a device in the LAN. 
All these options have different security and performance implications. 
If a solution claims to cover the whole of the IoT such as the solution proposed by Raza et al. (2013)~\cite{raza2013svelte} then the solutions needs to have a strategy for working in each of these scenarios, it is not simply enough to show how it solves just one of the restrictions (e.g. new protocols).
The location of an IDS also largely depends on what type of IDS it is. 
IDS types are commonly split into Host based (HIDS) and Network Based (NIDS); however, a further categorisations can be made when applied to the context of IoT Systems namely, Collaborative IDSs (CIDS) which involve the collaborations both a Network based and Host Based IDS; each of which could potentially be deployed across various networks/sub-networks.
Each of these approaches have different uses and scenarios.

\subsection{Network Intrusion Detection for IoT}

A NIDS is an IDS that monitors network traffic to detect remote attacks i.e. attacks carried out over a network connection.
These kinds of attacks work at the higher level of the stack often targeting vulnerabilities in protocols; however, their effect can have high impact on the device themselves. 
One of the most common network attack is a DoS attack, a DoS attack aims to target the availability of a device or network and due to the constraints of IoT devices they can be particularly effective.
Traditional safety measures that can be deployed to protect against these kinds of attacks i.e. security protocols, often cannot be ran on these devices due to their constraints~\cite{yadav2010itrust}. 

One of the core differences between most IoT systems and standard infrastructure is the protocols at play.
In the IoT there has been a shift from more traditional protocol to new technologies to cater to the constraints of devices.
Examples of this is the movement from IP to UDP, IPv4 to IPv6 and even more specialised protocols such as RPL which is specific for constrained networks.
With the complete change in the underlying infrastructure the current Network Based Intrusion Detection techniques may simply just not function on these kinds of systems.
To tackle these challenges various new techniques are developed specific to these new protocols.

Within the IoT we also have a raise in popularity of what are often referred to as unsupervised systems, these systems are deployed using the concept of machine to machine (M2M) interactions only, and therefore will operate largely unsupervised by humans. 
As an outcome of these scenarios a compromised node will often go unnoticed, allowing for an attacker to lunch attacks from within the network. 
These kinds of attacks cannot be blocked by an external facing firewall and therefore the deployment of an internal IDS is essential~\cite{yadav2010itrust,liu2018intrusion,khan2017trust}.
Not only is the issue with internal attacks and lack of authentication of devices, there are also new attacks raising in popularity due to the infrastructure of these systems.
One of the components of unsupervised systems is \textit{dynamic routing}, this means that routes are constructed based on shifts in the system, signal strengths and resource optimisations.
As an outcome of this it becomes beneficial for an attacker to disrupt this process and route packets in either a less than optimal way or a way that benefits his interests e.g. through a compromised node.
These routing attacks have raised hugely in popularity in WSN and IoT systems some of the most famous being, Wormhole attacks~\footnote{Definition available at: \url{https://www.sciencedirect.com/topics/computer-science/wormhole-attack}} and Sybil attacks~\footnote{Definition available at: \url{https://www.sciencedirect.com/topics/computer-science/sybil-attack}}, both of which rely heavily on the system being unsupervised and self adaptive.
These kinds of attacks have spawned various research papers specifically catered to detecting them, as we discuss in Sec.~\ref{sec:toolNIDS}.

Changes to the infrastructure do not only have negative effects on intrusion detection in the IoT.
The restrictive nature of these systems may come as an advantage from an attack detection perspective.
Due to the limited behaviours of the components of these systems new techniques that would not be suitable in standard networks are being developed and explored.
Work has developed using automatons~\cite{fu2017automata,misra2011learning} for behaviour modelling as the limited behaviours allows to avoid state explosion issue that traditionally prevented it.
Automata based approaches allow for system analysis as well as behaviour modelling making for more actionable IDSs.
The set of simple behaviours has also allowed for game theory based approaches~\cite{sedjelmaci2016lightweight}, to find the optimal behaviour of a simple system.
And the raise of popularity of clustering approaches~\cite{deng2018mobile,jiang2012dynamic,liu2018intrusion}, which rely on grouping similar behaving devices in a cluster to improve detection of misbehaving nodes.

Deployment strategies also play a big role in how intrusion detection is done, this does not only have to consider maximum coverage but resource constraints as well.
The vast majority of IDS systems for the IoT implement distributed IDSs.
This involves sharing of information between sub-networks, which adds complexity if there are contrasting information and around trust in reporting.
This also is reflected in the amount of data that needs to be processed by an IDS, in some cases with large systems it simple becomes infeasible.
A relatively new area of research has been focusing on reducing this overhead using dimensionality reduction techniques and smart data processing~\cite{liu2018intrusion}, to more effectively deal with alerts.
As each component in the system needs to report data in a distributed manner, authors have considered trust based scheme to ensure the quality of reported data is maintained~\cite{yadav2010itrust,cervantes2015detection,khan2017trust}.
All these changes in infrastructure, protocols and the types of system at play are key factors that need to be considered when designing a NIDS for IoT systems.

\subsection{Host Intrusion Detection for IoT}

A host based IDS refers to an IDS that monitors the activities on the device (i.e., the host) where it is deployed, to detect local attacks i.e. attacks executed by users of the targeted system or attacks directly impacting the device operations.
These types of IDS may monitor operations at a lower level than the network based IDS and they have access to much more details of the impact on the system itself.
Whilst a network base IDS might only see packets travelling through the system, the HIDS also have access to system logs and can monitor metrics of the device behaviour.
One of the recurring difficulties around converting IDS to work on IoT systems is around the low computational power of the devices.
This is particularly relevant in the scope of Host Based Systems, as these limitations greatly restrict how many operations can be done on the device itself, and the data overhead it can handle.

Devices in IoT systems often dedicate what little computing power they have to providing features or services. 
Design constraints, such as the need to increase performance or battery life, may restrict the ability of system designers to implement security effectively.
Furthermore traditional approaches such as signature based schemes are unsuitable for these kinds of systems due to the large data overhead.
Therefore, anomaly based detection methods, which attempt to identify deviations in measured statistics against a normal model of operation of a system, can be beneficial to use in resource constrained systems.

One of the most common advances in the field of HIDs is the way data is handled at the device level.
Most approaches optimise data handling to reduce state space and cleverly optimise data processing.
Data processing optimisation, can be achieved by feature engineering, this process is achieved by carefully selecting relevant features of data and reducing its dimensionality to decrease the number of operations needed to be conducted by the device. 
If done carefully, this approach is still able to preserve the patterns of behaviour within the dataset making it still useful for anomaly detection, whilst greatly reducing overhead.
One way feature engineering is applied in the IoT is through bit pattern matching~\cite{summerville2015ultra}, this allows to store a small chunk of the dataset in a lookup table, greatly reducing the number of comparison operations.

Further work takes advantage of the devices limited behaviour for effective anomaly detection.
Small resource constrained devices execute fewer and potentially less complex operations than general purpose computing platforms. 
This results in less complex patterns of communication and behaviours, making it easier to detect when such patterns have changed.
A raise in popularity in these cases is the use of immunity based techniques.
The immune system has been successfully applied to the information processing domain. 
In particular, it performs complex computations in parallel and decentralised patterns.
Furthermore, an immune system can learn new information and recall learned information.
Due to the decrease in complexity authors have been able to train these systems a lot more efficiently~\cite{liu2011research}.
Another further development has been the increase in use of pseudo or full modelling techniques.
Using Hidden Markov Models, IDSs are capable of learning the system behaviour an find anomalies.
To render this technique less resource intensive weak-HMMs are developed.
Using domain knowledge of system behaviour the formation of the HMM can be guided to make less memory intensive effective anomaly detection schemes~\cite{song2010weak}.

Host based systems in the IoT have some further challenges that need to be addressed.
A DoS attack against an IoT network has the potential to be significantly more detrimental than one against a standard network. 
This increased vulnerability is due in part to the low computational power and battery power characteristic of IoT devices.
This has lead to a raise in popularity of battery drain denial of service attacks~\cite{arnaboldi2017quantitative,arnaboldi2018generating}, these attacks target a device to perform power drain intensive operation to drain the battery.
If they are successful it requires human intervention to change the battery something that is to be avoided in large unsupervised systems.
Authors ave proposed battery monitoring techniques at the host level to detect these kinds of attacks~\cite{lee2014lightweight}.
This is an additional feature of interest that traditional IDS techniques do not consider.
However, as devices often do not have the ability to self monitor their battery drain~\cite{arnaboldi2018generating}, it becomes critical for an IDS to monitor these behaviours.

\subsection{Collaborative Intrusion Detection for IoT}

A collaborative IDS makes use of different IDSs together in a single system.
It may include both a network intrusion detection system as well as Host based ones.
Or as is often the case in IoT systems, the system spans across various subsystems composed of various devices and using different technologies, to safeguard these systems different kinds of IDSs need to be in place and they need to cooperate to prevent multi-protocol attacks and cross system intrusions.
A collaborative IDS may span various different sub-networks, communication protocols and even geo-locations, it may be all network based or contain several different types all linked together.
This distributed nature makes the design of an IDS particularly complex.

One of the main drivers behind the need of collaborative IDSs is the multi-protocol nature of these systems.
This scenario may render several current techniques less effective and be unsuitable to a single IDS scenario.
This is especially pertinent in the case of pattern matching IDSs as the same attack across different protocols may take a different form, and therefore bypass these techniques.
To cater to these difficulties literature has proposed data aggregation methods that allow for, non protocol restricted pattern matching and behaviour analysis~\cite{yu2008framework,coppolino2013applying}.
These types of approaches make use of local aggregator nodes collecting specific data to a sub-cluster or sub-component of a system, which then communicates to a central component (in a specific format) and centrally does the pattern matching~\cite{coppolino2013applying}.

A defining feature of IoT that is being considered across all configurations of IDS is their constrained nature.
Their restricted capabilities lead to the development of simpler detection engines. 
This is often fine as their behaviour is also restricted, however, a simpler engine is not capable of detecting attacks across multi protocols and multi systems, which in cooperation may exhibit vastly different behaviours.
These raises the need for several different systems to be deployed.
In these circumstances, the core difficulties are around how the data is communicated across the network.
If an anomaly detection finds an anomaly in a sub-part of the system it is difficult to understand how it will spread across the other components or if it would spread~\cite{zhang2011artificial}.
To do so authors propose specific risk assessments across the different layers, with efficient head nodes able to assess the impact of attacks at the different levels~\cite{zhang2011artificial}.

One of the biggest issues around collaborative IDSs is around the authenticity of communication.
When you have several different monitors and reporting mechanisms spread across a large network it is essential that the data is authentic.
Wrongful reporting, or illegitimate information may raise false alarms and cause meaningful damages to the system.
This is exacerbated in the IoT as these devices may struggle to implemt full cryptographic solutions that resolve these problems in standard systems.
To mitigate these circumstances authors have proposed new methods of  were behaviour of devices is checked against expected behaviours to validate if they are indeed communicating in a benign way~\cite{bostani2017hybrid,shreenivas2017intrusion}. 
These approaches are particularly effective in the context of insider attacks. 
As these are recognised components within the network they may cause a lot of damage.
Authors therefore propose metrics based on expected behaviours~\cite{shreenivas2017intrusion}, this creates a level of trust between the devices that may be altered as to replace traditional means of authentication.

\section{Techniques for use in Intrusion Detection Systems}
\label{sec:idstech}

Within the various different approaches to construct the IDSs, there are techniques and algorithms used for training the IDS and constructing rules.
These techniques vary from carefully curated databases to complex machine learning algorithms and may be used to construct any of the different types of IDS.

\subsection{Rule Based/Misuse Detection/Policy Based and Signature}

Misuse detection also referred to as Rule based and Policy based approaches generally rely on rules of the known intrusions taken from known attack databases. 
They are considered efficient and are used in most real time systems as they need less processing power than behaviour based detectors, and thus can handle large volumes of traffic without slowing down the normal activities. 
Their major criticism is the deficiency in detecting zero-day attacks i.e. the attacks which are previously unknown.
An often stronger type of detection, although still pattern based, is signature-based IDS.
Signature based detection refers to the detection of attacks by looking for specific patterns, such as byte sequences in network traffic, or known malicious instruction sequences used by malware. 
Although signature-based IDS can easily detect known attacks, it is difficult to detect new attacks, for which no pattern is available.

In the context of IoT, these approaches become particularly pertinent to very simple systems where known issues arise.
Particularly in the case of WSNs, there are several known attacks, specifically routing attacks, that may not be stopped by careful programming, as these kind of systems are vulnerable to these attacks by their very construction.
It is therefore desirable to deploy an IDS with preset rules for such attacks.
The downside of these approaches is that patterns may be easily bypassed by careful packet obfuscation or slight changes to known behaviour.
One approach that has been popularised to counter this is the usage of automated rule learning techniques~\cite{yu2008framework}.
The automated learning of rule has the advantage that it is more adaptable and thus better able to deal with new attacks.
However this is a backward facing approach as the rules are based on the current behaviour and only after the attack having taken place can a rule be automatically generated.

An area that has shown a lot of promise in the context of signature generation is through automatons and model based signatures.
The restricted behaviours of these devices allow the creation of modelling techniques to formalise the attack signatures~\cite{misra2011learning,fu2017automata}.
These approaches make use of the automata to specifically model the communication of a protocol or behaviour of a device, with each transition being a message exchange/action performed. 
The incoming data or behaviour of the device is then encoded and compared to the modelled approach and if deviating marked as anomalous.
This approach is not only a new way to describe behaviours, modelling has several key advantages, being able to reason, and calculate system wide properties could give security professionals key insights about their system that would not be possible with traditional approaches.
This approach is relatively new, but could be extended to not only manually create signatures, but our previous work has proposed modelling based techniques for the automated generation of stochastic attackers~\cite{arnaboldi2018generating,arnaboldi2018lisa}.
Although in early stages, this approach may be able to bypass a lot of the current downsides of signature based approaches to encompass unknown attacker behaviours.

\textit{Advantages:} Accurate in finding known attacks easy to deploy, good for known attacks, widely used.~\cite{di2008intrusion,weaver2013guide}

\textit{Disadvantages: } Cannot yet find unknown attacks, tends to be protocol specific, isn't customisable to your setup, may have high memory overload~\cite{axelsson2000base}.

\subsection{Anomaly/Statistical}

Anomaly/statistical-based intrusion detection systems were primarily introduced to detect unknown attacks, in part due to the rapid development of malware. 
The basic approach is to create a model of trustworthy activity, and then compare new behaviour against this model. 
Although this approach enables the detection of previously unknown attacks, it may suffer from false positives, which are previously unknown legitimate activity is classified as malicious. 

There are two strategies to anomaly based detection 1) Black Box, referring to training without domain knowledge of the system, often done using neural networks; and 2) White Box, in this case the semantics used by the detection system are an abstraction of the underlying system.
In the contexts of a lot of modern machine learning problems, black box approaches are very popular.
They allow a relatively simple push to go approach, that might find patterns in data or be quite accurate at classifying data, however they provide very little insight into \textit{why} they make decisions.
The problem is that black box detection system face the challenge of transferring their results into actionable reports~\cite{sommer2010outside}. 
Actionability however is a very difficult metric to measure and is not often considered in academic approaches~\cite{etalle2019network}.
In our review only a single paper considered actionability as a metric, Buennemeyer et al. (2008)~\cite{buennemeyer2008mobile}, in which the authors evaluated their IDS using a user study.
The second approach is white box detection, this approach requires much more curating than the black box system.
It requires the building of behaviour profiles either automatically or manually so that a malicious attack may be linked to the constructed profiles.
Whilst this is often impractical in highly complex systems, as we have discussed in previous discussion, the IoT allows for some easier behaviour modelling, greatly facilitating the usage of these methods.

Building system behaviour is an essential part of any effective anomaly detection technique. 
Statistical approaches, which are a subcategory of anomaly detection, are particularly effective in homogeneous systems such as WSN.
A statistical anomaly approach, is a white box approach that relies on knowledge of the devices behaviours.
Parameters are selected and specified for each device~\cite{lee2014lightweight}, or averaged across the network behaviour~\cite{cho2009attack}.
However, approaches using these techniques are prone to large amounts of false positives, in the work of Cho et al. (2009)~\cite{cho2009attack} the authors propose that an anomaly occurs when the packet size is above average, this kind of anomaly detection will obviously lead to issues, if for example the sensing environment changes.
To mitigate the lack of adaptability authors propose dynamic anomaly detection techniques which can dynamically change the expected \textit{normal} behaviour.
Authors using this approach~\cite{besson2009distributed,luo2018distributed}, use a central gateway to monitor system operation and update the expected behaviours of the various IDS components.
Specifically, Luo and Nagarajan (2018)~\cite{luo2018distributed}, make use of autoencoders as their primary detection mechanism, the output of the detection is then sent to a central node, which, given the results recomputes the autoencoder and sends the new version to the local detectors.
Whilst this approach tackled the downside of adaptability it is unable to provide insights like a white box detection would.
These difficulties are further complicated in heterogeneous systems, where white box behaviour is much harder to map.

Not only is it very difficult to have an anomaly detection approach that is both adaptable but also actionable. 
Further difficulties arise when trying to actually collect the behaviour.
Two main behaviours need to be considered, one is at the device level, and one at the system level.
At the device level (often done by HIDS), anomalies are categorised by changes in expected behaviour, which may be classified as insider attacks, device take over, or the device being targeted by an attack.
At the network level the general behaviour is evaluated against the expected operation of the system.
However, some of the characteristics of IoT systems are that, behaviours might change, devices may go offline, and devices might wake up after long periods of time, therefore, getting a baseline behaviour is difficult.
To construct baseline behaviours large amounts of data is needed and due to the lack of public data, researchers are forced to assemble their own datasets. 
However, in general this is not an easy task, as most lack access to appropriately sized networks or if done in production environment it may only collect a window of behaviour; as extensive testing is time consuming and disruptive to normal operations.
In the context of the IoT data collection also needs to be aggregated.
If working across networks and systems, results of analysis need to be transferred across components.

\textit{Advantage:} can detect unknown attacks, dynamic (if new data is collected)~\cite{patcha2007overview,kruegel2003anomaly}

\textit{Disadvantage}: time-consuming during detection proces, suffers from false positives, lack of datasets or realistic testing environments~\cite{tan2002undermining,lundin2000anomaly}.

\subsection{Stateful}

Stateful protocol analysis is the process of comparing predetermined profiles of generally accepted definitions of benign protocol activity for each protocol state against observed events to identify deviations. 
Unlike anomaly based detection, which uses host or network specific profiles, stateful protocol analysis relies on specifications of universal profiles that dictate how 
particular protocols should and should not be used. 
The ``stateful'' in stateful protocol analysis means that the IDS is capable of understanding and tracking the state of network, transport, and application protocols that have a notion of state.
A stateful detection methodology is very similar to signature based approaches, it differs in term of what is being detected.
Signature based approaches are looking for signature of attacks, conversely stateful detection is using the protocol exchange signature as the means of detecting good behaviour.

As most common stateful detection approaches are based on traditional protocols, bespoke stateful approaches are developed for IoT systems.
Further issues however arise in the context of constrained IoT networks as the unreliability of the communication and usage of UDP rather than TCP, may lead to a lot of false positives.
This approach suffers from a lot of the downsides of rule based detection as it is highly non adaptable.
It is likewise very actionable as the relatively simple state based description of the system allows to easily see where the pattern has deviated from expectation.
However, in practice, working from a specification of expected behaviour that doesn't consider implementation specifics has some downsides.
In the field of protocol analysis we see that specs often do not cover implementation specific behaviours or allow for multiple correct behaviours~\cite{arnaboldi2019AceOAuth}, further complicating the detection.
Furthermore, protocol specifications may be prone to attacks, and correct behaviour within the protocol may lead to data leakages or attacks that would not be detected by this approach.
In large IoT multi-protocol IoT systems, protocol specific detection may be too restricted and need for various different detection components, making a general behaviour based approach much more effective.

\textit{Advantage}: Identifies unexpected sequence of commands~\cite{scarfone2007guide,weaver2013guide}

\textit{Disadvantage:} very memory intensive, cannot detect attacks if they are within protocol behaviour, if protocol implemented different to specification it  may cause false positives~\cite{weaver2013guide}.

\subsection{Clustering}

Cluster analysis is defined as the technique of grouping data objects based on the information found in it that describes the objects and their relationships. 
The primary goal of clustering is to separate objects such that there is higher similarity within objects of a group and higher difference between the groups. 
By clustering a device within the systems together with other devices you can make general assessments about it e.g. if the device is clustered with known malicious devices it can be assumed to be malicious.
Clustering is a highly popular technique in constrained systems as their restrictive behaviours makes the cluster creation much easier, as there are few multi purpose devices fitting to multiple clusters.
This technique however is mostly used in WSNs and homogeneous systems~\cite{deng2018mobile}, as clustering heterogeneous systems may be much more complex.

Clustering techniques may extend beyond the clustering of devices.
Clustering techniques are used for message classification and also for efficient data processing.
As there is a huge amount of data collected in IoT systems, and the devices are less able to process it, clustering techniques have been proposed to more efficiently process it~\cite{liu2018intrusion}.
These clustering approaches are used to categorise data into normal behaviours and at risk behaviours as well as to perform dimensionality reduction.
This allows to greatly reduce overhead of the IDSs and allow for faster computation.
These techniques are emerging to deal with the issue of big data, and their approaches make it so that relatively low information loss is achieved whilst greatly reducing the dimensionality.
These techniques re very helpful hand in hand with other anomaly detection techniques as they help improve their efficiency significantly~\cite{liu2018intrusion,gupta2013computational,cervantes2015detection}.

\textit{Advantages:} Works well on static data, can find some very specific attacks~\cite{wang2005clustering,lin2015cann}

\textit{Disadvantages:} Not efficient on dynamic systems, works if data with strong correlations hence struggles with heterogeneous systems~\cite{vora2013survey,chormunge2015efficiency}

\subsection{Computational Intelligence} 

A system is computational intelligent when it: deals with only numerical (low-level) data, has pattern recognition components, does not use knowledge in the artificial intelligence sense; and additionally when it (begins to) exhibit (i) computational adaptivity, (ii) computational fault tolerance, (iii) speed approaching human-like turnaround, and (iv) error rates that approximate human performance~\cite{zurada1995review,craenen2002computational}.
Although there is not yet full agreement on what computational intelligence \textit{exactly} is~\cite{craenen2002computational}, there is a widely accepted view on which areas belong to CI: artificial neural networks, evolutionary computation, artificial immune systems, swarm intelligence, and soft computing. 
These approaches are capable of autonomously acquiring and integrating knowledge, and can be used in either supervised or unsupervised learning mode.
Computational intelligence differs from Artificial intelligence a field widely used in multiple other techniques listed above.

Artificial immune systems are a relatively recent advance in computational intelligence, their self adaptation and robustness makes them a very interesting approach for intrusion detection in the IoT.
Artificial Immune Systems are adaptive systems proposed by Leandro Nunes et al. (2002) ~\cite{castro2002artificial}.
They are inspired by theoretical immunology and observed immune functions, principles and models, which are applied to problem solving, as such they take specific components of the immune systems as inputs.
The main difficulty in the usage of this approach is the matching from network data to the input of the immunity algorithms.
Their ability to self adapt makes them particularly pertinent to systems where configurations might change and attack circumstances might be persistent, however will not be immediately useful for detection.
This approach may be particularly useful in the detection of physical intruders or changes of behaviours from a device.
Arrington et al. (2016)~\cite{arrington2016behavioral}, make use of these techniques to detect human intruders within a smart home, based on sensor data from the room.
The rise of popularity of this technique is in part due to the easier mapping of behaviours to the behaviours of a cell in constrained systems, as Liu et al. (2011)~\cite{liu2011research}, propose attack signature matching to ``antigents'' and normal behaviours to ``self elements'' which is in part possible due to restricted nature of IoT.

Another popular form of computational intelligence is, swarm intelligence.
Swarm intelligence is defined as a collective system capable of accomplishing difficult tasks in dynamic and varied environments, without any external guidance or control, and with no central coordination.
The definition is self explanatory as to why it would be beneficial, however very few approaches have explored its used in the context of intrusion detection.
The specific application of these techniques comes in the form of behaviour discovery.
As these systems are often unsupervised, routing patters are often pre-fixed, or dynamically discovered.
One technique that can be used to find optimum routing paths is Ant Colony Based Optimization (ACO).
ACO uses the concept of Stigmergy, which is the indirect communication via interaction with the environment. 
Each packet (or ant) leaves a pheromone trail along the path if a successful result is reached.
Paths with the most pheromone are considered the optimal paths.
This is of course adaptive as if paths are reconfigured new trails will be discovered as the new optimal.
This technique is used to find routing attacks by Arolkar et al. (2011)~\cite{arolkar2011ant}.
Routing attacks are characterised by malicious routing of packets to either drop them or decrease throughput, so if a large number of packets is being routed on unoptimal paths these attacks are easily detected.

\textit{Advantages:} Good for known attacks, can identify wide range of attacks given enough data, protocol agnostic (given enough data)~\cite{wu2010use}.

\textit{Disadvantages:} This approach relies heavily on pattern matches, whilst attack behaviour often only takes place once and varies significantly in between attacks~\cite{sabhnani2004machine} due to numeric analysis requires lot's of data pre-processing and therefore needs human supervision or doesn't scale well to different datasets.

\section{Evaluation of Intrusion Detection Systems for IoT}
\label{sec:eval-survey}   

The review we have conducted has found that almost all work in the area of intrusion detection focuses strongly on accuracy as means of evaluation.
What we also found is that, despite almost all papers claiming there is a need for new techniques and new approaches, they all present results with very high accuracy of prediction.
These two statements are contradictory, if the only metric of evaluation is that the IDSs are great at detecting attacks, then there would be no need for further IDS being developed.
We acknowledge that some of the proposed solutions aim to tackle very specific areas for which there were no previous existing IDS, however for the majority of the works this is not the case.
The real issue we have found is that to be a successful and more importantly a useful IDS, further characteristics are required than prediction accuracy.

In the remainder of this section we will present the current ways IDS are evaluated, we will discuss why we believe accuracy is not the sole metric of value and propose new evaluation criteria supported by literature and our own investigation.
We then propose a technique to evaluate IDSs for the IoT by these criteria and our case study evaluating the surveyed approaches following this methodology.

\subsection{Methods to Evaluate Intrusion Detection Systems for IoT}

With the increasing variety and complexity of IDSs, the ability to evaluate them becomes a key concern.
In order to ascertain which IDS is most suitable for a specific task a clear evaluation needs to take place, for instance one might wish to see which IDS is most suitable to detect attacks on their system, they might wish to test different configurations of an IDS to inspect performance. 
Whatever the need may be it is key that means are in place to evaluate multiple IDSs, as to allow for informed decisions.
There has been a lot of work in this area as discussed in the survey by  Milenkoski et al.(2015)~\cite{milenkoski2015evaluating}, the main differentiating factors in testing an IDS is whether one were to use a \textit{benign workflow}, a \textit{malicious workflow}, or a \textit{hybrid workflow} and any of these can either be \textit{live} tested or tested using \textit{trace} based approaches. 

A \textit{benign workflow}, is a purely normal system behaviour, these may involve CPU intensive workloads, I/O focused workloads, network intensive workloads and system wide workloads to test the hardware as well as the operating systems. 
Several tools are available for the testing of traditional IDS each with focus on different aspects of workload (e.g. httpbench~\footnote{Available at : \url{http://freecode.com/projects/httpbench}} for network testing using HTTP). 
However these tools may not be suitable for IoT systems as they are either using the wrong network protocols, be operating system dependant or simply not relevant since for example sensors might not do any I/O operations.

A \textit{malicious workflow}, is used to test the attack detection accuracy of IDSs. 
These workflows take the forms of attack scripts targeting the system, they may be manually constructed or already existing tools may be used. 
The manual process is very time consuming as it needs to collect various different attack scripts from many different available databases. 
Once the attack scripts are found, they may need to be adapted to fit the scenario, this may require expert code analysis and security knowledge, this is particularly compounded for scripts aimed at systems to test host based IDSs as tricky configurations may be needed~\cite{mell2003overview}.
The process of manual collection is also limited, not all attack scripts are available online and even though there are some excellent resources (e.g. expoitDB~\footnote{Available at : \url{https://www.exploit-db.com/}} and mitre~\footnote{Available at : \url{http://cve.mitre.org/}}), manually curating a list of of scripts is time consuming, complex and often bespoke to only one scenario.
To alleviate this issue ready made tools are available, these tools, referred to as penetration testing tools, are able to scan for vulnerabilities and launch various pre-set attacks on systems with much less need for expert configuration.
Popular pre-prepared attack environments such as Kali Linux~\footnote{Available at : \url{https://kali.org}} operating system and metasploit~\footnote{Available at : \url{https://metasploit.com}} pen-testing tool (available on Kali Linux) are able to act as \textit{jack of all trades}, able to target a variety of system but not specialising in any specific type or vulnerability.
Whilst this second approach is much less time intensive it is also less specialised.
Both these approaches suffer from the downside that an attacker may use multi step approaches and scout the system prior to attack, this complex behaviour as well as unknown attackers could not be covered by these techniques.
Finally, a \textit{hybrid workflow}, combines these two techniques to test both the system under \textit{benign} and \textit{malicious} behaviour.

There are two ways to examine the different kinds of workflow, the first way, \textit{trace} based, will use existing datasets or synthetically generated data to test the IDS in a non live environment; the second approach, \textit{execution} based, will test the IDS in a system using live attackers and/or data.
The generation of traces may take the form of an existing dataset, several standard datasets exist from a variety of sources, the most famous being the KDD datasets from DARPA~\footnote{Available at : \url{http://kdd.ics.uci.edu/databases/}}.
Despite these datasets being around since 1990/2000/2001 respectively, they are still widely used, they allow to test network based and host based IDSs making them attractive for a variety of research. 
Perhaps the most desirable aspect is that one can easily compare their results to those of other authors.
The downside is that they are quite outdated and would not truly be representative of modern systems~\cite{sommer2010outside}.
Alternatively one may wish to generate bespoke traces. 
One common approach is to use a testbed. 
A testbed has many advantages to get more pertinent traces, however construction of a realistic testbed may be expensive and challenging. 
If a testbed is too simplistic it may not capture activities of a real-life environment~\cite{sommer2010outside}.
Finally, one may use a real world production environments to generate traces. 
This is by far the most attractive option, however, it is often unfeasible.
This approach may also only capture snapshots of system behaviour, as it does not have the full flexibility to run different scenarios like a testbed could.
The second way is \textit{execution} based evaluation which can be run in a testbed or production environment. 
Its main attractiveness over a trace approach is that it is able to capture several more metrics than a trace evaluation.
Metrics can be run for performance, overhead, ability to scale to different devices and many more, making it the more complex but more precise option.
After a workload is determined evaluation should be performed based on metrics. 

Metrics are widely characterised as \textit{security metrics} and \textit{performance metrics}.
Performance metrics refer to how the IDS is able to cope with the inflow of messages and how the IDS impacts the ability for the system to operate effectively.
These may be evaluated on the basis of throughput, delays, computational cost etc.
However, these are difficult to benchmark and may be implementation and system dependent~\cite{milenkoski2015evaluating}.
In our review of 50+ IDSs for the IoT none ran performance based evaluation on execution based environments (although some did so using simulations).
The second and much more common evaluation metric is \textit{security}, these metrics are all about evaluating the IDS attack detection capabilities.
The core evaluations that can be performed are 1) False-negative rate, the percentage of attacks that are missed; 2) True positive, the percentage of correctly labelled attacks; 3) False positive, the percentage of non attacks which are labelled as attacks; 
To get a complete evaluation of accuracy these are all combined. 
Some more complex analysis may attempt to calculate costs of mislabelling (i.e. false negative/positive), based on the impact of the attacks.
As the reader may note, these metrics are exclusively evaluation of attack detection and do not consider several other more human aspects of security that should be considered just as valuable~\cite{etalle2017intrusion}.

An IDS is a complex system whose intent is to provide information to the security professional about potential intrusions to their system.
Naturally it is essential for this system to be accurate, however, several things are also needed to make the IDS \textit{useful}.
Papers usually indicate a detection rate of 90\% and above, using various evaluation methods, what would be more realistic in practice is if these same IDS could get a fraction of those results in realistic environments.
For usability purposed False Positives are the single most disrupting factor as they may cause the security professional to break legitimate workflows, lose time and money chasing non existing attacks and cause more disruption than a false negative could.
Furthermore, Etalle et al. (2017)~\cite{etalle2017intrusion}, propose three further metrics specific to the evaluation of IoT IDS: 1) Actionability, not unique to IoT, actionability refers to how descriptive the alert is, this metric takes one step forward in attack detection as, whilst it is useful to detect an attack, if the alert is without context it is impossible to fix it; 2) Adaptability, very specific to IoT systems, devices change continuously and for certain types of IDS the cost of adapting (such as anomaly based detection) may be very high; and 3) Scalability, a lot of traditional IDS are designed for single servers or small networks, however IoT systems could include thousands of devices and this needs to be a key consideration.
A combination of these metrics together with the evaluation of accuracy would make for a much stronger analysis and lead to better IDS solutions.

\section{Tools for Intrusion Detection in IoT Systems}
\label{sec:idssum}
As an outcome of the differences and challenges inherent with designing IDSs for the IoT, several authors have proposed new tools and mechanisms to defend it.
In the following section we produce an executive review of all new works tackling attack detection in constrained IoT environments since 2008.
We break down the work into network based tools, host based tools and collaborative tools.
An initial analysis and review is presented in Sec.~\ref{sec:summary},the summary of results are presented by means of heatmaps in Tables~\ref{tab:techniques}-~\ref{tab:simulations}. 
A full summary broken down by Network, Host and collaborative is presented in Sec.~\ref{sec:toolNIDS} for NIDS,~\ref{sec:toolHIDS} for HIDS, and Sec.~\ref{sec:toolCIDS} for CIDS, they provide short summaries of each paper in their category.
Summary tables are present in each subsections containing breakdowns of: Algorithms, Placement Strategies, Detection Capabilities and type of Evaluation performed are included - in Tables~\ref{tab:NIDSSum},~\ref{tab:HIDSSum}, and~\ref{tab:CIDSum} respectively.
A full representation of each paper and their techniques is presented in Fig.~\ref{fig:visualisation}.

\begin{amssidewaysfigure}
    \centering
    \includegraphics[width=\textwidth]{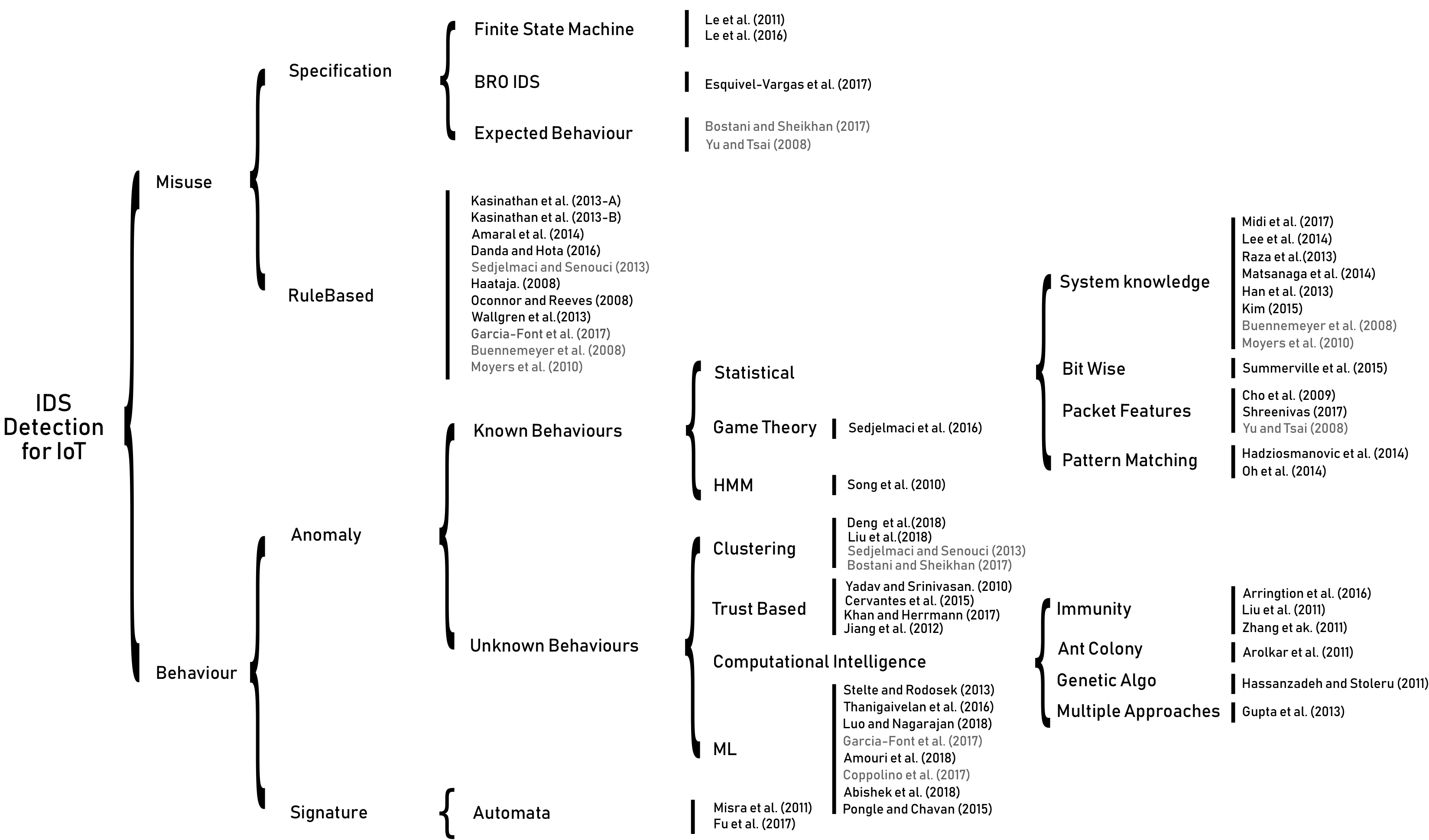}
    \caption{Full representation of the techniques used by different articles. Grey references denote hybrid approaches and the article is present in each of the hybrid techniques.}
    \label{fig:visualisation}
\end{amssidewaysfigure}

\subsection{Analysis and Summary of Proposed Tools}
\label{sec:summary}
So far this article has: 1) presented an overview of deployment strategies for IDSs in the IoT, 2) discussed that a constrained network may take many forms and locations of deployment, 3) presented techniques for the training of an IDS and detection of attacks, and 4) presented an overview attacks that are specifically pertinent to these scenarios.
We use these four core points of interest as the basis of our summary of the literature.
For our review we have surveyed 51 articles from the year 2008 to 2018, specifically looking for articles that propose intrusion detection mechanisms for IoT deployments.
Each collected article is broken down into how the IDS being presented deployed the IDS, what attacks were detected, which techniques were used and what evaluation was performed, in Sec.~\ref{subsec:tech}, Sec.~\ref{subsec:attacks}, Sec.~\ref{subsec:deploy} and Sec.~\ref{subsec:eval} respectively.
In Tables~\ref{tab:techniques}-\ref{tab:simulations}, heatmats break down the papers into categories. 
Heatmaps can be interpreted as follows: the tables are split into four columns, the first columns contains list of criteria evaluated, the subsequent three columns contain all the papers using that criteria divided by Network IDSs (col 2) Host IDSs (col 3) and Collaborative IDSs (col 4), colours represent the amount of papers who meet the criteria and a legend is presented underneath the tables.

\subsubsection{Breakdown of Techniques used}
\label{subsec:tech}
The surveyed papers used twelve unique detection techniques, across the different deployment types as show in in Table~\ref{tab:techniques}.

\begin{table}[htbp!]
    \centering
      \small
       \begin{tabular}{|cccc|}
        \cline{0-3}
        & NIDS & HIDS & CIDS\\
        \cline{0-3}
        \multicolumn{4}{c}{} \\
        \cline{0-3}
        Rule based  & \ApplyGradientTrick{3}{\cite{danda2016attack,haataja2008new,garcia2009anomaly}}{0}{7} & \ApplyGradientTrick{0}{}{0}{7} & \ApplyGradientTrick{2}{\cite{kasinathan2013denial,kasinathan2013ids}}{0}{7} \\
        Signature   & \ApplyGradientTrick{1}{\cite{sedjelmaci2013efficient}}{0}{7}& \ApplyGradientTrick{0}{}{0}{7}  &\ApplyGradientTrick{1}{~\cite{amaral2014policy}}{0}{7}  \\
        Anomaly     & \ApplyGradientTrick{7}{\textbf{A}}{0}{7} & \ApplyGradientTrick{2}{~\cite{summerville2015ultra,lee2014lightweight}}{0}{7} &\ApplyGradientTrick{5}{\textbf{B}}{0}{7} \\
        Statistical & \ApplyGradientTrick{3}{~\cite{cho2009attack,raza2013svelte,matsunaga2014low}}{0}{7} & \ApplyGradientTrick{3}{~\cite{kim2015physical,song2010weak,oh2014malicious}}{0}{7} & \ApplyGradientTrick{2}{~\cite{buennemeyer2008mobile,moyers2010multi}}{0}{7}  \\
        Stateful    & \ApplyGradientTrick{1}{~\cite{fu2017automata}}{0}{7} & \ApplyGradientTrick{0}{}{0}{7}& \ApplyGradientTrick{0}{}{0}{7}  \\
        Clustering  & \ApplyGradientTrick{3}{~\cite{deng2018mobile,liu2018intrusion,jiang2012dynamic}}{0}{7} & \ApplyGradientTrick{0}{}{0}{7}& \ApplyGradientTrick{0}{}{0}{7} \\
        CI          & \ApplyGradientTrick{1}{~\cite{arrington2016behavioral}}{0}{7} & \ApplyGradientTrick{1}{~\cite{liu2011research}}{0}{7} & \ApplyGradientTrick{4}{\textbf{C}}{0}{7} \\
        Specification & \ApplyGradientTrick{2}{~\cite{le2011specification,esquivel2017automatic}}{0}{7} & \ApplyGradientTrick{0}{}{0}{7} &\ApplyGradientTrick{2}{~\cite{yu2008framework,le2016specification}}{0}{7} \\
        Trust       & \ApplyGradientTrick{3}{~\cite{cervantes2015detection,khan2017trust,yadav2010itrust}}{0}{7} & \ApplyGradientTrick{0}{}{0}{7} &\ApplyGradientTrick{0}{}{0}{7}\\
        Automata    & \ApplyGradientTrick{2}{~\cite{misra2011learning,fu2017automata}}{0}{7}& \ApplyGradientTrick{0}{}{0}{7} &\ApplyGradientTrick{0}{}{0}{7} \\
        Game theory & \ApplyGradientTrick{1}{~\cite{sedjelmaci2016lightweight}}{0}{7} & \ApplyGradientTrick{0}{}{0}{7} &\ApplyGradientTrick{0}{}{0}{7}\\
        Misuse      & \ApplyGradientTrick{1}{~\cite{oconnor2008bluetooth}}{0}{7} & \ApplyGradientTrick{0}{}{0}{7} &\ApplyGradientTrick{1}{~\cite{coppolino2013applying}}{0}{7}\\
        \cline{0-3}
        \multicolumn{4}{c}{} \\
        \cline{0-3}
        Citations &\multicolumn{3}{l|}{
                    \begin{tabular}[c]{@{}l@{}}
                      \textbf{A} -~\cite{stelte2013thwarting,hadvziosmanovic2014through,thanigaivelan2016distributed,luo2018distributed,wallgren2013routing,garcia2009anomaly,han2013idsep}\\
                      \textbf{B}  -~\cite{shreenivas2017intrusion,pongle2015real,abhishek2018intrusion,amouri2018cross,yu2008framework}\\
                      \textbf{C} -~\cite{gupta2013computational,hassanzadeh2011towards,arolkar2011ant,zhang2011artificial}\\
        
                    \end{tabular}}\\\cline{0-3}
        \multicolumn{4}{c}{} \\
        \multicolumn{4}{c}{
        \begin{tabular}{ccccccccc}
           key  & \ApplyGradient{1}{0}{7} & \ApplyGradient{2}{0}{7} &\ApplyGradient{3}{0}{7} & \ApplyGradient{4}{0}{7}& \ApplyGradient{5}{0}{7}&\ApplyGradient{6}{0}{7} &\ApplyGradient{7}{0}{7}
        \end{tabular}
        }
        \end{tabular}
        \caption{Techniques Used to construct Intrusion Detection Systems in IoT}
        \label{tab:techniques}
\end{table}

The technique employed the most was by far anomaly based detection.
This is aligned with our presented hypothesis regarding it being difficult to determine rules or other strict specifications for IoT deployments.
This statistic is slightly skewed by the fact that unlike some other techniques, anomaly based detection might encompass a wide variety of approaches.
Another interesting thing to note, is that whilst anomaly is by far the most popular technique no work using this technique has been implemented on live environments, results mostly focusing on simulations and detection accuracy as the core metric. 
We hypothesise that this might be an outcome of the machine learning community being the main driving factor behind these IDSs, a field in which accuracy is often used as the main metric of interest.
Similarly to anomaly detection statistical techniques have also shown to be quite popular.
These technique whilst similar to anomaly detection, relies more domain specific knowledge and is therefore more suited to simpler systems, it is widely used in WSN systems.
Perhaps unexpectedly, Computational Intelligence (CI), a relatively new, and non traditional method has come in close third.
We note that this result is biased by our review being focused on somewhat recent literature with the earliest work from 2008, which unintentionally coincides with the first use of Immunity Based techniques (a CI technique) for intrusion detection~\cite{liu2008immunity}.
This technique is highly adaptable a strength which favours IoT deployments.
Its relative novelty also incites investigation in its feasibility in this area leading to more works being published.
Since the technique has been used in many tools since, as recently as 2018, its safe to say its been a good match.
Specification, Rule Based and to a lesser extent signature based techniques have come as clear favourites behind the discussed top three.
This is aligned with IDSs in more traditional systems and comes at no shock.
They are however more popular in simple IoT deployments, as they tend to perform poorly in dynamic environments. 
Making them less suitable for general IoT usage.
One further trend that is the lack of popularity of misuse detection.
Whilst misuse detection suffers from several downsides (see Sec.~\ref{sec:idstech}), it is vastly popular in commercial IDSs for traditional systems (in part due to its actionability), so it being the second least popular technique in academic literature is perhaps unexpected.

The very nature of how these systems are constructed, such as clustered networks, WSNs, and massively distributed systems, leads to the development of more novel techniques.
Techniques such as clustering, automatons, trust based and to some extent game theory are naturally more suited to these kinds of infrastructures and have consequently attracted attention from literature.
Due to the limited behaviours of IoT sensors clustering based technique have become more prominent especially in network based systems, with three papers.
Another new technique that is made possible largely in part to the fact that these systems operate unsupervised and interact machine to machine only is trust based detection.
Similarly taking advantage of these restricted behaviours two articles have proposed the use of Automaton based techniques, drawing from the many advantages of modelling but greatly reducing the complexity of the systems modelled (and therefore reducing state based issues).
One of the massive changes that have taken place as an outcome of the shift to the IoT is the threats that need to be detected.

\subsubsection{Breakdown of Attacks Detected}
\label{subsec:attacks}

In Table~\ref{tab:attacks-detec}, we see that papers are catered to detect 14 unique attacks.

\begin{table}[!htbp]
    \centering
    \small
    \begin{tabular}{|cccc|}
            \cline{0-3}
            & NIDS & HIDS & CIDS\\
            \cline{0-3}
            \multicolumn{4}{c}{} \\
            \cline{0-3}
            Scanning             & \ApplyGradientTrick{1}{~\cite{oconnor2008bluetooth}}{0}{11} & \ApplyGradientTrick{0}{}{0}{11}    & \ApplyGradientTrick{1}{~\cite{hassanzadeh2011towards}}{0}{11}   \\
            Web Exploits         
            & \ApplyGradientTrick{0}{}{0}{11}   & \ApplyGradientTrick{0}{}{0}{11}    & \ApplyGradientTrick{1}{~\cite{hassanzadeh2011towards}}{0}{11}   \\
            Routing Attacks      
            & \ApplyGradientTrick{11}{\textbf{A}}{0}{11}   & \ApplyGradientTrick{0}{}{0}{11}    & \ApplyGradientTrick{6}{\textbf{B}}{0}{11} \\
            Rank    
            & \ApplyGradientTrick{0}{}{0}{11}   & \ApplyGradientTrick{0}{}{0}{11}    & \ApplyGradientTrick{0}{}{0}{11}\\
            Information Theft    
            & \ApplyGradientTrick{1}{~\cite{oconnor2008bluetooth}}{0}{11}   & \ApplyGradientTrick{0}{}{0}{11}    & \ApplyGradientTrick{3}{~\cite{zhang2011artificial,arolkar2011ant,gupta2013computational}}{0}{11} \\
            MITM                
            & \ApplyGradientTrick{3}{~\cite{fu2017automata,esquivel2017automatic,oconnor2008bluetooth}}{0}{11}   
            & \ApplyGradientTrick{1}{~\cite{kim2015physical}}{0}{11}    & \ApplyGradientTrick{3}{~\cite{gupta2013computational,zhang2011artificial,arolkar2011ant}}{0}{11} \\
            Replay              
            & \ApplyGradientTrick{2}{~\cite{fu2017automata,esquivel2017automatic}}{0}{11}   & \ApplyGradientTrick{1}{~\cite{kim2015physical}}{0}{11}    & \ApplyGradientTrick{2}{~\cite{zhang2011artificial,gupta2013computational}}{0}{11}\\
            Spoofing             & \ApplyGradientTrick{3}{~\cite{fu2017automata,esquivel2017automatic}}{0}{11}   & \ApplyGradientTrick{1}{~\cite{kim2015physical}}{0}{11}    & \ApplyGradientTrick{2}{~\cite{zhang2011artificial,gupta2013computational}}{0}{11}\\
            Message Drop         & \ApplyGradientTrick{2}{~\cite{fu2017automata,esquivel2017automatic}}{0}{11}   & \ApplyGradientTrick{0}{}{0}{11}    & \ApplyGradientTrick{3}{~\cite{zhang2011artificial,gupta2013computational,abhishek2018intrusion}}{0}{11}\\
            DoS (resources)      
            & \ApplyGradientTrick{3}{~\cite{sedjelmaci2016lightweight,han2013idsep,haataja2008new}}{0}{11}   
            & \ApplyGradientTrick{1}{~\cite{lee2014lightweight}}{0}{11}    & \ApplyGradientTrick{7}{\textbf{C}}{0}{11}\\
            DDoS                
            & \ApplyGradientTrick{2}{~\cite{misra2011learning,cho2009attack}}{0}{11}   & \ApplyGradientTrick{1}{~\cite{oh2014malicious}}{0}{11}    & \ApplyGradientTrick{0}{}{0}{11} \\
            Worm                 & \ApplyGradientTrick{0}{}{0}{11}   & \ApplyGradientTrick{1}{~\cite{summerville2015ultra}}{0}{11}    & \ApplyGradientTrick{0}{}{0}{11}\\
            Injection            & \ApplyGradientTrick{0}{}{0}{11}   & \ApplyGradientTrick{1}{~\cite{summerville2015ultra}}{0}{11}    & \ApplyGradientTrick{0}{}{0}{11} \\
            Anomalous Behaviour  & \ApplyGradientTrick{3}{~\cite{thanigaivelan2016distributed,luo2018distributed,arrington2016behavioral}}{0}{11}   & \ApplyGradientTrick{0}{}{0}{11}    & \ApplyGradientTrick{1}{~\cite{shreenivas2017intrusion}}{0}{11}\\
            Cracking cred/passwd & \ApplyGradientTrick{1}{~\cite{yadav2010itrust}}{0}{11}   & \ApplyGradientTrick{0}{}{0}{11}    & \ApplyGradientTrick{0}{}{0}{11}\\
            KillerBee            & \ApplyGradientTrick{1}{~\cite{stelte2013thwarting}}{0}{11}   & \ApplyGradientTrick{0}{}{0}{11}    & \ApplyGradientTrick{0}{}{0}{11}\\
                \cline{0-3}
                \multicolumn{4}{c}{} \\
                \cline{0-3}
                Citations &\multicolumn{3}{l|}{
                            \begin{tabular}[c]{@{}l@{}}
                              \textbf{A} -~\cite{deng2018mobile,sedjelmaci2013efficient,wallgren2013routing,yu2008framework,raza2013svelte,matsunaga2014low,cervantes2015detection,garcia2009anomaly,khan2017trust,jiang2012dynamic,le2011specification}\\
                              \textbf{B}  -~\cite{le2016specification,bostani2017hybrid,arolkar2011ant,coppolino2013applying,yu2008framework,pongle2015real}\\
                              \textbf{C} -~\cite{kasinathan2013denial,kasinathan2013ids,midi2017kalis,buennemeyer2007battery,moyers2010multi,zhang2011artificial,gupta2013computational}\\
                            \end{tabular}}\\\cline{0-3}
                 \multicolumn{4}{c}{} \\
                \multicolumn{4}{c}{
                \begin{tabular}{ccccccccccccc}
                   key  & \ApplyGradient{1}{0}{11} & \ApplyGradient{2}{0}{11} 
                   &\ApplyGradient{3}{0}{11} & \ApplyGradient{4}{0}{11}
                   &\ApplyGradient{5}{0}{11} &\ApplyGradient{6}{0}{11}
                   &\ApplyGradient{7}{0}{11}
                   &\ApplyGradient{8}{0}{11}
                   &\ApplyGradient{9}{0}{11}
                   &\ApplyGradient{10}{0}{11}
                   &\ApplyGradient{11}{0}{11}
                \end{tabular}
                }
                \end{tabular}
                \caption{Attacks Detected by IoT IDSs}
    \label{tab:attacks-detec}
\end{table}

We also observe that 20\% of the literature focuses solely on the detection of routing attacks.
These attacks (including: blackhole, selective forwarding attacks, sinkhole attacks, Sybil attacks, wormhole attacks etc.) have risen in popularity due to their effectiveness against WSN networks and are a huge threat to the IoT.
As interconnectedness becomes more the norm, these attacks disrupt the ability for a system to achieve its goal greatly damaging the ability for it to function.
Another prominent threat across any internet infrastructure is DoS and DDoS attacks, this is even more so exacerbated as devices suffer from lack of resources and security measures to mitigate them.
Similarly, the combination of protocol related attacks such as MITM, Replay, Spoofing, Message Drop and Information Theft is a hot topic for many papers.
This is in contrast with the techniques in use as there is relatively little focus on techniques specific to these threats, such as stateful detection.
The general focus on research in the behaviour of these systems seems to instead have left secure protocol techniques behind, making attacks on these a huge issue.
Another rising trend is the detection of attacks bespoke to the specific system, such as IDSs specific for KillerBee a Zigbee vulnerability suite, and rank attacks which are specific to hierarchical RPL networks.

\subsubsection{Breakdown of Deployment Scenarios Covered}
\label{subsec:deploy}

In Table~\ref{tab:deploy}, we see IDSs are designed for 15 unique deployment scenarios.
As the definition of IoT is quite general it is very difficult to design an IDS solution that can bw suitable to its full scope.
However the majority of IDSs make this claim with 14 different papers designing general IoT IDS solutions.
Closely following these numbers are WSNs with 13 solutions, WSNs are much less abstract than general IoT systems, however they share a lot of constrains around computation power and connectivity.
It is however much easier to reason about these systems as they are homogeneous allowing for much easier behaviour analysis.

A raising area of focus is RPL, a protocol designed for constrained devices which has risen in popularity. 
This raise in popularity is in part due to the large number of new attacks on these systems, making IDS research for these very appealing and necessary.
Similarly relatively new technologies IPv6 and s6LowPan (a IPv6 subset for constrained devices) are focused by 7 papers.
What is exciting is the sheer range of different systems covered by these tools.
IDSs have been proposed for a variety of technologies including emerging paradigms such as Smart Cities, Smart Grid and even Healthcare.
This shows trends to design security for systems that may not even be a reality yet, and most definitely are not mainstream.
As the IoT will move to encompass large swathes of our lives its positive to see some security solutions are already being thought out for these specific circumstances.

\begin{table}[!htbp]
    \centering
   
    \begin{tabular}{|cccc|}
    \cline{0-3}
    & NIDS & HIDS & CIDS\\
    \cline{0-3}
    \multicolumn{4}{c}{} \\
    \cline{0-3}
    WSN         
    & \ApplyGradientTrick{6}{\textbf{A}}{0}{6}  & \ApplyGradientTrick{1}{\cite{song2010weak}}{0}{6}    & \ApplyGradientTrick{5}{\textbf{B}}{0}{6}\\
    IPv6        & \ApplyGradientTrick{1}{\cite{raza2013svelte}}{0}{6}  & \ApplyGradientTrick{0}{}{0}{6}           & \ApplyGradientTrick{1}{\cite{amaral2014policy}}{0}{6}\\
    6LowPan  
    & \ApplyGradientTrick{2}{\cite{cho2009attack,cervantes2015detection}}{0}{6}  & \ApplyGradientTrick{1}{\cite{lee2014lightweight}}{0}{6}    & \ApplyGradientTrick{2}{\cite{kasinathan2013denial,kasinathan2013ids}}{0}{6}\\
    IoT         & \ApplyGradientTrick{5}{\textbf{C}}{0}{6}  & \ApplyGradientTrick{3}{\cite{oh2014malicious,summerville2015ultra,liu2011research}}{0}{6}    & \ApplyGradientTrick{6}{\textbf{D}}{0}{6}\\
    RPL         & \ApplyGradientTrick{4}{\textbf{E}}{0}{6}  & \ApplyGradientTrick{0}{}{0}{6}           & \ApplyGradientTrick{3}{\cite{le2016specification,shreenivas2017intrusion,pongle2015real}}{0}{6}\\               
    SmartGrid   & \ApplyGradientTrick{0}{}{0}{6}         & \ApplyGradientTrick{0}{}{0}{6}           & \ApplyGradientTrick{1}{\cite{zhang2011artificial}}{0}{6}\\
    Relay Comm & \ApplyGradientTrick{0}{}{0}{6}         & \ApplyGradientTrick{1}{\cite{kim2015physical}}{0}{6}    & \ApplyGradientTrick{0}{}{0}{6}\\
    Smart Home  & \ApplyGradientTrick{1}{\cite{arrington2016behavioral}}{0}{6}  & \ApplyGradientTrick{0}{}{0}{6}           & \ApplyGradientTrick{0}{}{0}{6} \\
    ZigBee      & \ApplyGradientTrick{1}{\cite{stelte2013thwarting}}{0}{6}  & \ApplyGradientTrick{0}{}{0}{6}           & \ApplyGradientTrick{0}{}{0}{6} \\
    ICS         & \ApplyGradientTrick{1}{\cite{hadvziosmanovic2014through}}{0}{6}  & \ApplyGradientTrick{0}{}{0}{6}           & \ApplyGradientTrick{0}{}{0}{6}\\
    Bluetooth   & \ApplyGradientTrick{2}{\cite{haataja2008new,oconnor2008bluetooth}}{0}{6}  & \ApplyGradientTrick{0}{}{0}{6}           & \ApplyGradientTrick{0}{}{0}{6}\\
    Smart City  & \ApplyGradientTrick{1}{\cite{garcia2009anomaly}}{0}{6}  & \ApplyGradientTrick{0}{}{0}{6}           & \ApplyGradientTrick{0}{}{0}{6}\\
    Clustered   & \ApplyGradientTrick{1}{\cite{jiang2012dynamic}}{0}{6}  & \ApplyGradientTrick{0}{}{0}{6}           & \ApplyGradientTrick{0}{}{0}{6}\\
    BACNet      & \ApplyGradientTrick{1}{\cite{esquivel2017automatic}}{0}{6}  & \ApplyGradientTrick{0}{}{0}{6}           & \ApplyGradientTrick{0}{}{0}{6}\\
    Healthcare  & \ApplyGradientTrick{1}{\cite{khan2017trust}}{0}{6}  & \ApplyGradientTrick{0}{}{0}{6}           & \ApplyGradientTrick{0}{}{0}{6}\\
    \cline{0-3}
    \multicolumn{4}{c}{} \\
    \cline{0-3}
        Citations &\multicolumn{3}{l|}{
                 \begin{tabular}[c]{@{}l@{}}
                   \textbf{A} -~\cite{besson2009distributed,deng2018mobile,sedjelmaci2013efficient,yadav2010itrust,luo2018distributed,han2013idsep}\\
                   \textbf{B} -~\cite{bostani2017hybrid,gupta2013computational,arolkar2011ant,coppolino2013applying,yu2008framework}\\
                   \textbf{C}  -~\cite{liu2011research,sedjelmaci2016lightweight,misra2011learning,danda2016attack,fu2017automata}\\
                   \textbf{D} -~\cite{amouri2018cross,hassanzadeh2011towards,midi2017kalis,abhishek2018intrusion,buennemeyer2008mobile,moyers2010multi}\\
                  \textbf{E}-~\cite{le2011specification,thanigaivelan2016distributed,wallgren2013routing,matsunaga2014low}\\
                 \end{tabular}}\\\cline{0-3}
                 \multicolumn{4}{c}{} \\
                \multicolumn{4}{c}{
                \begin{tabular}{ccccccc}
                   key  & \ApplyGradient{1}{0}{6} & \ApplyGradient{2}{0}{6} 
                   &\ApplyGradient{3}{0}{6} & \ApplyGradient{4}{0}{6}
                   &\ApplyGradient{5}{0}{6} &\ApplyGradient{6}{0}{6}
                \end{tabular}
                }
     \end{tabular}
     \caption{Locations for Deployment of IDSs}
    \label{tab:deploy}
\end{table}

\subsubsection{Methods for Evaluation of IDS}
\label{subsec:eval}

In Table~\ref{tab:simulations}-left, we show the breakdown of evaluations used.
Out of the 51 surveyed tools, 12 authors didn't evaluate their IDS in any way. 
This was perhaps unexpected as when a paper is presenting a novel approach to detect attacks a lack of evaluation makes it hard to assess its effectiveness.
Although several different deployment areas are explored several authors made attempts to deploy live IDSs for evaluation, which is particularly powerful.
Whilst this technique can give the most insight in evaluation, it makes it very difficult to compare as it is unlikely other researchers will have a access to the same setup.
A more reusable approach is the use of trace evaluation which relies on datasets from real systems.
A dataset can be shared and is therefore useful to compare different IDSs performances.
It is worth noting that the IDSs may not always be comparable as some are very specific to some subsystems and protocols and therefore effective comparisons are unlikely.
The most common way by far to evaluate was the use of simulation tools, this comes as no surprise as the maturity of these tools has significantly escalated with several options available, either more general or system specific e.g. WSN~\cite{minakov2016comparative}.
Simulation tools  have the advantage of accessibility over real systems, and allow for comparative quick deployment.

A total of 24 tools were evaluated using simulators, a further breakdown of which tools are used by the different papers in Table~\ref{tab:simulations}-right.
The most common simulator used was Contiki/Cooja, a popular IoT simulation tool that focuses on communication and allows for design of various different setups.
Perhaps unsurprisingly this is closely followed by Matlab, a general purpose programming language very popular in various engineering fields for simulation.
Various other tools are used, with specific focus on Network Simulators, this is due to the IDSs being assessed being NIDS.
Interestingly it seems that other types of IDSs have many less tools available.
Contiki is used across the different approaches as it is a  multi purpose simulator for IoT, more analysis would have to be done to see if the same evaluation could be performed on other tools or if Cooja is the only option.
Likewise the fact most of these tools are network layer tools (although not all), means that several characteristics of the devices and the impact of the IDS and attacks cannot be evaluated.

\begin{table}
\parbox{.45\linewidth}{
\centering
\begin{tabular}{|cccc|}
        \cline{0-3}
        & NIDS & HIDS & CIDS\\
        \cline{0-3}
        \multicolumn{4}{c}{}  \\
        \cline{0-3}
None           & \ApplyGradientTrick{8}{\textbf{A}}{0}{13}    & \ApplyGradientTrick{0}{\cite{}}{0}{13}    & \ApplyGradientTrick{5}{\textbf{B}}{0}{13} \\
Mathematical   & \ApplyGradientTrick{3}{\cite{liu2008immunity,hadvziosmanovic2014through,han2013idsep}}{0}{13}    & \ApplyGradientTrick{3}{\cite{oh2014malicious,liu2008immunity,kim2015physical}}{0}{13}    & \ApplyGradientTrick{0}{\cite{}}{0}{13}\\
Simulation     & \ApplyGradientTrick{13}{\textbf{C}}{0}{13}   & \ApplyGradientTrick{2}{\cite{lee2014lightweight,song2010weak}}{0}{13}    & \ApplyGradientTrick{7}{\textbf{D}}{0}{13}\\
Trace          & \ApplyGradientTrick{4}{\cite{deng2018mobile,hadvziosmanovic2014through,luo2018distributed,esquivel2017automatic}}{0}{13}    & \ApplyGradientTrick{1}{\cite{oh2014malicious}}{0}{13}    & \ApplyGradientTrick{2}{\cite{coppolino2013applying,zhang2011artificial}}{0}{13} \\
Execution & \ApplyGradientTrick{3}{\cite{oconnor2008bluetooth,misra2011learning,esquivel2017automatic}}{0}{13}    & \ApplyGradientTrick{1}{\cite{summerville2015ultra}}{0}{13}    & \ApplyGradientTrick{6}{\textbf{E}}{0}{13} \\
Usability      & \ApplyGradientTrick{0}{}{0}{13}    & \ApplyGradientTrick{0}{}{0}{13}    & \ApplyGradientTrick{1}{\cite{buennemeyer2008mobile}}{0}{13} \\
\cline{0-3}
    \multicolumn{4}{c}{} \\
    \cline{0-3}
        Citations &\multicolumn{3}{l|}{
                 \begin{tabular}[c]{@{}l@{}}
                   \textbf{A} -~\cite{besson2009distributed,danda2016attack,le2011specification,haataja2008new,thanigaivelan2016distributed,wallgren2013routing,fu2017automata,jiang2012dynamic}\\
                   \textbf{B} -~\cite{gupta2013computational,amaral2014policy,arolkar2011ant,yu2008framework,moyers2010multi}\\
                   \textbf{D} -~\cite{amouri2018cross,hassanzadeh2011towards,le2016specification,shreenivas2017intrusion,bostani2017hybrid,abhishek2018intrusion,pongle2015real,zhang2011artificial}\\
                   \textbf{C}  -~\cite{arrington2016behavioral,cho2009attack,deng2018mobile,stelte2013thwarting,sedjelmaci2013efficient,sedjelmaci2016lightweight,yadav2010itrust}\\
                  \cite{raza2013svelte,matsunaga2014low,cervantes2015detection,garcia2017attack,khan2017trust,han2013idsep,wallgren2013routing}\\
                   \textbf{E} -~\cite{kasinathan2013denial,kasinathan2013ids,hassanzadeh2011towards,bostani2017hybrid,buennemeyer2008mobile,midi2017kalis}\\
                 \end{tabular}}\\\cline{0-3}
                 \multicolumn{4}{c}{} \\
                \multicolumn{4}{c}{
                \begin{tabular}{ccccccccc}
                   key  &  \ApplyGradient{2}{0}{13} 
                   & \ApplyGradient{4}{0}{13}
                    &\ApplyGradient{6}{0}{13}
                   &\ApplyGradient{8}{0}{13}
                   &\ApplyGradient{10}{0}{13}
                   &\ApplyGradient{12}{0}{13}
                   &\ApplyGradient{13}{0}{13}
                \end{tabular}
                }
     \end{tabular}
\caption*{Evaluation Performed on IoT IDSs}
}
\hfill
\parbox{.45\linewidth}{
\centering
\begin{tabular}{|cccc|}
        \cline{0-3}
        & NIDS & HIDS & CIDS\\
        \cline{0-3}
        \multicolumn{4}{c}{} \\
        \cline{0-3}
Contiki/Cooja & \ApplyGradientTrick{4}{\textbf{A}}{0}{5}    & \ApplyGradientTrick{0}{\cite{}}{0}{5}                      & \ApplyGradientTrick{4}{\textbf{B}}{0}{5}\\
Matlab        & \ApplyGradientTrick{2}{\cite{deng2018mobile,khan2017trust,zhang2011artificial}}{0}{5}                              & \ApplyGradientTrick{1}{\cite{song2010weak}}{0}{5}          & \ApplyGradientTrick{5}{\textbf{C}}{0}{5}\\
R             & \ApplyGradientTrick{1}{\cite{garcia2017attack}}{0}{5}                                                              & \ApplyGradientTrick{0}{\cite{}}{0}{5}                      & \ApplyGradientTrick{0}{\cite{}}{0}{5} \\
Avrora        & \ApplyGradientTrick{1}{\cite{stelte2013thwarting}}{0}{5}                                                           & \ApplyGradientTrick{0}{\cite{}}{0}{5}                      & \ApplyGradientTrick{0}{\cite{}}{0}{5}\\
Tossim        & \ApplyGradientTrick{2}{\cite{sedjelmaci2013efficient,sedjelmaci2016lightweight}}{0}{5}                             & \ApplyGradientTrick{0}{\cite{}}{0}{5}                      & \ApplyGradientTrick{0}{\cite{}}{0}{5}\\
OpenSim       & \ApplyGradientTrick{1}{\cite{arrington2016behavioral}}{0}{5}                                                       & \ApplyGradientTrick{0}{\cite{}}{0}{5}                      & \ApplyGradientTrick{0}{\cite{}}{0}{5}\\
Omnet++       & \ApplyGradientTrick{1}{\cite{yadav2010itrust}}{0}{5}                                                               & \ApplyGradientTrick{0}{\cite{}}{0}{5}                      & \ApplyGradientTrick{0}{\cite{}}{0}{5}\\
NS2           & \ApplyGradientTrick{1}{\cite{han2013idsep}}{0}{5}                                                                  & \ApplyGradientTrick{0}{\cite{}}{0}{5}                      & \ApplyGradientTrick{0}{\cite{}}{0}{5}\\
Qualnet       & \ApplyGradientTrick{0}{\cite{}}{0}{5}                                                                              & \ApplyGradientTrick{1}{\cite{lee2014lightweight}}{0}{5}    & \ApplyGradientTrick{0}{\cite{}}{0}{5}\\
Not Specified & \ApplyGradientTrick{1}{\cite{cho2009attack}}{0}{5}                                                                 & \ApplyGradientTrick{0}{\cite{}}{0}{5}                      & \ApplyGradientTrick{0}{\cite{}}{0}{5}\\
        \cline{0-3}
        \multicolumn{4}{c}{} \\
        \cline{0-3}
        Citations &\multicolumn{3}{l|}{
                    \begin{tabular}[c]{@{}l@{}}
                      \textbf{A} -~\cite{wallgren2013routing,raza2013svelte,matsunaga2014low,cervantes2015detection}\\
                      \textbf{B} -\cite{amouri2018cross,le2016specification,shreenivas2017intrusion,pongle2015real}\\
                      \textbf{C} -\cite{amouri2018cross,hassanzadeh2011towards,bostani2017hybrid,abhishek2018intrusion}\\
                    \end{tabular}}\\\cline{0-3}
                    
                 \multicolumn{4}{c}{} \\
                \multicolumn{4}{c}{
                \begin{tabular}{cccccc}
                   key  & \ApplyGradient{1}{0}{5} & \ApplyGradient{2}{0}{5} 
                   &\ApplyGradient{3}{0}{5} & \ApplyGradient{4}{0}{5}
                   &\ApplyGradient{5}{0}{5} 
                \end{tabular}
                }
        \end{tabular}
\caption*{Summary of simulators used to evaluate IDSs}
}
        \caption{IDS Evaluation and summary of used simulation tools}
    \label{tab:simulations}
\end{table}

\subsection{Proposed Network Intrusion Detection Systems for IoT}
\label{sec:toolNIDS}
Network intrusion detection has gained a lot of traction in the context of IoT as the distributed nature, its heterogeneity and the amount of new protocols raise several new challenges. 
In this section we summarise the current state of NIDS in the IoT and provide descriptions of each paper. 
The full analysis summary is presented in Table~\ref{tab:NIDSSum}.

\begin{table}[htbp!]
\caption{Summary of Network Intrusion Detection Systems Proposed for IoT}
\begin{tabular}{|l|c|c|c|c|c|}
\hline
Paper & Placement & Algorithm & Env. & Eval. & Detection \\
\hline
\multicolumn{6}{c}{} \\
\hline
Arrington et al.~\cite{arrington2016behavioral}& central  &  immunity  & smart-home & $\circ$      & internal$^2$     \\
Besson et al.~\cite{besson2009distributed} & distributed & - & WSN & $\Crossedbox$ & - \\
Cho et al.~\cite{cho2009attack}&   central   &  statistical    &   6LowPAN &   $\circ$   & DDoS     \\
Danda et al.~\cite{danda2016attack}& central  &  rule-based    &   testbed    &  $\Crossedbox$    &  -\\
Deng et al ~\cite{deng2018mobile}& -     & clustering     & WSN      & $\circ,\delta$     &  Sybil    \\
Le et al.~\cite{le2011specification}  & distributed &  specification & RPL  & $\Crossedbox$ & topology \\
Liu et al~\cite{liu2018intrusion}   & - &  clustering   & IoT &  $\Sigma$   &  -    \\
Stelte et al.~\cite{stelte2013thwarting} &  distributed &  anomaly & ZigBee & $\circ$ & KillerBee   \\
Had{\v{z}}iosmanovi{\'c} et al.~\cite{hadvziosmanovic2014through}& - & anomaly   &  ICS    &  $\Sigma,\delta$    &  -    \\
Sedjelmaci at al.~\cite{sedjelmaci2013efficient}& central & signature &   WSN    &   $\circ$   &  routing$^3$    \\
Sedjelmaci et al.~\cite{sedjelmaci2016lightweight}&   central   & game-theory    &  IoT     &  $\circ$    &  DoS    \\
Haataja et al.~\cite{haataja2008new}&   central   &  rule-based    &  Bluetooth    &  $\Crossedbox$    &  $\beta^4$    \\
Yadav et al.~\cite{yadav2010itrust}&  distributed    &   trust-based   &  WSN    &  $\circ$    & $\beta^5$    \\
OConnor et al.~\cite{oconnor2008bluetooth}& central  & misuse & Bluetooth & $\checkmark$ &  $\beta^6$    \\
Thanigaivelan et al.~\cite{thanigaivelan2016distributed}& distributed     &   anomaly   &RPL&   $\Crossedbox$   &   internal$^2$   \\
Luo et al.~\cite{luo2018distributed}&  distributed  & anomaly  & WSN      & $\delta$     & internal$^2$ \\
Wallgren et al.~\cite{wallgren2013routing}&  distributed  &  hybrid  &  RPL    &  $\circ$     & routing$^3$     \\
Raza et al.~\cite{raza2013svelte}&  distributed & statistical & IPv6 & $\circ$     &  routing$^3$    \\
Matsanaga et al. ~\cite{matsunaga2014low}&  distributed    &    statistical  &   RPL   &$\circ$      &    routing$^3$  \\
Cervantes et al.~\cite{cervantes2015detection}&  distributed    & trust &  6LowPAN    & $\circ$ & sinkhole      \\
Misra et al.~\cite{misra2011learning}& centralised & automaton   & IoT &  $\checkmark$   &  DDoS    \\
Garcia-Font et al.~\cite{garcia2017attack}& central & hybrid &   smart-city    & $\circ $&  routing   \\
Fu et al.~\cite{fu2017automata}&  central  &  automata    &  http    &  $\Crossedbox$    &    protocol$^1$  \\
Khan et al.~\cite{khan2017trust}& distributed &   trust-based   &  healthcare   &   $\circ$  & routing$^3$ \\
Han et al.~\cite{han2013idsep}&    central  &  anomaly    &   WSN   & $\Sigma,\circ$      & energy-DoS      \\
Jiang et al.~\cite{jiang2012dynamic} &  central   &  clustering&    clustered &  $\Crossedbox$    &  routing$^3$     \\
Esquivel-Vargas et al.~\cite{esquivel2017automatic} &  central   &  specification &    BACnet  &  $\delta,\checkmark$    &  protocol$^1$     \\

\hline
\multicolumn{6}{c}{} \\
\hline
Evaluation &\multicolumn{5}{l|}{
            \begin{tabular}[c]{@{}l@{}}
              $\Crossedbox$ - no evaluation performed\\
              $\Sigma$ - mathematical evaluation performed\\
              $\circ$ - simulation performed\\
              $\delta$ - trace evaluation performed\\
              $\checkmark$ - execution evaluation performed
            \end{tabular}}\\\hline
Detection &\multicolumn{5}{l|}{
            \begin{tabular}[c]{@{}l@{}}
              $\beta$ - custom attack list\\
              $^1$MITM, replay, spoofing, message dropping etc.\\
              $^2$unusual behaviour by an agent within the system\\
              $^3$routing attacks, e.g. Sinkhole, Wormhole, Blackhole\\
              $^4$resource drain DoS, pin cracking, and spoofing\\
              $^5$collision attack, Hello Flood, Selective forwarding attack \\
              $^6$reconnaissance, DoS, and information theft attacks\\
            \end{tabular}}\\
\hline
\end{tabular}
\label{tab:NIDSSum}
\end{table}

Le et al. (2011)~\cite{le2011specification}, propose a new IDs for the Routing Protocol for Low Powered and Lossy Networks (RPL), a new protocol specific for IoT systems.
Their work focuses on detecting topology attacks, which are attacks which attempt to change the system configuration, to cause delays and non optimal system behaviour.
They present two novel topology attacks on these systems: 1) rank attack, a rank attack implies the attacker has compromised some nodes in the network, the attacker will then change the routes of the network by skipping the rank checks on the packages and rerouting the packets; since each node has the responsibility to ensure their rank is greater than the previous rank, this fake routing causes wrong rankings of nodes and consequently wrong pathings; and 2) local repair attack, by performing a local repair, the compromised node changes it's unique ID, this will cause all his children to synchronise themselves; by repeatedly performing this action the attacker is able to disrupt part of the network operation.
The authors propose the use of a specification based IDS to specifically look for these behaviours, as these are legitimate operations they suggest that behaviour based intrusion detection may not suffice.
The authors propose a architecture for their approach using monitor nodes, each monitor covers a subnetwork and the nodes within it each send messages through the monitors.
The monitors will process all incoming messages and look for signatures of the attacks.
The specification is done through a finite state machine to monitor the flow of behaviours for each component, no evaluation is performed.

Liu et al. (2018)~\cite{liu2018intrusion}, propose a new algorithm for intrusion detection using suppressed fuzzy clustering.
The approach first clusters all incoming messages to reduce the load on the IDS.
The clustering is done on high-risk data and low-risk data.
Given a data matrix, the algorithm first performs a pre-judgement, using clustering they identify data from risk targets and non-risk targets.
Then a dimensionality reduction is performed to reduce the data size, using Pearson Correlation Analysis (PCA).
PCA has the advantage that it maintains patterns and information even after reduction.
A further processing is done to remove features of low statistical influence and further reduce the dataset size.
This approach was compared with two other machine learning approaches, Neural Networks and Bayesian, their purely mathematical simulation showed that their approach had better accuracy and faster results than the other approach.

Stelte and Rodosek (2014)~\cite{stelte2013thwarting}, designed a IDS for ZigBee and WSNs.
Almost all available commercial and research sensor nodes are equipped with ZigBee transceiver chips, and thus making ZigBee the de-facto standard in WSN communication
Their proposed anomaly based IDS is specifically made to defend against the KillerBee attacks.
The KillerBee attack framework is a set of tools specifically designed to exploit vulnerabilities for ZigBee.
To this end the authors develop an IDS specifically tailored to "thwart" these attacks.
The proposed IDS, includes a cooperation between the WSN nodes, there are three elements 1) local node 2) near neighbours and 3) network wide evaluation.
Each node monitors the parameters for the own communications. 
By analysing the communication between other sensor nodes in the neighbourhood, also attacks targeted onto other node scan be detected. 
This information can be sent to the whole network so that information can be spread.. 
If a sensor detects a misbehaving node in his neighbourhood, the node will be ignored in the further communication process.
The authors evaluated their IDS through a simulator, Avrora~\footnote{Available at : \url{http://tinyos.stanford.edu/tinyos-wiki/index.php/Avrora}}, a tool for the simulation of WSN.

Arrington et al. (2016)~\cite{arrington2016behavioral}, proposed a new IDS for behavioural anomalies, behavioural modelling intrusion detection system (BMIDS).
This approach leverages a different factor of the IoT which is it's ability to collect and sense data to create smart environments.
The authors specifically make use of a smart home scenario with various sensors collecting data about its occupants and room usage.
Contrary to most other work in the area they observe attacks as real human intruders in the environments using a centralised IDS approach.
This new technique is evaluated in a virtual reality environment OpenSim, where the scenario is created and the solution tested.

Deng et al. (2019)~\cite{deng2018mobile}, discuss current limitations of IDSs in IoT and discuss their new approach for tackling the issues of heterogenuity and distributed nature of IoT.
The authors first presents the security features and security issues of IoT, discusses the security architecture of IoT, and some security key technologies, especially focusing on the key management and routing technologies.
Specifically they focus on detecting Sybyl Attacks.
A Sybil Attack is a type of attack seen in peer-to-peer networks in which a node in the network operates multiple identities actively at the same time and undermines the authority/power in reputation systems.
Their approach uses transfer learning, adapting the Sybil intrusion prevention system to be used as an intrusion detection system.
Their approach uses the PCA algorithm, to make detection more lightweight like the work of Liu et al. with the addition of FCM algorithm for clustering of data for quicker processing.
Their evaluation is performed both using matlab simulation and using datasets from KDD cup99.

Had{\v{z}}iosmanovi{\'c} et al. (2014)~\cite{hadvziosmanovic2014through}, specifically targeted industrial control systems, and Programmable Logic Controllers, as the environment to protocol trough their IDS.
Their white box approach constructs patterns of behaviour for each process of the devices.
This helps create a benchmark of behaviour for their anomaly based IDS.
The approach has three stages:
1) data extraction, data extraction phase is a preprocessing step that distils the values of process variables out of ICS network traffic.
It consists of two subparts: (i) parsing the application-layer network protocol to extract the relevant commands, including all their parameters; and (ii) constructing shadow memory maps inside the analysis system that track the current state of all observed process variables, providing us with a mirror of the PLCs’ internal memory.
2) characterisation, separates PLC variables into different categories.
(i) control : variables for configuring plant operation (ii) reporting: variables for reporting alarms and events (iii) measurement: variables reflecting readings from field devices, and (iv) program state: variables holding internal PLC state such as program counters, clocks, and timeouts. 3) modelling and detection, to model the behaviour they use autoregressive modelling, which creates a time series of action which are dependent on the previous state with a predictive error, through this technique they are able to predict future states as well. 
To detect issues the check against the series and raise an alert of the value sits outside the of the enumeration set.
To evaluate their approach the authors set up a small testbed consisting of a water purification plant, consisting of one PLC and one operation workstation from an ICS vendor.
They evaluated their approach using Bro (now Zeek) a common tool for implementing IDSs using their bespoke language and custom C++.
Their evaluation was not performed on attack detection but rather on their methodology, as they explain there is no database of attacks available and their injections would be trivial and direct.
Their evaluation was based on the datasets collected from their testbed.
Their approach has the advantage that it is quite actionable as observing what is raising the alarm is traceable to the previous steps.

Cho et al. (2009)~\cite{cho2009attack}, specifically focused on detection DDoS attacks on 6LowPAN networks.
They assume a central gateway will receive all traffic in the network and will act as the base of the IDS.
Their detection approach is based on a simple observation that statistically attack packets are longer than the normal behaviour packets.
So if they detect abnormal sized packets an alarm is raised.
The evaluation is done using a ``simulation'', however it is not specified how it is done, implemented, or how they achieved the results.

Sedjelmaci and Senouci (2013)~\cite{sedjelmaci2013efficient}, present ELID, a lightweight IDS for WSN.
WSN are similar to the IoT in terms of their constrains, unlike the IoT, they are homogeneous.
Due to the highly homogeneous nature of WSN, the authors base their detection on the fact all nodes within a cluster should behave similarly, if a node within a cluster differs in behaviour they flag it as malicious.
They specifically aim to detect routing attacks, namely: Black Hole, Sinkhole, Wormhole and DoS.
The authors add specific rules for these attacks looking for the behaviours within a cluster of devices. 
The IDS is constructed with the following components 1) Intrusion Detection Agent, this is a random agent within the network elected to monitor the communication and perform the following actions i) data collection, collects all packets within its range, ii) intrusion detection module, this component applies the policy to the data collected, and iii) reaction module, if an anomaly is found it is reported and if multiple reports are done the node is changed. 
2) IDA election module, to maintain energy equilibrium through the system if the current IDA is low on energy a new IDA is elected, furthermore if the IDA is determined to be malicious it is ejected.
The IDS was evaluated using TOSSIM, a simulator of tinyOS nodes, their simulator checks for both attack detection accuracy as well as battery drainage of the system

Sedjelmaci et al. (2016)~\cite{sedjelmaci2016lightweight}. design an IDS specific for constrained IoT networks.
Their approach makes use of game theory to improve on anomaly based techniques. 
Specifically, the authors wish to balance the equilibrium between energy consumption and detection. 
This is because of the high energy consumption traditional anomaly detection has, so they wish to decrease the overhead, and only trigger the detection at the point of equilibrium.
Their approach is to have an IDS agent on each device that monitors all nearby neighbours. 
Their hybrid approach combines pre-set rule specifications with anomaly detection, each time the anomaly detection finds an alarm its specification is stored as an attack signature.
The game matrix works in the following manner, each time a node activates its anomaly technique it either detects an attack and adds it's signature which results in a positive payoff or a negative payoff if it doesn't. 
Conversely the attacker starts an attack if it is caught his payoff goes down if undetected payoff goes up.
The authors stipulate that this approach leads to optimal rule generation based on their payoff model the node will only add the rules when the payoff (the chance of detecting attacks with it) is maximised.
The authors evaluate their IDS using TOSSIM simulator checking for detection accuracy and battery consumption as their core parameters, the results report a slightly lower accuracy but much better energy consumption compared to an un-named hybrid detection system.

Danda and Hota (2016)~\cite{danda2016attack}, propose a simplistic IoT system composed of an android device, a router and a arduino board and build a rule based IDS to protect the system.
Their IDS is deployed centrally on the router, and the board is programmed to do a simple operation. 
Their approach extracts communication patterns from the network traces of the arduino device and then manually constructs rules about the communication.
The authors then construct setup SNORT IDS on the router and construct SNORT rules out of the pattern.
No evaluation is performed on their testbed.

Besson and Leleu (2009)~\cite{besson2009distributed}, propose an architecture for intrusion detection for WSN.
Their proposal is the design and characteristics needed for an intrusion detection system including architecture specifics.
Their design is, 1) autonomous, can adapt to changes in the network without human intervention, 2) lightweight, 3) distributed, and 4) ensured trustworthiness through encryption.
The data collection is done collaboratively between nodes, as resources are constrained they propose the use of metrics to ensure minimal data is collected.
Only a subset of nodes act as NIDS as the heterogeneous nature of the system might not permit some nodes from performing the required actions.
The unique aspect of their approach is that they ensure that the communication between the devices is always authenticated through use of encryption keys.
This is something no other approach addresses even though the authors do not present details of how they intent to ensure this or what security protocols would be in place.
No implementation details are provided nor any evaluation is in place.
We felt it was worth while to report on this work as it did bring forth the unique challenge of inter node authentication which no other work brought up and is indeed a significant consideration for any distributed IDS design.

Haataja (2008)~\cite{haataja2008new}, focuses on developing an IDS/IPS specifically for Bluetooth networks. 
The IDS is a rather simple set of 15 rules which check for specific set of intrusions: 1) Unusually many repeated failed authentication attempts, 2) Unusually many repeated successful authentications and disconnections, 3) Unusually many NAK transmissions, 4) unusually long delays, 5) unusually many POLL packets, 6) unusually high BER, 7) unusually high communication between two parties, 8) sudden increase in transmit powers, 9) two identical BD\_ADDR in range, 10) HV1 (High-quality Voice 1) SCO link established with the piconet master when other type of SCO or eSCO
link could also have been used, 11) L2CAP level request for the highest possible data rate or the smallest possible latency, 12) Surprising connection attempts and data transfer requests from unknown Bluetooth devices, 13)A Bluetooth device requests that the length of an encryption key must be shorter than 128 bits, 14)RF signature mismatch, and 15) SSP’s Just Works association model activated between such devices that could use a more secure option (e.g.Numeric Comparison or OOB).
The authors see these rules as comprehensive for the range of attacks on Bluetooth networks.
The IDS architecture is a central node, that monitors all traffic, if a rule is broken an administrator is alerted.
Their automated IPS proposes for the malicious device to be disconnected as means of prevention.
No implementation is provided nor evaluation is performed.

Yadav and Srinivasan (2010)~\cite{yadav2010itrust}, propose a trust based intrusion detection system for WSN.
Due to the often hostile environment in which WSN networks are deployed and their autonomous nature it is difficult to use traditional techniques, as a compromised node might be very difficult to detect.
Specifically for these environments, the authors propose a trust based approach in which nodes evaluate each others behaviour to create a trust mechanism.
This assigned trust valued is shared within the network and untrusted individuals may be blacklisted.
The evaluation is performed using simulator OMNet++~\footnote{Available at : \url{https://omnetpp.org/}}, however due to the fact this simulator is not able to model attacker nodes, the authors have extended the tool.
Their experiment deployed 25 nodes with 4 monitor nodes that cover the full network. 
Their evaluation checks for the detection accuracy of their algorithm against, collision attacks hello flood attack, and selective forwarding attack as well as the performance of their trust based scheme.

OConnor and Reeves (2008)~\cite{oconnor2008bluetooth}, bring forth a misuse based intrusion detection system for Bluetooth technologies.
Similarly to the work of Haataja (2008), the authors propose a set of rules to detect specific attacks on Bluetooth networks.
The IDS performs packet inspection looking for specific attack signatures.
Unlike previous work the rule generation process is automated based on pre-training and is expanded for better detection.
The generated rules are automatically generate din SNORT format for use within the existing IDS.
The authors have implemented their infrastructure in a tool and evaluated it in a a testbed.
This testbed constituted an 1) attacker node running BackTrack2 live Linux distribution from remote-exploit.org which includes more than 300 different security tools, 2) 4 different target devices, including 3 phones and one bluetooth headset, 3) a defense components, running the Merlin LeCroy Protocol Analyzer~\footnote{Available at : \url{https://teledynelecroy.com/support/softwaredownload/psg\_swarchive.aspx}} to capture packets.
Evaluation was performed live on the testbed and 20 attacks were carried out.
Metrics for evaluation included, response time, coverage and detection accuracy.

Thanigaivelan et al. (2016)~\cite{thanigaivelan2016distributed}, propose an internet of thing internal IDS to detect misbehaving/malicious nodes within the network.
Their distributed cooperative approach makes each node in the system be responsible for its neighbours within 1-hob.
Each node learns the \textit{normal} behaviour of its neighbours and reports any anomaly.
The authors devise a new type of message that allows effective reporting to gateway nodes, about inactivity or malicious behaviour and the gateway makes decisions about what to do.
The decisions are done on a grading system.
No further details about implementation or evaluation are provided.

Luo and Nagarajan (2018)~\cite{luo2018distributed}, propose an autoencoder approach to anomaly detection in WSN networks.
An autoencoder is a powerful machine learning technique that provides all the accuracy advantages of a neural network with a further layer of adaptability.
Their IDS architecture has two parts 1) sensor, each sensor runs one copy of the autoencoder and provides the results of its autoencoder to the cloud as training data and secondly it uses its autoencoder to perform anomaly detection, and 2) cloud, this is represented by a gateway which relays data to the sensors, it also uses the outputs of all the sensors to train and update the autoencoder locally and then sends the updated parameters to the sensors for an adaptive anomaly detection.
The authors constructed a WSN testbed to collect trace data for evaluation.
Their testbed consist of 8 sensor nodes collecting humidity every 2 minutes. They collected data for 3 months.
To simulate anomalies they created a scenario of spikes, which are sudden increases in humidity followed by a sharp decline; and, bursts which are continuous raises for a finite period of time.
Their evaluation was performed on the accuracy of the detection of these bursts/spiked and not on any attacks.

Wallgren et al. (2013)~\cite{wallgren2013routing}, present a comprehensive analysis of attacks in RPL connected internet of things.
The authors present and implement several routing attacks against 6LowPAN, in the Cooja~\footnote{Available at : \url{https://anrg.usc.edu/contiki/index.php/Cooja\_Simulator}} simulator.
They show how the new features available in the IPv6 protocol can be used to the benefit of a RPL IDS, and simulate the results.
Their proposed IDS architecture takes a hybrid approach with sensor nodes deploying signature based detection collaboratively with anomaly based detection on more powerful nodes. 
The authors stipulate that this approach will decrease overhead whilst maximising detection.
Once nodes are detected as malicious they are eliminated from the network.
The authors find that due to the wide mandatory deployment in IPv6, this protected ICMPv6 messages can be means for a lightweight solution to selective forwarding attacks that may be difficult to detect in IPv4 networks.
Their evaluation was performed as a simulation in cooja, and they implemented their new attacks in the simulator.
Evaluation was done on performance implications of the attacks but no performance was done on their IDS.

Raza et al. (2013)~\cite{raza2013svelte}, create SVELTE, a lightweight intrusion detection system for IPv6, that is not only constrained to WSN networks.
The authors of SVELTE propose a hybrid placement approach distributing IDS modules on the gateways as well as the devices themselves.
the architecture of SVELTE is trifold: 1) 6LoWPAN Mapper, placed on the gateways it is used to collect traffic and construct the network, 2) IDS components, this analyses the collected data and detects intrusions, and 3) a mini firewall deployed across the system which filters unwanted traffic before in enters the constrained network.
The intrusion detection operates using a statistical approach, based on the network graph it observes if there are any statistical inconsistencies, if wrong information is sent or there are incorrect data packets this is flagged.The role of the gateways is to analyse all the individual network graphs and alerts from the nodes, if various nodes are reporting inconsistencies for a components than that node is disconnected.
Both the IDS components and SVELTE are implemented in Contiki (Cooja), for their experiment the authors make use of Tmote Sky nodes~\footnote{Available at : \url{https://wirelesssensornetworks.weebly.com/blog/tmote-sky}}, they assume the gateways are non constrained devices such as a laptop.
Experiments show that SVELTE has good detection accuracy however presents quite high false positive rate, as SVELTE is non adaptable this cannot be fixed live.
The authors also run analysis on energy overhead and memory usage using Contiki Powertrace~\footnote{Available at : \url{https://github.com/contiki-os/contiki/blob/master/apps/powertrace/powertrace.c}} and Memory usage showing some promising results.

Matsanaga et al. (2014)~\cite{matsunaga2014low}, propose a new scheme for RPL connected internet of things with low false alarm rate.
The authors expand on the previous work of SVELTE and attempt to fix the problem of the high false alert rates caused by natural rank adjustments and the lossy nature of RPL connected systems.
To fix these issues the authors make two core adjustments to the messages used in the system.
Firstly each node sends a rank at the time that the DIO message is broadcast, so as to collect all ranks at the same time.
Te DIO (DODAG Information Object) message is used to set up the structure of the network.
Secondly, each node attaches a timestamp to the time it receives the DIO message, and sends it to the root in order to sort out timing inconsistencies.
To evaluate their approach the authors implement their extension in Cooja expanding on the previous SVELTE version.
In both their scenarios with the two extensions they show that the false alarm rates goes down than the standard SVELTE implementation.

Cervantes et al. (2015)~\cite{cervantes2015detection}, propose Instrusion detection for SiNkhole attacks over 6LoWPANfor InterneTof ThIngs (INTI).
SINKI, combines watchdog, reputation and trust strategies for detection of attackers by analysing the behaviour of each node.
INTI runs on four modes of operation: 1) cluster configuration,t his module defines a leader-based hierarchy establishing node clustering to ensure scalability and extend the lifetime of the network; 2)route monitoring, which defines a monitoring module to count the transmission number of input and output performed by a node responsible for forwarding messages; 3) attack detection, this performs two kinds of evaluations trust based and reputation based, reputation is the perception each node establishes after iterations based on actions and information exchanged, this value is shared with the rest of the network; and 4) attack isolation, this isolates a sinkhole node after its detection.
The INTI system was evaluated in the simulator Cooja.
The evaluation scenario consists of 50 nodes, some fixed and others mobile.
Evaluation was performed on: 1) detection rate, 2) precision (false positive and false negative), and 3) delivery of packets.
INTI was evaluated against SVELTE~\cite{raza2013svelte}, showing better false positives and false negative rates and thus being more precise for sinkhole attacks.

Misra et al. (2011)~\cite{misra2011learning}, propose a learning automata based solution for preventing DDoS in IoT.
Their designed setup follows a Service Oriented Architecture (SOA) approach, a SOA is a platform independent of the underlying system for developers to work on the underlying system.
The learning automaton revolves around the notion of an ``automaton'' which is a  self-operating machine or a mechanism that responds to a sequence of instructions in a certain way, so as to achieve a certain goal.
The automaton either responds to a pre-determined set of rules, or adapts to the environmental dynamics in which it operates.
The IDS works in the following manner, once a DDoS is detected against the SOA, a DoS Alert (DALERT) message is propagated across the physical network neighbour to neighbour.
To identify the intruder a sampling is done on all the devices raising alerts and once the attacker is identified the information is sent back up to the SOA.
Once an attacker has been identified all nodes are made aware of their identity.
The learning automaton comes into play to decide the rate of sampling.
Sampling each message within the system is not feasible and therefore a cost is associated with sampling, and a reward with a successful alert.
The automate will therefore find the optimised sampling rate to maximise the reward function.
To evaluate their approach the authors create a physical testbed, consisting of five objects: 1) a internet router, 2) a bridge forwarding requests from the serving object, 3) three serving objects.
They then launched a DDoS attack on the application layer using non specific tooling, a total of 83 service requests were sent by this attacker, one should note this would not traditionally be considered a DDoS, but maybe a flooding DoS (and not a very strong one even against constrained devices).
Their evaluation shows that the learning automaton is able to improve the performance of power consumption compared to a naive sampling approach.

Garcia-Font et al. (2017)~\cite{garcia2017attack}, investigate the feasibility of constructing an IDS for the context of smart cities.
Due to the large distributed nature of smart cities, various WSN components are deployed across it to get data.
Their proposed IDS approach uses an attack classification schema based on anomaly detection to help city administrator pin point the source of attacks.
Their IDS is composed of two components, a rule based engine for finding patters at the local level and an anomaly based engine to monitor the network traffic as a whole at the city level.
Their attack classification approach resides in the way attack assessment is done on the greater network.
There is less emphasis on the individual WSN but rather it observes how the attack spreads across the city, in terms of geo-location, transmission medium and type of attack.
The signature based IDS placed within the sub-network allow for a lot more actionability and information about the attacks.
Their attack classification has the following attacks: 1) vertical attacks, attacks that show vertical attack traces (i.e., from a group of node leaves to the base station) on a single WSN. The main aims of these attacks is to obstruct one or more paths in order to increase the arrival time of the packets from the target leaves, to crash intermediate routing nodes, to decrease node batteries or to provoke a general DoS, 2) Transmission medium attacks, attacks that affect nearby nodes using the same frequency bands or MAC protocols. Other bands or protocols are not affected. Basically, attackers take advantage of the transmission medium in order to prevent the proper delivery or reception of packets from certain nodes. 
These attacks applied to routing nodes also hamper the correct communication of other nodes outside of the attacker’s direct influence area, 3) Locally-dispersed attacks, attacks that affect dispersed nodes from a single WSN with the main goal of creating delays, dropping packets and depleting node batteries, 4) Widely-dispersed attacks, attacks that affect dispersed nodes from several WSNs. 
Attackers aim to reduce the proper operation of one or several WSNs. 
In this case, attackers do not use constant attack techniques, which would be more effective, in order to cover up their intentions and delay the moment of their discovery, 5) Widely-intensive attacks, Attacks that affect a great percentage of nodes from several WSNs using the same gateway. Attackers use these techniques to completely stop the service provided by one or more WSNs, 6) Local service alteration attacks, Attacks that affect several nearby nodes from the same WSN. The main goal is to alter application information from an area. The attackers either drop application packets or send false information, and 7) Single node attacks, Attacks that aim at depleting the batteries of a single node. This becomes very critical when attackers aim at an important router node in network areas with few paths to the sink. Several of these attacks on each path divide the network.
Their rule based classification establishes which classification an attack is based on a set of parameters.
The authors evaluate their work by building a proof of concept in R, a network is mimicked using geolocation as a core value, and a selective forwarding attack is simulated by forwarding 20\% of packets to a single gateway.
Their approach was analysis based on detection accuracy.

Fu et al (2017)~\cite{fu2017automata}, propose an automata based IDS that focuses specifically on protocol based attacks for IOT systems.
They make use of the automata to specifically model the communication steps of a protocol, with each transition being a message exchange.
The way the automaton works is that when data is incoming n the network it is abstracted as an event in the automaton and deviations from the expected protocol are seen as malicious.
The way this is done is by ignoring the contents of messages and simply looking at type of message in the context of HTTP e.g. ack, fin etc.
The intrusion detection has two phases 1) based on a database of malicious actions, if the sequence of actions matches one of the malicious one it is matched as malicious, and 2) anomaly based detection, if the sequence of actions is unexpected it is labelled as malicious.
Their experiment is evaluated in a simple physical testbed composed of: i) two raspberry pis 2) a android phone, and a wireless router and 3) a server hosted on a pc.
The server contains a MySQL database with a dataset of malicious event traces for the three attacks : 1) replay attack, 2) jam attack and 3) fake message injection.
Their evaluation is focused on the translation of messages to automatons and creates a graphical representation which allows to get a good view of the network behaviour.
Whilst no results on the IDS detection are presented their analysis shows how their tool behaves and constructs the event traces.

Khan and Herrmann (2017)~\cite{khan2017trust}, propose a trust based IoT IDS, with specific focus on the healthcare domain.
Their IDS is meant to operate in 6LowPAN environments and focuses on the detection of routing attacks, a common threat in these scenarios.
Their trust based technique allows each individual node to assess the trustworthiness of its neighbour devices, depending on positive and negative experiences.
The trust is build using subjective logic, this increments the trust based on positive experiences, decrements it due to negative experiences but also contains a uncertainty value that can change the amount of trust gained or lost.
The authors propose a set of operations which gain trust and a set of operations which loose trust.
The weighting of these values is also adjusted depending on how often the behaviour is performed.
These trust values are composed for each node and transmitted to the central gateway, where a reputation score is measured for each component based on his trust results.
If a bad reputation value indicates a node as a potential intruder, the border router removes it from the network and notifies the operators of the network.
The authors simulate this algorithm using MATLAB experimenting with different trust gain algorithms.
The simulation consists of 1000 nodes randomly deployed in a network, several different number of intruders was assessed (from 0 to 30\%).
Evaluation was performed on how many of the intruders were detected after the simulation had ended as well as accuracy of detection.
Results showed that this approach had a high detection accuracy of intruders with as high as 95\% intruder detection, and relatively low overhead.

Han et al. (2013)~\cite{han2013idsep}, recognise a new kind of DoS attack that affects sensors based system, battery drain DoS, and propose a detection schema for these attacks in cluster based networks.
Their schema is based on an expected energy consumption calculation, if the actual consumption of a node exceeds expectation it is assumed under attack.
Their specific attack models assume an internal compromised node targeting other members of the network.
Their assumption is that the malicious nodes will use abnormally high levels of energy.
To predict the energy consumption a markov chain is constructed with fixed energy consummation per action.
Given a probability of performing an action it is possible to calculate the average energy consumption of a node.
For detecting attacks, initially the IDS is aware of all energy levels and expected consumption for each node.
After each message is received by a node it sends its residual energy value to the central IDS node.
We note that the authors make the assumption that these nodes have the ability to calculate this value and do not calculate this as potential overhead, which may not always be the case.
If the energy consumption received by the IDS is not consistent with the expected amount the node is marked as malicious.
The authors also propose a method to classify the kind of attack based on how much extra consumption is performed by the malicious node.
Each DoS attack is classified with a specific energy threshold and the increase is matched to one of the attacks.
Evaluation is performed both theoretically, showing the reasoning behind why it would be able to detect attacks and simulations are run in NS2~\footnote{Available at : \url{https://www.isi.edu/nsnam/ns/}}.
The authors evaluate their detection ration, energy consumption for their IDS as well as the increase in throughput as an outcome of malicious nodes being stopped.

Jiang et al. (2012)~\cite{jiang2012dynamic}, present work on an intrusion detection scheme specific for clustered wireless networks.
Their anomaly based technique uses a clustering algorithm to create their standard behaviour.
Cluster based Wireless Sensor Networks are organised into clusters by a cluster formation algorithm. 
For a cluster, there is a cluster head and several member nodes.
Their IDS structure is the following, each cluster head will be in charge of detecting attacks within the cluster, it will base an anomaly on the following features 1) rate of packets received, 2) rate of packets sent, 3) rate of data packets matched, and 4) rate of data packets dropped.
This is made easily possible to detect due to a fixed distance clustering algorithm which will ensure each node is at one hop from the cluster head.
The authors do not evaluate their work by experimentation, but provide intuitions as to why it might be effective naming low energy consumption and the ability for real time detection as key factors.

Esquivel-Vargas et al.~\cite{esquivel2017automatic}, propose a means to create specification based IDSs for Building Automation Systems (BAS) using the BACnet protocol.
Their approach focuses on the assumptions that BAS systems behaviour is well understood and therefore easy to specify.
This is in part due to the BACnet protocol standard specifying that each BACnet implementation needs to provide a PICS (Protocol Implementation Conformance Statement) detailing its behaviours and subsets of the standard that have been used.
The authors work aims to simplify the generation of specification based IDSs for such systems, by providing an automated approach to generate them based on the PICS of the system.
Firstly the authors use a device fingerprinting technique to discover the system setup, this allows them to discover what devices re in the network and how they are interconnected.
These are connected to the PICS document to relate to the expected device behaviours.
Then each device behaviour is matched, using network traffic analysis to construct a specification of system behaviour automatically,
The approach is tested firstly on traces of real device behaviour from a real BAS setup, and then on a implementation of a BACnet testbed consisting of two BACnet devices and a router device.
Their results show high levels of accuracy for any attacks which violates expected protocol behaviour.

\subsection{Proposed Host Based Intrusion Detection Systems for IoT}
\label{sec:toolHIDS}
With the IoT there has been a significant shift in the computation power of the devices present in systems.
Along with these constraints come computational challenges, specifically it becomes challenging to make use of existing host based IDS techniques on devices with limited resources.
As most of the current HIDS are too computationally intensive for the IoT, literature has proposed several new approaches for intrusion detection, this section provides a review of these approaches and the full summary can be seen in Table~\ref{tab:HIDSSum}.

\begin{table}[htbp!]
\caption{Summary of Host Intrusion Detection Systems Proposed for IoT}
\begin{tabular}{|l|c|c|c|c|}
\hline
Paper & Algorithm & Env. & Eval. & Detection \\
\hline
\multicolumn{5}{c}{} \\
\hline
Oh et al.~\cite{oh2014malicious}& pattern-matching  & IoT & $\Sigma,\delta$  &    DDoS, virus signatures   \\
Summerville et al.~\cite{summerville2015ultra}& bit-wise anomaly & IoT & $\checkmark$ & worm, code injection\\
Liu et al.~\cite{liu2011research}& immunity & IoT &   $\Sigma$   & -     \\
Kim ~\cite{kim2015physical}& likeliood ratio& relay coms &  $\Sigma$    & message integrity     \\
Lee et al.~\cite{lee2014lightweight} & anomaly & 6LowPAN& $\circ$ & battery DoS\\
Song et al.~\cite{song2010weak}& W-HMM & WSN  &  $\circ$   &  - \\
\hline
\multicolumn{5}{c}{} \\
\hline
Evaluation &\multicolumn{4}{l|}{
            \begin{tabular}[c]{@{}l@{}}
              $\Crossedbox$ - no evaluation performed\\
              $\Sigma$ - mathematical evaluation performed\\
              $\circ$ - simulation performed\\
              $\delta$ - trace evaluation performed\\
              $\checkmark$ - execution evaluation performed
            \end{tabular}}\\\hline
\end{tabular}
\label{tab:HIDSSum}
\end{table}

Oh et al. (2014)~\cite{oh2014malicious}, conscious of the low computational power of IoT device, propose two new lightweight pattern matching algorithms for intrusion detection that decrease overhead by reducing the number of matching operations.
he Wu–Manber (WM) algorithm is one of the fastest multiple pattern-matching algorithms and is widely used as a malicious pattern-detection engine.
It is composed of two stages. 
The first is a preprocessing stage to construct three required tables, the shift, hash, and prefix tables. 
The second stage is a pattern-matching stage to perform matching operations using these tables. The second stage accounts for most of the execution time.
The authors propose an auxiliary shift value to increase the efficiency of the second stage.
Traditional approaches scan the entries one by one shifting to the right one step, the authors propose a way to calculate a more efficient shift distance to improve speed and number of transactions required.
The second suggested algorithms is based on early decisions, they improve the traditional algorithm by sorting the data entries in alphabetical order, they then instead of searching by word search by prefix allowing to quickly discard matches.
This allows for smart look ups avoiding starting the match base don prefix rather than word values.
The authors perform a theoretical analysis of the work and evaluated the impact of their changes on efficiency and time performance.
They show that their approach has the same complexity as the traditional WM approach however performs many less operations.
The authors also implement their algorithm on a real device and compare it to traditional approaches.
They perform analysis on a setup constituting of a raspberry pi, they extract datasets from SNORT IDS and ClamAV~\footnote{Available at : \url{https://www.clamav.net/}} antivirus as attack tests.
Their results shows that their overhead is significantly decreased at the cost of 5\% detection accuracy.

Summerville et al. (2015)~\cite{summerville2015ultra}, propose a deep packet anomaly inspection that can operate on constrained embedded devices.
Their approach can run both on the network stack or at the device level.
Firstly a bit wise feature extraction is performed on each packet.
Network payloads are treated as a sequence of bytes.
Feature extraction operates on overlapping tuples of bytes, called n-grams.
A three dimensional feature space is constructed out of the ngrams.
After the three dimensional representation is constructed out of the training data it is used as means for anomaly detection.
The authors evaluated the proposed approach against traffic obtained from two Internet-enabled devices: a weather station and an interactive networked video camera. These were chosen, respectively, to represent an IoT device that primarily outputs data, a sensor, and one that primarily receives control commands, an actuator.
They trained a detector for each of their devices on the specific network behaviour and tested various attacks against including worm propagation and code injections.
Their results show that due to the very basic actions of the devices their very lightweight approach is able to easily detect all the attacks with high accuracy, at a very low overhead.

Liu et al. (2011)~\cite{liu2011research}, in their work present an immunity based detection system for IoT.
Artificial immune systems are a recent advance in computational intelligence, their self adaptation and robustness makes them a very interesting approach for intrusion detection in the IoT.
Artificial Immune Systems are adaptive systems are proposed by Leandro Nunes et al. ~\cite{castro2002artificial}.
They are inspired by theoretical immunology and observed immune functions, principles and models, which are applied to problem solving, as such they take specific components of the immune systems as inputs.
The main difficulty in the usage of this approach is the matching from network data to the input of the immunity algorithms.
The authors propose the following matching 1) Antigen $\longrightarrow$ signature of IoT datagram, 2) Self element $\longrightarrow$ normal datagram, 3) non-self element $\longrightarrow$ attack data, 4) immune cell $\longrightarrow$ detector, 5) immature immune cell $\longrightarrow$ immature detector, 6) mature immune cell $\longrightarrow$ mature detector, 7) memory immune cell $\longrightarrow$ memory detector, and 8) immune cells recognised harmful antigens $\longrightarrow$ detector matches datagram with attacks.
Each packet in the system is converted to a bit string.
The evaluation is performed mathematically analysing how the detector evolves to match the malicious antigens per traditional immune system approach.

Kim (2015), proposes a physical-layer technique for checking the integrity of information provided by an intermediate node(relay) in cooperative relay communication systems.
Cooperative relays are a way to enhance the coverage of communications in wireless sensor networks.
As the relay depends on cooperation between nodes, almost like the game of Chinese whispers, if a single node acts maliciously it will impact the whole system negatively.
The authors propose a detection methodology to evaluate whether a received message has been tampered with and its integrity maintained.
Integrity is the trustworthiness of information. 
The proposed approach exploits the physical layer signal to detect the presence of false data in the received packet at the destination. 
Once the packet provided by the relay is determined to be modified, then it is discarded to avoid incorrect decoding at the destination.
The author defines the proposed system as a mathematical model, they  consider 1) a cooperative relay communication system in which a source wishes to send its message to a destination with the assistance of a relay,2) assume that the relay is not able to transmit and receive simultaneously, and 3) that time synchronisation between the source and relay is maintained.
Evaluation is performed mathematically and get results from simulation.
The evaluated the approach on 1) Detection Error Probability, Average Detection Error, Outage Probability, and finally they use their formula to find the optimum detection threshold balancing the outage vs accuracy.

Lee et al. (2013)~\cite{lee2014lightweight}, analyse the problem of battery drain DoS attacks on wireless 6LowPAN networks.
The authors assume that each node knows their own energy consumption and a central gateway manages communication between nodes.
The central gateway pols for energy readings every 0.5s, if the energy consumption increases above 30\% of the expected energy consumption for any given node an alarm is raised, after the second alarm is raised the node is removed from the routing table.
The authors evaluated their work through simulation using Qualnet~\footnote{Available at : \url{https://www.scalable-networks.com/qualnet-network-simulation-software-tool}}.
No analysis was done for accuracy of detection the authors simple showed that under DoS the energy consumption will go up significantly and infer that therefore their approach will be 100\% accurate.
The authors fail to provide convincing evaluation for their approach not to mention that their proposed scheme in itself with polling and energy consumption calculation every 0.5 seconds for each device wold have massive overhead in itself likely speeding up the drain, they also assume that devices can self calculate energy consumption which for IoT devices this is not the case and finally their solution of disconnecting the device if they consume more energy than expected is in itself dossing the system and doing the attacker a favour.

Song et al. (2010)~\cite{song2010weak}, make use of Hidden Markov Models (HMM) as means for anomaly detection in wireless sensor networks.
Since HMMs have been proven to be too resource intensive for sensor devices the authors adapt the traditional approach to remove transition probabilities to be replaced with rules of reachability called Weak-HMMs (W-HMM).
Their approach works because when a HMM is functioning normally it will be orbiting within the states following the rules, however, if the network is under attack it will deviate from ordinary orbit into an unknown rule, this will be flagged as an anomaly.
Their algorithm is broken into two stages, 1) training,this is where the ordinary orbit model is constructed, and 2) detection, the surveillance nodes collect changes in system parameters and a deviation distance is calculated, if above the allowed threshold it is marked as an attack.
Note that the do not specified every how often the HMM is recalculated.
The authors simulate their approach in MATLAB and evaluate detection accuracy and false alarms to show the validity of their approach.

\subsection{Proposed Collaborative Intrusion Detection Systems for IoT}
\label{sec:toolCIDS}

As more and more devices are connected to the internet, IoT systems grow from tens of devices to potentially thousands.
These vast systems may spawn across various infrastructures, networks, protocols and geo-locations.
To properly monitor these systems various techniques and systems need to be in place and work together collaboratively to detect attacks.
These new types of systems have increased the need for researchers to consider approaches that can work together and tackle different intrusions across the layers.
This section provides a review of these approaches and the full summary can be seen in Table~\ref{tab:CIDSum}.

\begin{table}[htbp!]
\caption{Summary of Collaborative Intrusion Detection Systems Proposed for IoT}
\begin{tabular}{|l|c|c|c|c|c|}
\hline
Paper & Placement & Algorithm & Env. & Eval. & Detection \\
\hline
\multicolumn{6}{c}{} \\
\hline
Gupta et al. ~\cite{gupta2013computational} & distributed & CI (multi)  & wireless & $\Crossedbox$ & protocol$^1$    \\
Amaral et al.~\cite{amaral2014policy} & distributed & signature   & IPv6 &   $\Crossedbox$   &  -    \\
Kasinathan et al.~\cite{kasinathan2013denial} & hybrid  & rule-based &   6LoWPAN    &  $\checkmark$    & DoS     \\
Kasinathan et al.~\cite{kasinathan2013ids}&  hybrid    & rule-based & 6LowPAN      & $\checkmark$     &   DoS   \\
Amouri et al.~\cite{amouri2018cross} & cross-layer & behaviour & IoT & $\circ$ & -\\
Hassanzadeh et al.~\cite{hassanzadeh2011towards} & distributed & genetic-algo.  & IoT & $\circ,\checkmark$ & $\beta^1$ \\
Le et al.~\cite{le2016specification}& hybrid      & specification     & RPL     & $\circ$     &    topology$^2$  \\
Midi et al.~\cite{midi2017kalis}&  centralised    &  adaptable  &   IoT   & $\checkmark$     &   DoS   \\
Shreenivas et al.~\cite{shreenivas2017intrusion} & hybrid & anomaly &  RPL    &  $\circ$     &  insider\\
Bostani et al.~\cite{bostani2017hybrid}& distributed  &  hybrid    & WSN      &  $\circ,\checkmark$    &  routing    \\
Arolkar et al.~\cite{arolkar2011ant}&  distributed    &  ant-colony    & WSN     &   $\Crossedbox$   & $\beta^3$     \\
Coppolino et al.~\cite{coppolino2013applying}&   hybrid   & hybrid  & WSN     &  $\delta$    &   $\beta^4$   \\
Abhishek et al.~\cite{abhishek2018intrusion}&  distributed    &   anomaly   &   IoT   &  $\circ$     &  $\beta^5$   \\
Pongle et al.~\cite{pongle2015real} &   distributed   & anomaly     & RPL     & $\circ$      &  wormhole     \\
Zhang et al.~\cite{zhang2011artificial}& distributed&  immunity     & smart-grid     & $\delta,\circ$  &   $\beta^6$   \\
Yu et al.~\cite{yu2008framework}&  mixed    &  hybrid    &   WSN   & $\Crossedbox$     &  routing  \\
Buennemeyer et al.~\cite{buennemeyer2008mobile} & distributed & hybrid & IoT& $\checkmark,\alpha$ & battery-DoS \\
Moyers et al.~\cite{moyers2010multi} &distributed &hybrid&IoT&$\Crossedbox$&-\\ 

\hline
\multicolumn{6}{c}{} \\
\hline
Evaluation &\multicolumn{5}{l|}{
            \begin{tabular}[c]{@{}l@{}}
              $\Crossedbox$ - no evaluation performed\\
              $\Sigma$ - mathematical evaluation performed\\
              $\circ$ - simulation performed\\
              $\delta$ - trace evaluation performed\\
              $\checkmark$ - execution evaluation performed\\
              $\alpha$ - usability evaluated
            \end{tabular}}\\\hline
            Detection &\multicolumn{5}{l|}{
            \begin{tabular}[c]{@{}l@{}}
              $\beta$ - custom attack list\\
              $^1$Flooding, Port Scanning, Web Exploits\\
              $^2$Sinkhole, Rank, Local repair, DIS, Neighbour\\
              $^3$Sinkhole, Misdirection, passive information gathering\\
              $^4$Sinkhole, sleep exhaustion\\
              $^5$Compromised Gateways\\
              $^6$MITM, replay, spoofing, message dropping, DoS, data leakage\\
            \end{tabular}}\\\hline
\end{tabular}
\label{tab:CIDSum}
\end{table}

In the work of Gupta et al. (2013)~\cite{gupta2013computational}, the authors propose a general IDS for use in wireless and pervasive computing triggered by the increasing growth of the IoT and CPS. 
The authors present their view that due to the increase in connectivity and adaptability the best option for intrusion detection is to use computational intelligence techniques as means of detection.
The authors present a review of attacks and vulnerabilities specific of these networks, namely: 1) Access Control Attacks, Access control attacks are an attempt to penetrate into a wireless  network  by  evading  weakly  laid  WLAN  access  control measures, such as 802.1X port access controls and AP filters 2) Confidentiality attacks, an attempt to illegally intercept the private and confidential information sent over wireless associations, both in unencrypted as well as encrypted form using 802.11 or other encryption protocols, 3) Integrity attacks, forged  data  frames  over  wireless  networks to either mislead the recipient or to form a ground to facilitate launching another type of attack, 4) Authentication Attacks, make use of these attacks to steal legitimate user identities and credentials to access otherwise confidential and private networks and services, and 5) Availability Attacks, a class of Denial of Service (DoS) attacks restricting a wireless server from providing services to authorised clients because of resource exhaustion by  unauthorised clients.
They then proposed a three tiered framework to operate on such wireless systems as an intrusion monitor against these attacks: 1) Monitoring the network, 2) Computational Intelligence, 3) clustering and reporting, this approach allows for lightweight and comprehensible detection. 
The authors do not provide an evaluation of their approach.

Amaral et al. (2014)~\cite{amaral2014policy}, propose a collaborative IDS for the internet of things. 
Their proposed collaborative architecture distributes a mixture of NIDS and HIDS across the system.
Each NIDS performs local traffic snooping, each node in the neighbourhood is in itself a host based IDS. 
At the network layer a set of rules is configures for detection purposes.
As their proposed environment is IoT they assume a heterogeneous system and therefore customised rules are present on each of the NIDS.
The authors do not specify how the host based IDSs are configured.
Each of the NIDS report to a central workstation that collates the data and performs assessment.
They construct an IDS application, that is now available on TinyOS repositories working on UDP and 6LowPAN.
Despite the application being tested as running successfully on TinyOS, no analysis of accuracy or evaluation of performance is done.

Kasinathan et al. (2013-A)~\cite{kasinathan2013denial}, propose a DoS detection architecture for 6LoWPAN.
The authors embed their security solution in ebbits~\footnote{Available at : \url{https://vicinity2020.eu/vicinity/content/ebbits}}, a business oriented services architecture for the IoT.
They create a collaborative attack detection mechanism composed of a DoS protection manager at the network level, combined with distributed IDS probes within the physical layer.
Their centralised approach assumes there is a cloud layer above the physical network running a network manager (ebbits), their DoS protection resides as a module within the network manager. 
Within the physical layer the NIDS probes monitor all traffic witin their subnetwors.
The actual IDS components within the probes is Suricatta~\footnote{Available at : \url{https://suricata-ids.org/}}, a widely popular open source IDS.
The alerts from the IDS are sent onward to the DoS detection engine in the could.
The authors claimed that due to their detection engine residing outside the physical network it adds a layer of resilience and makes it in itself immune to DoS attacks.
To evaluate their approach the authors implement a physical testbed of devices, a Linux penetration testing machine using metasploit as well as a Linux host running suricata.
Their pen-testing machine made use of UDP flooding to target the network, and the suricata was setup with a rule to detect large amounts of traffic and consequently rat limit the traffic.
Results showed that the IDS successfully alerted of attacks even with multiple target machines, and by implementing their attack on a more powerful machine was not overwhelmed by the attack (of five other machines).

Kasinathan et al. (2013-B)~\cite{kasinathan2013ids}, the authors expand on their previous DoS detection technique by construstion 6LowPAN decoders to work directly with suricata.
As suricata was never designed to work with 6LoPan these advancements go a long way to make traditional tools function in new IoT environements.
A similar testbed is constructed as per the previous work to test the new architecture.
The authors made use of Scapy~\footnote{Available at : \url{https://scapy.net/}} to construct malicious flooding packets and evaluated it on their system.
The authors proposed approach of making existing mainstream IDS function and adapt to IoT environments is a valuable first step to the advancement of the IoT IDS field as a whole.

Amouri et al. (2018)~\cite{amouri2018cross}, propose a collaborative and distributive approach to intrusion detection in the IoT.
Dedicated sniffers perform detection at the local level whilst a super node performs system wide detection based on data from the sniffers.
The IDS consists of a two-stage detection process, local and global. 
The local detection is achieved via a the sniffers located at opposite ends of a system. 
The packet counts from MAC and network layer are the features used by the classifier to generate classified instances. 
Each sniffer will calculate the classified instances at set intervals. 
The instances are communicated to the super node.
The super node applies an iterative linear regression to generate a time-based profile called the accumulated measure of fluctuation for malicious and normal nodes.
A profile of a malicious and a normal node is obtained, and an anomaly is detected using this.
The authors evaluate their system using a simulation, using Cooja and MATLAB, performance evaluation is done on how efficiently the nodes are labelled by the sniffers and how the supernodes perform the algorithms.

Hassanzadeh and Stoleru (2011)~\cite{hassanzadeh2011towards}, design a multi objective IDS that  optimise on different circumstances and variable factors.
Models are developed for individual objectives, such as node energy, permissible delay in reporting data, information contained in reports, and network coverage.
Their approach uses a genetic algorithm , employing a penalised function, to solve the multi-objective optimisation problem.
The system consists of a resource constrained wireless network with a single base station. 
The network is loosely time synchronised and all nodes know their locations. 
The base station periodically collects network information and uses this information to decide which IDS functions to run, and where to run them. 
This decision is done periodically, or when network conditions change.
The GA is used to optimise a specific parameter and its output can be used as a predictive function.
The authors evaluated their algorithm through MATLAB simulations, analysing time for convergence of the algorithm, number of parameters, and efficiency of generation for different nodes.
They authors evaluated the proposed IDS on a physical testbed, this consisted of a system of 6 laptops, with constrained resources, the central IDS was running SNORT and they evaluated performance and detection accuracy.
Results show that their optimisation approach reduced overhead whilst their information exchange was sufficient for the central node to detect attacks with high levels of accuracy. 

Le et al. (2016)~\cite{le2016specification}, propose a new specification based IDS for RPL, with a new methodology to create behaviour specifications.
The authors make use of Cooja network simulator to create traces of network behaviour, specifically focusing on route establishment and maintenance.
The authors then devise two algorithm to extract a state transition system from these network traffic datasets.
These high level transitions are then used to represent a specification of positive behaviours in the system.
The authors then place the specification based detectors across the system, with both HIDS on every node as well as NIDS across the network for full coverage.
A clustering architecture is form with both the NIDS and HIDS cooperating to communicate with a central sink.
To avoid high overhead the system is not set on promiscuous mode but rather reporting is done periodically.
As nodes report information of its neighbours, a cross-check of the collected data is required.
The authors evaluated their approach using cooja, with an experiment consisting of 100 nodes in a 600x600 m area, with transmission ranges of 50m.
Evaluation was performed on detection accuracy including and precision, as well as energy usage.
Their results show a small overhead of 6.3\% increase in energy consumption per node, with high accuracy of detection including the ability to find new attacks thanks to their adaptive specification algorithm.

Midi et al. (2017)~\cite{midi2017kalis}, bring forth Kalis, a system for knowledge-driven adaptable intrusion detection for the IoT.
The authors proposed approach envisions for their IDS to adapt based on the knowledge gathered on the system.
The Architecture for Kali is the following:
1) Communication System, the Communication System interfaces with the external world. Specialised sub-components take care of interacting with traffic on different protocols.
2) Data Store: the Data Store listens for events from the Communication System on newly captured packets, manages the history of recent traffic for modules to access, and logs all traffic on disk or memory, if required by the user
3) Knowledge Base and Collective Knowledge Management:This component stores all the available information about the features of the monitored entities and networks in a unique, centralised place, and makes this information available to all the parties requiring it, such as Detection Modules and the Module Manager.
The core component of Kalis is the way they model the knowledge about the system.
The knowledge in the system may take various different forms and is consequently labelled with different types, e.g. number of components in a cluster is an int whilst whether the cluster is multi-hoop is a boolean.
Each knowledge component has a creator ID associated with the creating node and an entity field for whom it refers to.
These knowledge components may be multi-level and collaboratively enhanced.
The collective knowledge of the system is the combination of all te elements created by the various components of the system.
This knowledge is used by the IDS to make informed decisions about relevant information.
Knowledge elements can be static or adaptable (e.g. signal strength).
The final component of the architecture is 4) modules, in Kalis any network feature-specific or attack-specific functionality is implemented as an independent module. 
The Module Manager component coordinates all the modules, activating/deactivating them as needed, depending on changes in the Knowledge Base,
Modules are divided into \textit{sensing modules}, which are the core of the knowledge discovery mechanism and \textit{detection modules} which analyse the captured traffic together with the knowledge elements to look for, they implement this on a TelosB board using TinyOS.
For their experiment they create a WSN of 6 of these nodes, a smart thermostat a smart lock, a smart light bulb an arlo security system~\footnote{Available at : \url{https://www.arlo.com/uk/products/arlo/default.aspx}} and an amazon dash button.
The Kalis node is placed near the centre of the WSN to overhear all traffic.
They perform the following evaluations 1) breath of coverage, 2) performance of their detection, 3) adaptability of their system to changes, and, 4) benefits of the collaborative mechanism.
To test the system a ICMP dos attack flood is used.
Their evaluation shows good results even when an initial environment is not specified showing the adaptability, good detection accuracy compared to SNORT IDS and low overhead for their knowledge based algorithm.

Shreenivas et al. (2017)~\cite{shreenivas2017intrusion}, propose new extensions to their previous IDS SVELTE to detect insider attacks, specifically routing attacks, in unsupervised systems.
To achieve this aim the authors make used of ETX (Expected Transmissions) metric, ETX is a link reliability metric and monitoring the ETX value can prevent an intruder from actively engaging 6LoWPAN nodes in malicious activities.
A  type of attack that is possible in these kinds of networks is for a malicious insider to advertise a fake ETX link to trick other nodes in believing it has a strong connection, therefore modifying routing.
The authors therefore propose an IDS based on expected ETX values, their approach estimates the value for each node and values which are strongly different get counted as anomalies.
As their algorithm relies heavily on inter node communication it does not cater to multiple insiders.
To adapt to this they propose an extra caveat of geographical hints, to get geographical insights clusters are made and a knowledge base of neighbourhoods is created so if several nodes are misbehaving from the previous knowledge base it can be spotted.
Similarly to the precious approach the authors implement their IDS in Cooja for evaluation.
They preform analysis on power consumption, true positive rate and memory usage.
Results show that for this specific attack the detection has improved, with a slightly higher overhead.

Bostani and Sheikhan (2017)~\cite{bostani2017hybrid}, propose a hybrid IDS, both using anomaly and specification based detection, for the IoT.
Their unsupervised clustering approach is used for the anomaly detection, whilst specifications are used for known attack patterns.
To deal with large amount of data they make use of Googles MapReduce, this approach reduces computation into smaller tasks on different machines and is ran simultaneously.
Their IDS has the following stages: 
Stage 1) identifying malicious nodes using specification, the goal of this stage is identifying the suspicious nodes that may cause sinkhole and selective-forwarding attacks.
To detect Sinkhole attacks, the authors propose a specification indicating that when a sinkhole is happening the rate of parent change for a node is increased, if this happens then a Sinkhole is occurring.
To detect selective forwarding attacks, the authors use the following signature: when a malicious node wants to launch a selective-forwarding attack, it selectively forwards packets to the root; so, a preferred parent node can identify the suspicious node by knowing the approximate number of packets received from each node.
Stage 2) anomaly detection, the anomaly based IDS creates a sample for each source node by extracting four traffic-related features from the raw received packet of the source node at each time-slot (a) packet receiving rate; (b) packet dropping rate; (c) average latency; and (d) maximum hop-count.
The MapReduce approach is used to optimise this process, a sample is sent to each source node to compute the features in parallel.
Stage 3) anomaly detection decision based on a voting mechanism, in this stage, the proposed framework employs the first stage results to make a decision about abnormalities detected in the second stage.
The authors create their own WSN simulator for RPL using .NET and c\#.
They also implemented their IDS using MATLAB on a standard PC receiving the trace data from the simulator.
Several simulations were run using different network parameters and detection accuracy was evaluated.
As an experiment several other RPL attacks were evaluated to examine the flexibility of their IDS.
As a final consideration the authors further evaluate their work on a real-world scenario.
They implemented their a system using Waspmote Mote Runner~\footnote{Available at : \url{http://www.libelium.com/products/waspmote-mote-runner-6lowpan/}} for gateways and nodes.
Their setup mimicked a smart city including, smart lighting, acoustic noise-maps, air quality monitoring ans waste management.
This is available as a smart kit from WaspMote.
The IDSs are implemented in the gateways and evaluation is performed.
No results are discussed on the real-word implementation but rather a feasibility and implementation discussion is provided.

Arolkar et al. (2011)~\cite{arolkar2011ant}, focus on proposing a new biologically inspired algorithm for Intrusion detection is WSN.
Their proposed ant colony based architecture has the following for stages 1) Formation of cluster without malicious node, To form clusters, the WSN is divided into different geographical  regions G. Within each geographical region G a node N is chosen randomly and a parameter L that indicates the level of neighbour in the cluster is decided.
2)Identification of Cluster Head within Clusters, three random nodes are chosen within the cluster. 
Whichever node has with highest resources is then selected as a cluster head.
3) Deployment of Ant Pheromones on Cluster Heads, cluster heads deploy initial pheromone  values using ant agents on the neighbour cluster heads. The pheromone values are refreshed based on preset time intervals.
4) Routing and Detecting Malicious Activities based on Pheromone Values, once the pheromone is deployed over all the cluster heads, the 
routing process can begin. 
The routing process is now a two stage 
process. 
In the first stage, data is sent to respective cluster head by the source node.
In the second stage, the source cluster head, based on the  pheromone value, then analyses possible threats like sinkhole  attack, misdirection and passive information gathering. 
A sinkhole attack is detected if any cluster head responds with a very high pheromone value before the completion of the pheromone refresh time interval.
Misdirection is detected if the received pheromone value is very low even after it is recently 
refreshed
Passive information gathering is said to be done if energy and pheromone values do not decrease proportionally.
No evaluation is performed.

Coppolino et al. (2013)~\cite{coppolino2013applying}, propose an adaptive IDS for attacks on constrained networks.
The architecture of the proposed IDS comprises a Central Agent(CA), which is connected to a number of Local Agents (LAs)running on the WSN nodes. The Central Agent performs misuse-based detection activities, while lighter, anomaly-based detection techniques are used by the Local Agents to spot local symptoms of an intrusion.
On the contrary of other similar approaches that use a fixed anomaly threshold, the authors adopted a more flexible approach to threshold setting that consists in estimating the threshold value based on the specific characteristics of the WSN being monitored.
This allows to decrease communication overhead, which are highly energy consuming for WSN motes.
The Central Agent is located outside the network on a server Local Agents perform local agents perform a preliminary analysis within the network.
When one or more Local Agents detect a potential attack, an alert is raised and propagated to the Central Agent.
Local Agents have three components 1) Packet Monitor, 2) Data Collector and 3)Detection Engine.
The Local Agents have threshold based anomaly as lightweight detectors.
The Central Agent uses a Decision Tree based algorithm.
The IDS has two stages 1) profiling, this is where the Central Agent maps out the network topology and then \textit{normal} behaviours are mapped based on reports from the local agents.
The normal behaviour is set by a set of parameters each with ranges of values.
The central agent stores each of these parameters and its accepted ranges.
Stage 2) each Local Agent performs anomaly-based intrusion detection based on the normal profile built in the previous stage. 
A local alert is raised whenever a parameter of the monitored node exceeds the normal profile threshold; 
the Central Agent performs misuse detection for the whole WSN and validates local alerts through the multidimensional representation of parameters it create din the previous step.
The authors specifically focus on sinkhole attack as well as sleep deprivation attack, an attack whose focus is to exhaust the resources of a node by not allowing it to enter sleep mode.
Evaluation was performed with different decision tree approaches on static data purely on false positive and false negative rates. 

Abishek et al. (2018)~\cite{abhishek2018intrusion}, propose an IDS specifically for detecting compromised gateways something which is quite difficult in  the context of systems using clusters.
The proposed system uses the packet drop probability as a means to monitor the gateways.
As part of this technique all  the  IoT  devices  will  track  the  number of packets dropped at their respective ends.
This is then reported periodically to the end point.
Based on deviations from expected drops the authors are able to detect anomalies using the Maximum Likelihood estimation.
The approach is evaluated using a simulation in MATLAB, using RMSE they estimate the effectiveness of the algorithm.

Pongle and Chavan (2015)~\cite{pongle2015real}, propose an IDS specifically for the detection of wormhole attacks.
Their architecture envisions two types of modules, centralised and distributed.
They both perform the following operations: 
i) Neighbor Validation.
In this module we are collecting the neighbour information from all sensor nodes storing them. 
The stored neighbours information are verified based on the distance between the node and that neighbour.
If the distance found to be more than transmission range of node then this module send the victim packet to the information sender node and to the neighbour whose distance is more than transmission range.
ii) Distance RSSI.
This  module  calculates  the  distance  be-tween two geographical coordinates. It also convert the RSSI (Received Signal Strength Indicator) value to distance and vice versa.
iii) RSSI collection.
The calculated distances are sent to the central nodes for collection.
iv) Attacker detection.
Once the distances are calculated they are used by the central node to detect attacks.
To evaluate this approach the authors implement their algorithm as well as a wormhole attacker in Cooja, using Tmote Sky nodes, with a non constrained central component running a Linux pc. 
The experiments are done on 8, 16, and 24 nodes.
Evaluation is performed on energy overhead as well as true positive detection.

As the tradional grid moved towards a \textit{smart grid}, several new security challenges arise, to cater to this, Zhang et al. (2011)~\cite{zhang2011artificial} propose an IDS to be able to detect attacks on smart grids.
Most of these changes will occur as an Internet-like communications network is superimposed on top of the current power grid using wireless technologies including 802.15.4, 802.11, and the Zigbee protocol. 
Each of these will expose the power grid to cyber security threats.
As the smart grid infrastructure is expansive and covers several components, including power plants, homes and the transfer network itself, the authors propose a distributed collaborative approach with different IDSs at every layer of the smart infrastructure.
Firstly the authors consider a network architecture with three key components, 1) home area networks (HAN), 2) neighbourhood area networks (NAN), and 3) wide area networks (WAN). 
Each of these types networks will contain a different kind of IDS.
At the HAN level the main component is the service module with provides real time energy costs and consumption to the costumer and record usage statistics for the grid.
The HAN IDS will take the form of a Host Based IDS, to detect malicious behaviour on the sensing module itself.
At the NAN level, we see a large metering system which combines all the HAN networks in the area.
Each HAN will report to its nearest HAN and all output from the individual HIDS is sent to the NAN to process.
The NAN contains a Network based IDS to detect network attacks which may come from within the network our aimed at attacking the metering infrastructure.
Finally the WAN, is in charge of the whole metering infrastructure for the grid, receiving data from all the NANs, the job of the NAN IDS is to provide a barrier between the outside network data and the SCADA controllers in the grid to ensure that the energy and service cooperation can be successful.
At the network layer attacks taken in consideration are DoS attacks, man in the middle attacks and data leakage of the costumer data.
Each of the different IDS will incorporate an Immunity based algorithm for intrusion detection.
The authors specifically choose the AIRS2Parallel~\cite{watkins2004exploiting} and CLONALG~\cite{brownlee2005clonal} algorithm, and evaluate both approaches.
To evaluate their approach the authors firstly train their algorithms using trace data from the KDD data set, and secondly they run a simulation in MATLAB, where they construct a hierarchical architecture with each of their components systems.
Each of their different IDS are tested using the two algorithms and results show AIRS2Parallel having better results, the only evaluating having been done on detection and not performance.  

Yu and Tsai (2008)~\cite{yu2008framework}, propose a hybrid approach to intrusion detection combining both a anomaly based technique and specification based technique, with the unique approach that the behaviours and specifications are automatically computed.
In their proposed architecture, each node is equipped with an intrusion detection agent (IDA), however there is no collaboration to avoid insider attacks.
Unlike traditional HIDS where it is possible to analyse all features in the host in constrained systems it is quite possible for a device to be wiped, so for their IDA they propose two approaches.
The authors propose a Local Intrusion Detection Components (LIDC) to detect whether a node is being attacked by other nodes within the network.
This is then combined with a Packet based Intrusion Detection component (PIDC), which monitors network packets sent within the neighbourhood of the node.
Any intrusion alert is then send to the base station to process using a anomaly based IDS.
The authors then propose a comprehensive list of features that are of interest to the positive behaviour of a WSN, such as transmission rate and packet drop rate.
Using these features the authors propose the use of SLIPPER~\cite{cohen1999simple} a machine learning algorithm which automatically generates rules out of data.
Using the features collected by the nodes in the networks the central IDS is able to automatically generate rules for the IDS without prior intervention.
Each time an alarm is sent the model is tuned in case of false alarms, and the IDAs are updated.
No evaluation is performed on their approach.

Buennemeyer et al. (2008),~\cite{buennemeyer2008mobile}, provide a device profiling methodology able to detect battery exhausting attacks on IoT devices.
The IDS works on WiFi and Bluetooth and makes use of a battery profiling engine CIDE to view the battery drain.
CIDE uses a dynamic threshold calculation algorithm that is able to detect unusual patters of battery drain that might correlate to attacks.
Each host is equipped with an IDS able to raise alarms regarding battery consumption, this is combined with the integration to a Network Based IDS to whom alerts are sent.
For their specific work the authors make use of SNORT, and the alerts are correlated to rules in the central IDS.
To minimise the number of alerts and false positives the battery thresholds are dynamically tuned during operation.
Each device in the system must be configured with a smart battery whose parameters are tracked by the CIDE installed on the device.
A CIDE will also be present on a server, this provides a graphical user interface able to let the user configure the device characteristics as well as monitor the battery fluctuations.
To evaluate the best reporting rate for the IDSs on the device the users implemented their system on, Ten Dell Axim X51 PDAs, all with the same settings and initial battery levels.
Further experimentation was done on several other devices and they were able to determine optimal polling rates to diminish battery rates, with the caveat that in some cases due to high optimal polling rate it let them vulnerable to timing attacks within the poll interval.
As an additional evaluation the author performed a usability study with 31 participants, collecting feedback on usability of the tool, quality of reporting and several other metrics, the expert feedback found their hybrid approach to be very useful.
The authors also showed changes made in accordance with this usability study.

Moyers et al. (2010)~\cite{moyers2010multi}, propose Multi-Vector Portable Intrusion Detection System (MVP-IDS).
In this work the authors expand on the work of B-SIPS by introducing an element of network monitoring to the battery drain evaluation performed by previous work.
They integrate Snort rules within the HOST IDS to perform more pre-detection at the device level directly.
The MVP-IDS approach is to be able to correlate specific changes in battery usage to network traffic incoming to the divide level.
The architecture of the system has four components 1) B-SIPS client, for current anomaly triggering, 2) SNORT WiFi module for WiFi attack detection, 3) SNORT Bluetooth attack detection system for Bluetooth attacks, and 4) CIDE server correlation response and holistic view of the system.
In Buennemeyer’s B-SIPS project, B-SIPS clients sent device status reports to the CIDE server, but never received any response because the system was implemented with uni-directional communication.  
An improvement that was added in the creation of MVP-IDS was changing the system to operate in a bi-directional manner.  
Not only do B-SIPS clients send device status report to the CIDE server, but the CIDE server also then replies when action needs to be taken to secure the device and alert the user of malicious activity.
No evaluation was performed on their approach  and no discussion on how their new approach would improve upon the existing IDS.

\section{Collecting Tools for Unified Evaluation}
\label{sec:tool-collection}
To to gather a comprehensive evaluation of the capabilities of IDSs for IoT we firstly needed to gather implementations for the tools.
To do so we firstly classified the tools into four categories, 
i) tools which had code available online, 
ii) tools which didn't have code online, however discussed implementations and therefore should have code, 
iii) tools which had implemented in simulations or theoretically analysed, as perhaps they has since prepared an implementation, and, 
iv) tools which had no evaluation nor discussed implementation details.

Secondly we needed to get a updated list of contact details for all the authors, as some of the papers were quite old and due to the nature of academic positions moving around a lot, many of the emails on the papers were outdated.
To get up to date emails we first looked up to see if the author had moved to a new institution and had new contact details, this allowed us to find several new emails.
In the case of the authors leaving academia as was often the case for PhD and masters students, we looked for Linkedin profiles for emails, even though this was not always possible.
Whilst it was often the case for professors or senior members of staff to have remained in the same place, our intuition was that it was worth contacting the full author list as it is often the case that the implementation is performed by the less senior members of staff and therefore it is more likely to get access from them.
For some authors we were unable to identify their new email or whether the current email was still valid.
In total 136, email addressed were collected, to the best of our knowledge with the most up to date details.

The third step was contacting the authors for their tools.
We constructed an email template, customisable for each tool paper, as can be seen in Appendix~A and Appendix~B.
The email has seven components, 1) a custom greeting to the specific authors addressing them by their title, 2) a short introduction about who I am as well as all my contact details, 3) an introduction to the work I am working on, 3) reasons for contacting them with specific reference to their paper, 4) a personalised comment on their work, manually generated, 5) request for the tool code, 6) a query asking how authors would like to be cited in case of publications and if any licenses applied to their tools and 7) a closing statement thanking them and letting them know that if they may have any question they may feel free to contact me.

Emails were sent out, in two stages, the first stage was sent out to any author in category ii of classification, except for tools that were not suitable to test in our designed evaluation IoT tesbed, an example of this was tools that looked at physical intruders in a home.
A second stage was sent out to authors in category iii and iv of classification whose proposed solutions may reasonably be implemented (and may have been since), this included approaches who implemented their evaluation through a network simulator.
A total of 84 authors were contacted via email in the two stages.

The resulting collection totalled to, three tools in category iii, whose code was available on GitHub, of which all three were based on CONTIKI simulator, and non suitable to our testing environment. 
One tool pertaining to category ii, which responded to my emails and provided a code implementation (not publicly available).
A single tool in category i whose implementation was available online.
And no responses in other categories.
The net collection resulted in 4/51 papers with code shared publicly, 1/84 authors who responded to a request for implementation, and 46/51 papers with no code accounted for.
We do not claim that our inability to find these tools means they were unavailable but merely that following our protocol, these were the ones available to us.
We therefore relegate the initial goal of evaluating the available tools under the same environment to a future work.

\subsection{Analysis}

The review we have conducted has shown there is a dire need for better evaluation of these systems.
Whilst several of these approaches make use of simulation, which one would presume provides more ease in comparing results, there is still a lack of disclosure of both datasets and the tools themselves.
We observe that this issue is not unique to the area of IoT, as even on traditional IDS papers, there is a lack of available datasets, with many evaluations still relying on outdated datasets such as KDD 99, almost two decades later.

The metrics of evaluation are something that we choose as a core feature of our approach.
Whilst accuracy is very important for any prediction system, perhaps not as much as one would think.
The ability for an IDS to perform, run, and be analysed is just as if not more important.
Consequently, simulation might not be enough to realistically evaluate the system.
There are of course limitations in terms of constraints of access to realistic test beds for implementation of the system.

\section{Conclusion}
\label{sec:chaptConc}

In this work we have extensively analysed and summarised the current state of the art in intrusion detection in the IoT.
We have reviewed and summaries 51 proposed tools and technologies and characterised them.
Our analysis has shown several new insights into how research in this field is developing, and potential gaps for future works.
We have identified a major gap for the reproducibility of the tools evaluation and also discussed shortcomings in the current evaluation process, namely 1) single focus on detection accuracy; 2) lack of unified evaluation methodology; and 3) disregard for evaluation regarding usability of these tools.
Our findings indicate that perhaps the very large numbers of papers in this area is in part due to the inability to build upon the software and the lack of research into tools that can be monitored and supervisable.
We see a need for a shift to focus away from development of new techniques to a focus on making these tools that could scale and be usable across IoT deployments.
The analysis shows that this is in part due to \emph{how} the tools have been evaluated and we suggest the usage of metrics from Etalle, S (2019)~\cite{etalle2017intrusion} as a better set of metrics against which to decide the suitability of an IDS for IoT systems. 
As a final consideration we see a thorough comparative evaluation by these tools through this evaluation a beneficial and necessary task, although we reserve this for future work as there currently is not enough access to the proposed tools.

\bibliographystyle{ACM-Reference-Format}
\bibliography{bibliography}


\begin{thebibliography}{93}


\ifx \showCODEN    \undefined \def \showCODEN     #1{\unskip}     \fi
\ifx \showDOI      \undefined \def \showDOI       #1{#1}\fi
\ifx \showISBNx    \undefined \def \showISBNx     #1{\unskip}     \fi
\ifx \showISBNxiii \undefined \def \showISBNxiii  #1{\unskip}     \fi
\ifx \showISSN     \undefined \def \showISSN      #1{\unskip}     \fi
\ifx \showLCCN     \undefined \def \showLCCN      #1{\unskip}     \fi
\ifx \shownote     \undefined \def \shownote      #1{#1}          \fi
\ifx \showarticletitle \undefined \def \showarticletitle #1{#1}   \fi
\ifx \showURL      \undefined \def \showURL       {\relax}        \fi
\providecommand\bibfield[2]{#2}
\providecommand\bibinfo[2]{#2}
\providecommand\natexlab[1]{#1}
\providecommand\showeprint[2][]{arXiv:#2}

\bibitem[\protect\citeauthoryear{Abhishek, Lim, Sikdar, and Tandon}{Abhishek
  et~al\mbox{.}}{2018}]%
        {abhishek2018intrusion}
\bibfield{author}{\bibinfo{person}{Nalam~Venkata Abhishek},
  \bibinfo{person}{Teng~Joon Lim}, \bibinfo{person}{Biplab Sikdar}, {and}
  \bibinfo{person}{Anshoo Tandon}.} \bibinfo{year}{2018}\natexlab{}.
\newblock \showarticletitle{An intrusion detection system for detecting
  compromised gateways in clustered IoT networks}. In
  \bibinfo{booktitle}{\emph{2018 IEEE International Workshop Technical
  Committee on Communications Quality and Reliability (CQR)}}. IEEE,
  \bibinfo{pages}{1--6}.
\newblock


\bibitem[\protect\citeauthoryear{Amaral, Oliveira, Rodrigues, Han, and
  Shu}{Amaral et~al\mbox{.}}{2014}]%
        {amaral2014policy}
\bibfield{author}{\bibinfo{person}{Jo{\~a}o~P Amaral},
  \bibinfo{person}{Lu{\'\i}s~M Oliveira}, \bibinfo{person}{Joel~JPC Rodrigues},
  \bibinfo{person}{Guangjie Han}, {and} \bibinfo{person}{Lei Shu}.}
  \bibinfo{year}{2014}\natexlab{}.
\newblock \showarticletitle{Policy and network-based intrusion detection system
  for IPv6-enabled wireless sensor networks}. In \bibinfo{booktitle}{\emph{2014
  IEEE International Conference on Communications (ICC)}}. IEEE,
  \bibinfo{pages}{1796--1801}.
\newblock


\bibitem[\protect\citeauthoryear{Amouri, Alaparthy, and Morgera}{Amouri
  et~al\mbox{.}}{2018}]%
        {amouri2018cross}
\bibfield{author}{\bibinfo{person}{Amar Amouri}, \bibinfo{person}{Vishwa~T
  Alaparthy}, {and} \bibinfo{person}{Salvatore~D Morgera}.}
  \bibinfo{year}{2018}\natexlab{}.
\newblock \showarticletitle{Cross layer-based intrusion detection based on
  network behavior for IoT}. In \bibinfo{booktitle}{\emph{2018 IEEE 19th
  Wireless and Microwave Technology Conference (WAMICON)}}. IEEE,
  \bibinfo{pages}{1--4}.
\newblock


\bibitem[\protect\citeauthoryear{Arnaboldi and Morisset}{Arnaboldi and
  Morisset}{2017}]%
        {arnaboldi2017quantitative}
\bibfield{author}{\bibinfo{person}{Luca Arnaboldi} {and}
  \bibinfo{person}{Charles Morisset}.} \bibinfo{year}{2017}\natexlab{}.
\newblock \showarticletitle{Quantitative analysis of dos attacks and client
  puzzles in iot systems}. In \bibinfo{booktitle}{\emph{International Workshop
  on Security and Trust Management}}. Springer, \bibinfo{pages}{224--233}.
\newblock


\bibitem[\protect\citeauthoryear{Arnaboldi and Morisset}{Arnaboldi and
  Morisset}{2018a}]%
        {arnaboldi2018generating}
\bibfield{author}{\bibinfo{person}{Luca Arnaboldi} {and}
  \bibinfo{person}{Charles Morisset}.} \bibinfo{year}{2018}\natexlab{a}.
\newblock \showarticletitle{Generating Synthetic Data for Real World Detection
  of DoS Attacks in the IoT}. In \bibinfo{booktitle}{\emph{Federation of
  International Conferences on Software Technologies: Applications and
  Foundations}}. Springer, \bibinfo{pages}{130--145}.
\newblock


\bibitem[\protect\citeauthoryear{Arnaboldi and Morisset}{Arnaboldi and
  Morisset}{2018b}]%
        {arnaboldi2018lisa}
\bibfield{author}{\bibinfo{person}{Luca Arnaboldi} {and}
  \bibinfo{person}{Charles Morisset}.} \bibinfo{year}{2018}\natexlab{b}.
\newblock \showarticletitle{LISA: Predicting the Impact of DoS Attacks on
  Real-World Low Power IoT Systems}.
\newblock \bibinfo{journal}{\emph{Foundations of Computer Security Workshop}}
  (\bibinfo{year}{2018}).
\newblock


\bibitem[\protect\citeauthoryear{Arnaboldi and Tschofenig}{Arnaboldi and
  Tschofenig}{2019}]%
        {arnaboldi2019AceOAuth}
\bibfield{author}{\bibinfo{person}{Luca Arnaboldi} {and}
  \bibinfo{person}{Hannes Tschofenig}.} \bibinfo{year}{2019}\natexlab{}.
\newblock \showarticletitle{{A Formal Model for Delegated Authorization of IoT
  Devices Using ACE-OAuth}}. In \bibinfo{booktitle}{\emph{OAuth Security
  Workshop}}.
\newblock


\bibitem[\protect\citeauthoryear{Arolkar, Sheth, and Tamhane}{Arolkar
  et~al\mbox{.}}{2011}]%
        {arolkar2011ant}
\bibfield{author}{\bibinfo{person}{Harshal~A Arolkar},
  \bibinfo{person}{Shraddha~P Sheth}, {and} \bibinfo{person}{Vaidehi~P
  Tamhane}.} \bibinfo{year}{2011}\natexlab{}.
\newblock \showarticletitle{Ant colony based approach for intrusion detection
  on cluster heads in WSN.}. In \bibinfo{booktitle}{\emph{ICCCS}}.
  \bibinfo{pages}{523--526}.
\newblock


\bibitem[\protect\citeauthoryear{Arrington, Barnett, Rufus, and
  Esterline}{Arrington et~al\mbox{.}}{2016}]%
        {arrington2016behavioral}
\bibfield{author}{\bibinfo{person}{Briana Arrington}, \bibinfo{person}{LiEsa
  Barnett}, \bibinfo{person}{Rahmira Rufus}, {and} \bibinfo{person}{Albert
  Esterline}.} \bibinfo{year}{2016}\natexlab{}.
\newblock \showarticletitle{Behavioral modeling intrusion detection system
  (bmids) using internet of things (iot) behavior-based anomaly detection via
  immunity-inspired algorithms}. In \bibinfo{booktitle}{\emph{2016 25th
  International Conference on Computer Communication and Networks (ICCCN)}}.
  IEEE, \bibinfo{pages}{1--6}.
\newblock


\bibitem[\protect\citeauthoryear{Axelsson}{Axelsson}{2000}]%
        {axelsson2000base}
\bibfield{author}{\bibinfo{person}{Stefan Axelsson}.}
  \bibinfo{year}{2000}\natexlab{}.
\newblock \showarticletitle{The base-rate fallacy and the difficulty of
  intrusion detection}.
\newblock \bibinfo{journal}{\emph{ACM Transactions on Information and System
  Security (TISSEC)}} \bibinfo{volume}{3}, \bibinfo{number}{3}
  (\bibinfo{year}{2000}), \bibinfo{pages}{186--205}.
\newblock


\bibitem[\protect\citeauthoryear{Besson and Leleu}{Besson and Leleu}{2009}]%
        {besson2009distributed}
\bibfield{author}{\bibinfo{person}{Lionel Besson} {and}
  \bibinfo{person}{Philippe Leleu}.} \bibinfo{year}{2009}\natexlab{}.
\newblock \showarticletitle{A distributed intrusion detection system for ad-hoc
  wireless sensor networks: the AWISSENET distributed intrusion detection
  system}. In \bibinfo{booktitle}{\emph{2009 16th International Conference on
  Systems, Signals and Image Processing}}. IEEE, \bibinfo{pages}{1--3}.
\newblock


\bibitem[\protect\citeauthoryear{Bostani and Sheikhan}{Bostani and
  Sheikhan}{2017}]%
        {bostani2017hybrid}
\bibfield{author}{\bibinfo{person}{Hamid Bostani} {and}
  \bibinfo{person}{Mansour Sheikhan}.} \bibinfo{year}{2017}\natexlab{}.
\newblock \showarticletitle{Hybrid of anomaly-based and specification-based ids
  for internet of things using unsupervised opf based on mapreduce approach}.
\newblock \bibinfo{journal}{\emph{Computer Communications}}
  \bibinfo{volume}{98} (\bibinfo{year}{2017}), \bibinfo{pages}{52--71}.
\newblock


\bibitem[\protect\citeauthoryear{Brownlee}{Brownlee}{2005}]%
        {brownlee2005clonal}
\bibfield{author}{\bibinfo{person}{Jason Brownlee}.}
  \bibinfo{year}{2005}\natexlab{}.
\newblock \showarticletitle{Clonal selection theory \& clonalg-the clonal
  selection classification algorithm (csca)}.
\newblock \bibinfo{journal}{\emph{Swinburne University of Technology}}
  (\bibinfo{year}{2005}), \bibinfo{pages}{38}.
\newblock


\bibitem[\protect\citeauthoryear{Buennemeyer, Gora, Marchany, and
  Tront}{Buennemeyer et~al\mbox{.}}{2007}]%
        {buennemeyer2007battery}
\bibfield{author}{\bibinfo{person}{Timothy~K Buennemeyer},
  \bibinfo{person}{Michael Gora}, \bibinfo{person}{Randy~C Marchany}, {and}
  \bibinfo{person}{Joseph~G Tront}.} \bibinfo{year}{2007}\natexlab{}.
\newblock \showarticletitle{Battery exhaustion attack detection with small
  handheld mobile computers}. In \bibinfo{booktitle}{\emph{Portable Information
  Devices}}.
\newblock


\bibitem[\protect\citeauthoryear{Buennemeyer, Nelson, Clagett, Dunning,
  Marchany, and Tront}{Buennemeyer et~al\mbox{.}}{2008}]%
        {buennemeyer2008mobile}
\bibfield{author}{\bibinfo{person}{Timothy~K Buennemeyer},
  \bibinfo{person}{Theresa~M Nelson}, \bibinfo{person}{Lee~M Clagett},
  \bibinfo{person}{John~P Dunning}, \bibinfo{person}{Randy~C Marchany}, {and}
  \bibinfo{person}{Joseph~G Tront}.} \bibinfo{year}{2008}\natexlab{}.
\newblock \showarticletitle{Mobile device profiling and intrusion detection
  using smart batteries}. In \bibinfo{booktitle}{\emph{Proceedings of the 41st
  Annual Hawaii International Conference on System Sciences (HICSS 2008)}}.
  IEEE, \bibinfo{pages}{296--296}.
\newblock


\bibitem[\protect\citeauthoryear{Butun, Morgera, and Sankar}{Butun
  et~al\mbox{.}}{2014}]%
        {butun2014survey}
\bibfield{author}{\bibinfo{person}{Ismail Butun}, \bibinfo{person}{Salvatore~D
  Morgera}, {and} \bibinfo{person}{Ravi Sankar}.}
  \bibinfo{year}{2014}\natexlab{}.
\newblock \showarticletitle{A survey of intrusion detection systems in wireless
  sensor networks}.
\newblock \bibinfo{journal}{\emph{IEEE communications surveys \& tutorials}}
  \bibinfo{volume}{16}, \bibinfo{number}{1} (\bibinfo{year}{2014}),
  \bibinfo{pages}{266--282}.
\newblock


\bibitem[\protect\citeauthoryear{Castro, De~Castro, and Timmis}{Castro
  et~al\mbox{.}}{2002}]%
        {castro2002artificial}
\bibfield{author}{\bibinfo{person}{Leandro~Nunes Castro},
  \bibinfo{person}{Leandro~Nunes De~Castro}, {and} \bibinfo{person}{Jonathan
  Timmis}.} \bibinfo{year}{2002}\natexlab{}.
\newblock \bibinfo{booktitle}{\emph{Artificial immune systems: a new
  computational intelligence approach}}.
\newblock \bibinfo{publisher}{Springer Science \& Business Media}.
\newblock


\bibitem[\protect\citeauthoryear{Cervantes, Poplade, Nogueira, and
  Santos}{Cervantes et~al\mbox{.}}{2015}]%
        {cervantes2015detection}
\bibfield{author}{\bibinfo{person}{Christian Cervantes}, \bibinfo{person}{Diego
  Poplade}, \bibinfo{person}{Michele Nogueira}, {and} \bibinfo{person}{Aldri
  Santos}.} \bibinfo{year}{2015}\natexlab{}.
\newblock \showarticletitle{Detection of sinkhole attacks for supporting secure
  routing on 6LoWPAN for Internet of Things}. In \bibinfo{booktitle}{\emph{2015
  IFIP/IEEE International Symposium on Integrated Network Management (IM)}}.
  IEEE, \bibinfo{pages}{606--611}.
\newblock


\bibitem[\protect\citeauthoryear{Cho, Kim, and Hong}{Cho et~al\mbox{.}}{2009}]%
        {cho2009attack}
\bibfield{author}{\bibinfo{person}{Eung~Jun Cho}, \bibinfo{person}{Jin~Ho Kim},
  {and} \bibinfo{person}{Choong~Seon Hong}.} \bibinfo{year}{2009}\natexlab{}.
\newblock \showarticletitle{Attack model and detection scheme for botnet on
  6LoWPAN}. In \bibinfo{booktitle}{\emph{Asia-Pacific Network Operations and
  Management Symposium}}. Springer, \bibinfo{pages}{515--518}.
\newblock


\bibitem[\protect\citeauthoryear{Chormunge and Jena}{Chormunge and
  Jena}{2015}]%
        {chormunge2015efficiency}
\bibfield{author}{\bibinfo{person}{Smita Chormunge} {and}
  \bibinfo{person}{Sudarson Jena}.} \bibinfo{year}{2015}\natexlab{}.
\newblock \showarticletitle{Efficiency and Effectiveness of Clustering
  Algorithms for High Dimensional Data}.
\newblock \bibinfo{journal}{\emph{International Journal of Computer
  Applications}} \bibinfo{volume}{125}, \bibinfo{number}{11}
  (\bibinfo{year}{2015}).
\newblock


\bibitem[\protect\citeauthoryear{Cohen and Singer}{Cohen and Singer}{1999}]%
        {cohen1999simple}
\bibfield{author}{\bibinfo{person}{William~W Cohen} {and}
  \bibinfo{person}{Yoram Singer}.} \bibinfo{year}{1999}\natexlab{}.
\newblock \showarticletitle{A simple, fast, and effective rule learner}.
\newblock \bibinfo{journal}{\emph{AAAI/IAAI}}  \bibinfo{volume}{99}
  (\bibinfo{year}{1999}), \bibinfo{pages}{335--342}.
\newblock


\bibitem[\protect\citeauthoryear{Coppolino, DAntonio, Garofalo, and
  Romano}{Coppolino et~al\mbox{.}}{2013}]%
        {coppolino2013applying}
\bibfield{author}{\bibinfo{person}{Luigi Coppolino}, \bibinfo{person}{Salvatore
  DAntonio}, \bibinfo{person}{Alessia Garofalo}, {and} \bibinfo{person}{Luigi
  Romano}.} \bibinfo{year}{2013}\natexlab{}.
\newblock \showarticletitle{Applying data mining techniques to intrusion
  detection in wireless sensor networks}. In \bibinfo{booktitle}{\emph{2013
  Eighth International Conference on P2P, Parallel, Grid, Cloud and Internet
  Computing}}. IEEE, \bibinfo{pages}{247--254}.
\newblock


\bibitem[\protect\citeauthoryear{Craenen and Eiben}{Craenen and Eiben}{2002}]%
        {craenen2002computational}
\bibfield{author}{\bibinfo{person}{B Craenen} {and} \bibinfo{person}{A Eiben}.}
  \bibinfo{year}{2002}\natexlab{}.
\newblock \showarticletitle{Computational intelligence. Encyclopedia of Life
  Support Sciences}.
\newblock \bibinfo{journal}{\emph{EOLSS, EOLSS Co. Ltd}}
  (\bibinfo{year}{2002}).
\newblock


\bibitem[\protect\citeauthoryear{Danda and Hota}{Danda and Hota}{2016}]%
        {danda2016attack}
\bibfield{author}{\bibinfo{person}{Jagan Mohan~Reddy Danda} {and}
  \bibinfo{person}{Chittaranjan Hota}.} \bibinfo{year}{2016}\natexlab{}.
\newblock \showarticletitle{Attack identification framework for IoT devices}.
\newblock In \bibinfo{booktitle}{\emph{Information Systems Design and
  Intelligent Applications}}. \bibinfo{publisher}{Springer},
  \bibinfo{pages}{505--513}.
\newblock


\bibitem[\protect\citeauthoryear{Deng, Li, Yao, Cox, and Wang}{Deng
  et~al\mbox{.}}{2018}]%
        {deng2018mobile}
\bibfield{author}{\bibinfo{person}{Lianbing Deng}, \bibinfo{person}{Daming Li},
  \bibinfo{person}{Xiang Yao}, \bibinfo{person}{David Cox}, {and}
  \bibinfo{person}{Haoxiang Wang}.} \bibinfo{year}{2018}\natexlab{}.
\newblock \showarticletitle{Mobile network intrusion detection for IoT system
  based on transfer learning algorithm}.
\newblock \bibinfo{journal}{\emph{Cluster Computing}} (\bibinfo{year}{2018}),
  \bibinfo{pages}{1--16}.
\newblock


\bibitem[\protect\citeauthoryear{Di~Pietro and Mancini}{Di~Pietro and
  Mancini}{2008}]%
        {di2008intrusion}
\bibfield{author}{\bibinfo{person}{Roberto Di~Pietro} {and}
  \bibinfo{person}{Luigi~V Mancini}.} \bibinfo{year}{2008}\natexlab{}.
\newblock \bibinfo{booktitle}{\emph{Intrusion detection systems}}.
  Vol.~\bibinfo{volume}{38}.
\newblock \bibinfo{publisher}{Springer Science \& Business Media}.
\newblock


\bibitem[\protect\citeauthoryear{Echeverria, Seitz, Klinedinst, and
  Lewis}{Echeverria et~al\mbox{.}}{2019}]%
        {secheverria-disadvantaged}
\bibfield{author}{\bibinfo{person}{Sebastian Echeverria},
  \bibinfo{person}{Ludwig Seitz}, \bibinfo{person}{Dan Klinedinst}, {and}
  \bibinfo{person}{Grace Lewis}.} \bibinfo{year}{2019}\natexlab{}.
\newblock \bibinfo{booktitle}{\emph{{ACE Clients in Disadvantaged Networks}}}.
\newblock \bibinfo{type}{Internet-Draft}
  draft-secheverria-ace-client-disadvantaged-00. \bibinfo{institution}{Internet
  Engineering Task Force}.
\newblock
\urldef\tempurl%
\url{https://datatracker.ietf.org/doc/html/draft-secheverria-ace-client-disadvantaged-00}
\showURL{%
\tempurl}
\newblock
\shownote{Work in Progress.}


\bibitem[\protect\citeauthoryear{Elrawy, Awad, and Hamed}{Elrawy
  et~al\mbox{.}}{2018}]%
        {elrawy2018intrusion}
\bibfield{author}{\bibinfo{person}{Mohamed~Faisal Elrawy},
  \bibinfo{person}{Ali~Ismail Awad}, {and} \bibinfo{person}{Hesham~FA Hamed}.}
  \bibinfo{year}{2018}\natexlab{}.
\newblock \showarticletitle{Intrusion detection systems for IoT-based smart
  environments: a survey}.
\newblock \bibinfo{journal}{\emph{Journal of Cloud Computing}}
  \bibinfo{volume}{7}, \bibinfo{number}{1} (\bibinfo{year}{2018}),
  \bibinfo{pages}{21}.
\newblock


\bibitem[\protect\citeauthoryear{Esquivel-Vargas, Caselli, and
  Peter}{Esquivel-Vargas et~al\mbox{.}}{2017}]%
        {esquivel2017automatic}
\bibfield{author}{\bibinfo{person}{Herson Esquivel-Vargas},
  \bibinfo{person}{Marco Caselli}, {and} \bibinfo{person}{Andreas Peter}.}
  \bibinfo{year}{2017}\natexlab{}.
\newblock \showarticletitle{Automatic deployment of specification-based
  intrusion detection in the BACnet protocol}. In
  \bibinfo{booktitle}{\emph{Proceedings of the 2017 Workshop on Cyber-Physical
  Systems Security and PrivaCy}}. \bibinfo{pages}{25--36}.
\newblock


\bibitem[\protect\citeauthoryear{Etalle}{Etalle}{2017}]%
        {etalle2017intrusion}
\bibfield{author}{\bibinfo{person}{Sandro Etalle}.}
  \bibinfo{year}{2017}\natexlab{}.
\newblock \showarticletitle{From intrusion detection to software design}. In
  \bibinfo{booktitle}{\emph{European Symposium on Research in Computer
  Security}}. Springer, \bibinfo{pages}{1--10}.
\newblock


\bibitem[\protect\citeauthoryear{Etalle}{Etalle}{2019}]%
        {etalle2019network}
\bibfield{author}{\bibinfo{person}{Sandro Etalle}.}
  \bibinfo{year}{2019}\natexlab{}.
\newblock \showarticletitle{Network Monitoring of Industrial Control Systems:
  The Lessons of SecurityMatters}. In \bibinfo{booktitle}{\emph{Proceedings of
  the ACM Workshop on Cyber-Physical Systems Security \& Privacy}}.
  \bibinfo{pages}{1--1}.
\newblock


\bibitem[\protect\citeauthoryear{Fu, Yan, Cao, Kon{\'e}, and Cao}{Fu
  et~al\mbox{.}}{2017}]%
        {fu2017automata}
\bibfield{author}{\bibinfo{person}{Yulong Fu}, \bibinfo{person}{Zheng Yan},
  \bibinfo{person}{Jin Cao}, \bibinfo{person}{Ousmane Kon{\'e}}, {and}
  \bibinfo{person}{Xuefei Cao}.} \bibinfo{year}{2017}\natexlab{}.
\newblock \showarticletitle{An automata based intrusion detection method for
  internet of things}.
\newblock \bibinfo{journal}{\emph{Mobile Information Systems}}
  \bibinfo{volume}{2017} (\bibinfo{year}{2017}).
\newblock


\bibitem[\protect\citeauthoryear{Garcia-Font, Garrigues, and
  Rif{\`a}-Pous}{Garcia-Font et~al\mbox{.}}{2017}]%
        {garcia2017attack}
\bibfield{author}{\bibinfo{person}{Victor Garcia-Font}, \bibinfo{person}{Carles
  Garrigues}, {and} \bibinfo{person}{Helena Rif{\`a}-Pous}.}
  \bibinfo{year}{2017}\natexlab{}.
\newblock \showarticletitle{Attack classification schema for smart city WSNs}.
\newblock \bibinfo{journal}{\emph{Sensors}} \bibinfo{volume}{17},
  \bibinfo{number}{4} (\bibinfo{year}{2017}), \bibinfo{pages}{771}.
\newblock


\bibitem[\protect\citeauthoryear{Garcia-Teodoro, Diaz-Verdejo,
  Maci{\'a}-Fern{\'a}ndez, and V{\'a}zquez}{Garcia-Teodoro
  et~al\mbox{.}}{2009}]%
        {garcia2009anomaly}
\bibfield{author}{\bibinfo{person}{Pedro Garcia-Teodoro},
  \bibinfo{person}{Jesus Diaz-Verdejo}, \bibinfo{person}{Gabriel
  Maci{\'a}-Fern{\'a}ndez}, {and} \bibinfo{person}{Enrique V{\'a}zquez}.}
  \bibinfo{year}{2009}\natexlab{}.
\newblock \showarticletitle{Anomaly-based network intrusion detection:
  Techniques, systems and challenges}.
\newblock \bibinfo{journal}{\emph{computers \& security}} \bibinfo{volume}{28},
  \bibinfo{number}{1-2} (\bibinfo{year}{2009}), \bibinfo{pages}{18--28}.
\newblock


\bibitem[\protect\citeauthoryear{Gupta, Pandey, Shukla, Dadhich, Mathur, and
  Ingle}{Gupta et~al\mbox{.}}{2013}]%
        {gupta2013computational}
\bibfield{author}{\bibinfo{person}{Abhishek Gupta}, \bibinfo{person}{Om~Jee
  Pandey}, \bibinfo{person}{Mahendra Shukla}, \bibinfo{person}{Anjali Dadhich},
  \bibinfo{person}{Samar Mathur}, {and} \bibinfo{person}{Anup Ingle}.}
  \bibinfo{year}{2013}\natexlab{}.
\newblock \showarticletitle{Computational intelligence based intrusion
  detection systems for wireless communication and pervasive computing
  networks}. In \bibinfo{booktitle}{\emph{2013 IEEE International Conference on
  Computational Intelligence and Computing Research}}. IEEE,
  \bibinfo{pages}{1--7}.
\newblock


\bibitem[\protect\citeauthoryear{Haataja}{Haataja}{2008}]%
        {haataja2008new}
\bibfield{author}{\bibinfo{person}{Keijo~MJ Haataja}.}
  \bibinfo{year}{2008}\natexlab{}.
\newblock \showarticletitle{New efficient intrusion detection and prevention
  system for Bluetooth networks}. In \bibinfo{booktitle}{\emph{Proceedings of
  the 1st international conference on MOBILe Wireless MiddleWARE, Operating
  Systems, and Applications}}. ICST (Institute for Computer Sciences,
  Social-Informatics and~…, \bibinfo{pages}{16}.
\newblock


\bibitem[\protect\citeauthoryear{Had{\v{z}}iosmanovi{\'c}, Sommer, Zambon, and
  Hartel}{Had{\v{z}}iosmanovi{\'c} et~al\mbox{.}}{2014}]%
        {hadvziosmanovic2014through}
\bibfield{author}{\bibinfo{person}{Dina Had{\v{z}}iosmanovi{\'c}},
  \bibinfo{person}{Robin Sommer}, \bibinfo{person}{Emmanuele Zambon}, {and}
  \bibinfo{person}{Pieter~H Hartel}.} \bibinfo{year}{2014}\natexlab{}.
\newblock \showarticletitle{Through the eye of the PLC: semantic security
  monitoring for industrial processes}. In
  \bibinfo{booktitle}{\emph{Proceedings of the 30th Annual Computer Security
  Applications Conference}}. ACM, \bibinfo{pages}{126--135}.
\newblock


\bibitem[\protect\citeauthoryear{Han, Jiang, Shen, Shu, and Rodrigues}{Han
  et~al\mbox{.}}{2013}]%
        {han2013idsep}
\bibfield{author}{\bibinfo{person}{Guangjie Han}, \bibinfo{person}{Jinfang
  Jiang}, \bibinfo{person}{Wen Shen}, \bibinfo{person}{Lei Shu}, {and}
  \bibinfo{person}{Joel Rodrigues}.} \bibinfo{year}{2013}\natexlab{}.
\newblock \showarticletitle{IDSEP: a novel intrusion detection scheme based on
  energy prediction in cluster-based wireless sensor networks}.
\newblock \bibinfo{journal}{\emph{IET Information Security}}
  \bibinfo{volume}{7}, \bibinfo{number}{2} (\bibinfo{year}{2013}),
  \bibinfo{pages}{97--105}.
\newblock


\bibitem[\protect\citeauthoryear{Hassanzadeh and Stoleru}{Hassanzadeh and
  Stoleru}{2011}]%
        {hassanzadeh2011towards}
\bibfield{author}{\bibinfo{person}{Amin Hassanzadeh} {and}
  \bibinfo{person}{Radu Stoleru}.} \bibinfo{year}{2011}\natexlab{}.
\newblock \showarticletitle{Towards optimal monitoring in cooperative ids for
  resource constrained wireless networks}. In
  \bibinfo{booktitle}{\emph{Computer Communications and Networks (ICCCN), 2011
  Proceedings of 20th International Conference on}}. IEEE,
  \bibinfo{pages}{1--8}.
\newblock


\bibitem[\protect\citeauthoryear{Hui Suo Jiafu~Wan}{Hui Suo Jiafu~Wan}{2014}]%
        {IOTSecSurv2014}
\bibfield{author}{\bibinfo{person}{Caifeng Zou Jianqi~Liu Hui Suo Jiafu~Wan}.}
  \bibinfo{year}{2014}\natexlab{}.
\newblock \showarticletitle{Security in the Internet of Things: A Review}.
\newblock \bibinfo{journal}{\emph{International Journal of Computer
  Applications}} (\bibinfo{year}{2014}).
\newblock


\bibitem[\protect\citeauthoryear{Iera and Morabito}{Iera and Morabito}{2010}]%
        {IOTSurv2010}
\bibfield{author}{\bibinfo{person}{Atzori Luigi~Antonio Iera} {and}
  \bibinfo{person}{Giacomo Morabito}.} \bibinfo{year}{2010}\natexlab{}.
\newblock \showarticletitle{The Internet of Things: A Survey}.
\newblock \bibinfo{journal}{\emph{Computer Networks}} (\bibinfo{year}{2010}).
\newblock


\bibitem[\protect\citeauthoryear{Jiang, Wang, and Yu}{Jiang
  et~al\mbox{.}}{2012}]%
        {jiang2012dynamic}
\bibfield{author}{\bibinfo{person}{Tingyao Jiang}, \bibinfo{person}{Gangliang
  Wang}, {and} \bibinfo{person}{Heng Yu}.} \bibinfo{year}{2012}\natexlab{}.
\newblock \showarticletitle{A dynamic intrusion detection scheme for
  cluster-based wireless sensor networks}. In \bibinfo{booktitle}{\emph{World
  Automation Congress 2012}}. IEEE, \bibinfo{pages}{259--261}.
\newblock


\bibitem[\protect\citeauthoryear{Kai~Zhao}{Kai~Zhao}{2013}]%
        {IOTSecSurv2013}
\bibfield{author}{\bibinfo{person}{Lina~Ge Kai~Zhao}.}
  \bibinfo{year}{2013}\natexlab{}.
\newblock \showarticletitle{A Survey on the Internet of Things Security}.
\newblock \bibinfo{journal}{\emph{Computational Intelligence and Security
  (CIS)}} (\bibinfo{year}{2013}).
\newblock


\bibitem[\protect\citeauthoryear{Kasinathan, Costamagna, Khaleel, Pastrone, and
  Spirito}{Kasinathan et~al\mbox{.}}{2013a}]%
        {kasinathan2013ids}
\bibfield{author}{\bibinfo{person}{Prabhakaran Kasinathan},
  \bibinfo{person}{Gianfranco Costamagna}, \bibinfo{person}{Hussein Khaleel},
  \bibinfo{person}{Claudio Pastrone}, {and} \bibinfo{person}{Maurizio~A
  Spirito}.} \bibinfo{year}{2013}\natexlab{a}.
\newblock \showarticletitle{An ids framework for internet of things empowered
  by 6lowpan}. In \bibinfo{booktitle}{\emph{Proceedings of the 2013 ACM SIGSAC
  conference on Computer \& communications security}}. ACM,
  \bibinfo{pages}{1337--1340}.
\newblock


\bibitem[\protect\citeauthoryear{Kasinathan, Pastrone, Spirito, and
  Vinkovits}{Kasinathan et~al\mbox{.}}{2013b}]%
        {kasinathan2013denial}
\bibfield{author}{\bibinfo{person}{Prabhakaran Kasinathan},
  \bibinfo{person}{Claudio Pastrone}, \bibinfo{person}{Maurizio~A Spirito},
  {and} \bibinfo{person}{Mark Vinkovits}.} \bibinfo{year}{2013}\natexlab{b}.
\newblock \showarticletitle{Denial-of-Service detection in 6LoWPAN based
  Internet of Things}. In \bibinfo{booktitle}{\emph{2013 IEEE 9th international
  conference on wireless and mobile computing, networking and communications
  (WiMob)}}. IEEE, \bibinfo{pages}{600--607}.
\newblock


\bibitem[\protect\citeauthoryear{Khan and Herrmann}{Khan and Herrmann}{2017}]%
        {khan2017trust}
\bibfield{author}{\bibinfo{person}{Zeeshan~Ali Khan} {and}
  \bibinfo{person}{Peter Herrmann}.} \bibinfo{year}{2017}\natexlab{}.
\newblock \showarticletitle{A trust based distributed intrusion detection
  mechanism for internet of things}. In \bibinfo{booktitle}{\emph{2017 IEEE
  31st International Conference on Advanced Information Networking and
  Applications (AINA)}}. IEEE, \bibinfo{pages}{1169--1176}.
\newblock


\bibitem[\protect\citeauthoryear{Kim}{Kim}{2015}]%
        {kim2015physical}
\bibfield{author}{\bibinfo{person}{Sang~Wu Kim}.}
  \bibinfo{year}{2015}\natexlab{}.
\newblock \showarticletitle{Physical integrity check in cooperative relay
  communications}.
\newblock \bibinfo{journal}{\emph{IEEE Transactions on Wireless
  Communications}} \bibinfo{volume}{14}, \bibinfo{number}{11}
  (\bibinfo{year}{2015}), \bibinfo{pages}{6401--6413}.
\newblock


\bibitem[\protect\citeauthoryear{Kruegel and Vigna}{Kruegel and Vigna}{2003}]%
        {kruegel2003anomaly}
\bibfield{author}{\bibinfo{person}{Christopher Kruegel} {and}
  \bibinfo{person}{Giovanni Vigna}.} \bibinfo{year}{2003}\natexlab{}.
\newblock \showarticletitle{Anomaly detection of web-based attacks}. In
  \bibinfo{booktitle}{\emph{Proceedings of the 10th ACM conference on Computer
  and communications security}}. ACM, \bibinfo{pages}{251--261}.
\newblock


\bibitem[\protect\citeauthoryear{Le, Loo, Chai, and Aiash}{Le
  et~al\mbox{.}}{2016}]%
        {le2016specification}
\bibfield{author}{\bibinfo{person}{Anhtuan Le}, \bibinfo{person}{Jonathan Loo},
  \bibinfo{person}{Kok Chai}, {and} \bibinfo{person}{Mahdi Aiash}.}
  \bibinfo{year}{2016}\natexlab{}.
\newblock \showarticletitle{A specification-based IDS for detecting attacks on
  RPL-based network topology}.
\newblock \bibinfo{journal}{\emph{Information}} \bibinfo{volume}{7},
  \bibinfo{number}{2} (\bibinfo{year}{2016}), \bibinfo{pages}{25}.
\newblock


\bibitem[\protect\citeauthoryear{Le, Loo, Luo, and Lasebae}{Le
  et~al\mbox{.}}{2011}]%
        {le2011specification}
\bibfield{author}{\bibinfo{person}{Anhtuan Le}, \bibinfo{person}{Jonathan Loo},
  \bibinfo{person}{Yuan Luo}, {and} \bibinfo{person}{Aboubaker Lasebae}.}
  \bibinfo{year}{2011}\natexlab{}.
\newblock \showarticletitle{Specification-based IDS for securing RPL from
  topology attacks}. In \bibinfo{booktitle}{\emph{2011 IFIP Wireless Days
  (WD)}}. IEEE, \bibinfo{pages}{1--3}.
\newblock


\bibitem[\protect\citeauthoryear{Lee, Wen, Chang, Chiang, and Hsieh}{Lee
  et~al\mbox{.}}{2014}]%
        {lee2014lightweight}
\bibfield{author}{\bibinfo{person}{Tsung-Han Lee}, \bibinfo{person}{Chih-Hao
  Wen}, \bibinfo{person}{Lin-Huang Chang}, \bibinfo{person}{Hung-Shiou Chiang},
  {and} \bibinfo{person}{Ming-Chun Hsieh}.} \bibinfo{year}{2014}\natexlab{}.
\newblock \showarticletitle{A lightweight intrusion detection scheme based on
  energy consumption analysis in 6LowPAN}.
\newblock In \bibinfo{booktitle}{\emph{Advanced Technologies, Embedded and
  Multimedia for Human-centric Computing}}. \bibinfo{publisher}{Springer},
  \bibinfo{pages}{1205--1213}.
\newblock


\bibitem[\protect\citeauthoryear{Lin, Ke, and Tsai}{Lin et~al\mbox{.}}{2015}]%
        {lin2015cann}
\bibfield{author}{\bibinfo{person}{Wei-Chao Lin}, \bibinfo{person}{Shih-Wen
  Ke}, {and} \bibinfo{person}{Chih-Fong Tsai}.}
  \bibinfo{year}{2015}\natexlab{}.
\newblock \showarticletitle{CANN: An intrusion detection system based on
  combining cluster centers and nearest neighbors}.
\newblock \bibinfo{journal}{\emph{Knowledge-based systems}}
  \bibinfo{volume}{78} (\bibinfo{year}{2015}), \bibinfo{pages}{13--21}.
\newblock


\bibitem[\protect\citeauthoryear{Liu, Yang, Chen, Zhang, and Zeng}{Liu
  et~al\mbox{.}}{2011}]%
        {liu2011research}
\bibfield{author}{\bibinfo{person}{Caiming Liu}, \bibinfo{person}{Jin Yang},
  \bibinfo{person}{Run Chen}, \bibinfo{person}{Yan Zhang}, {and}
  \bibinfo{person}{Jinquan Zeng}.} \bibinfo{year}{2011}\natexlab{}.
\newblock \showarticletitle{Research on immunity-based intrusion detection
  technology for the internet of things}. In \bibinfo{booktitle}{\emph{2011
  Seventh International Conference on Natural Computation}},
  Vol.~\bibinfo{volume}{1}. IEEE, \bibinfo{pages}{212--216}.
\newblock


\bibitem[\protect\citeauthoryear{Liu, Xu, Zhang, and Wu}{Liu
  et~al\mbox{.}}{2018}]%
        {liu2018intrusion}
\bibfield{author}{\bibinfo{person}{Liqun Liu}, \bibinfo{person}{Bing Xu},
  \bibinfo{person}{Xiaoping Zhang}, {and} \bibinfo{person}{Xianjun Wu}.}
  \bibinfo{year}{2018}\natexlab{}.
\newblock \showarticletitle{An intrusion detection method for internet of
  things based on suppressed fuzzy clustering}.
\newblock \bibinfo{journal}{\emph{EURASIP Journal on Wireless Communications
  and Networking}} \bibinfo{volume}{2018}, \bibinfo{number}{1}
  (\bibinfo{year}{2018}), \bibinfo{pages}{113}.
\newblock


\bibitem[\protect\citeauthoryear{Liu and Yu}{Liu and Yu}{2008}]%
        {liu2008immunity}
\bibfield{author}{\bibinfo{person}{Yang Liu} {and} \bibinfo{person}{Fengqi
  Yu}.} \bibinfo{year}{2008}\natexlab{}.
\newblock \showarticletitle{Immunity-based intrusion detection for wireless
  sensor networks}. In \bibinfo{booktitle}{\emph{2008 IEEE International Joint
  Conference on Neural Networks (IEEE World Congress on Computational
  Intelligence)}}. IEEE, \bibinfo{pages}{439--444}.
\newblock


\bibitem[\protect\citeauthoryear{Lundin and Jonsson}{Lundin and
  Jonsson}{2000}]%
        {lundin2000anomaly}
\bibfield{author}{\bibinfo{person}{Emilie Lundin} {and} \bibinfo{person}{Erland
  Jonsson}.} \bibinfo{year}{2000}\natexlab{}.
\newblock \showarticletitle{Anomaly-based intrusion detection: privacy concerns
  and other problems}.
\newblock \bibinfo{journal}{\emph{Computer networks}} \bibinfo{volume}{34},
  \bibinfo{number}{4} (\bibinfo{year}{2000}), \bibinfo{pages}{623--640}.
\newblock


\bibitem[\protect\citeauthoryear{Luo and Nagarajan}{Luo and Nagarajan}{2018}]%
        {luo2018distributed}
\bibfield{author}{\bibinfo{person}{Tie Luo} {and} \bibinfo{person}{Sai~G
  Nagarajan}.} \bibinfo{year}{2018}\natexlab{}.
\newblock \showarticletitle{Distributed anomaly detection using autoencoder
  neural networks in wsn for iot}. In \bibinfo{booktitle}{\emph{2018 IEEE
  International Conference on Communications (ICC)}}. IEEE,
  \bibinfo{pages}{1--6}.
\newblock


\bibitem[\protect\citeauthoryear{Matsunaga, Toyoda, and Sasase}{Matsunaga
  et~al\mbox{.}}{2014}]%
        {matsunaga2014low}
\bibfield{author}{\bibinfo{person}{Takumi Matsunaga}, \bibinfo{person}{Kentaroh
  Toyoda}, {and} \bibinfo{person}{Iwao Sasase}.}
  \bibinfo{year}{2014}\natexlab{}.
\newblock \showarticletitle{Low false alarm rate RPL network monitoring system
  by considering timing inconstancy between the rank measurements}. In
  \bibinfo{booktitle}{\emph{2014 11th International Symposium on Wireless
  Communications Systems (ISWCS)}}. IEEE, \bibinfo{pages}{427--431}.
\newblock


\bibitem[\protect\citeauthoryear{Mell, Hu, Lippmann, Haines, and Zissman}{Mell
  et~al\mbox{.}}{2003}]%
        {mell2003overview}
\bibfield{author}{\bibinfo{person}{Peter Mell}, \bibinfo{person}{Vincent Hu},
  \bibinfo{person}{Richard Lippmann}, \bibinfo{person}{Josh Haines}, {and}
  \bibinfo{person}{Marc Zissman}.} \bibinfo{year}{2003}\natexlab{}.
\newblock \bibinfo{title}{An overview of issues in testing intrusion detection
  systems}.
\newblock
\newblock


\bibitem[\protect\citeauthoryear{Midi, Rullo, Mudgerikar, and Bertino}{Midi
  et~al\mbox{.}}{2017}]%
        {midi2017kalis}
\bibfield{author}{\bibinfo{person}{Daniele Midi}, \bibinfo{person}{Antonino
  Rullo}, \bibinfo{person}{Anand Mudgerikar}, {and} \bibinfo{person}{Elisa
  Bertino}.} \bibinfo{year}{2017}\natexlab{}.
\newblock \showarticletitle{Kalis—A system for knowledge-driven adaptable
  intrusion detection for the Internet of Things}. In
  \bibinfo{booktitle}{\emph{2017 IEEE 37th International Conference on
  Distributed Computing Systems (ICDCS)}}. IEEE, \bibinfo{pages}{656--666}.
\newblock


\bibitem[\protect\citeauthoryear{Milenkoski, Vieira, Kounev, Avritzer, and
  Payne}{Milenkoski et~al\mbox{.}}{2015}]%
        {milenkoski2015evaluating}
\bibfield{author}{\bibinfo{person}{Aleksandar Milenkoski},
  \bibinfo{person}{Marco Vieira}, \bibinfo{person}{Samuel Kounev},
  \bibinfo{person}{Alberto Avritzer}, {and} \bibinfo{person}{Bryan~D Payne}.}
  \bibinfo{year}{2015}\natexlab{}.
\newblock \showarticletitle{Evaluating computer intrusion detection systems: A
  survey of common practices}.
\newblock \bibinfo{journal}{\emph{ACM Computing Surveys (CSUR)}}
  \bibinfo{volume}{48}, \bibinfo{number}{1} (\bibinfo{year}{2015}),
  \bibinfo{pages}{12}.
\newblock


\bibitem[\protect\citeauthoryear{Minakov, Passerone, Rizzardi, and
  Sicari}{Minakov et~al\mbox{.}}{2016}]%
        {minakov2016comparative}
\bibfield{author}{\bibinfo{person}{Ivan Minakov}, \bibinfo{person}{Roberto
  Passerone}, \bibinfo{person}{Alessandra Rizzardi}, {and}
  \bibinfo{person}{Sabrina Sicari}.} \bibinfo{year}{2016}\natexlab{}.
\newblock \showarticletitle{A comparative study of recent wireless sensor
  network simulators}.
\newblock \bibinfo{journal}{\emph{ACM Transactions on Sensor Networks (TOSN)}}
  \bibinfo{volume}{12}, \bibinfo{number}{3} (\bibinfo{year}{2016}),
  \bibinfo{pages}{1--39}.
\newblock


\bibitem[\protect\citeauthoryear{Misra, Krishna, Agarwal, Saxena, and
  Obaidat}{Misra et~al\mbox{.}}{2011}]%
        {misra2011learning}
\bibfield{author}{\bibinfo{person}{Sudip Misra}, \bibinfo{person}{P~Venkata
  Krishna}, \bibinfo{person}{Harshit Agarwal}, \bibinfo{person}{Antriksh
  Saxena}, {and} \bibinfo{person}{Mohammad~S Obaidat}.}
  \bibinfo{year}{2011}\natexlab{}.
\newblock \showarticletitle{A learning automata based solution for preventing
  distributed denial of service in Internet of things}. In
  \bibinfo{booktitle}{\emph{2011 International Conference on Internet of Things
  and 4th International Conference on Cyber, Physical and Social Computing}}.
  IEEE, \bibinfo{pages}{114--122}.
\newblock


\bibitem[\protect\citeauthoryear{Moyers, Dunning, Marchany, and Tront}{Moyers
  et~al\mbox{.}}{2010}]%
        {moyers2010multi}
\bibfield{author}{\bibinfo{person}{Benjamin~R Moyers}, \bibinfo{person}{John~P
  Dunning}, \bibinfo{person}{Randolph~C Marchany}, {and}
  \bibinfo{person}{Joseph~G Tront}.} \bibinfo{year}{2010}\natexlab{}.
\newblock \showarticletitle{The multi-vector portable intrusion detection
  system (MVP-IDS): a hybrid approach to intrusion detection for portable
  information devices}. In \bibinfo{booktitle}{\emph{2010 IEEE International
  Conference on Wireless Information Technology and Systems}}. IEEE,
  \bibinfo{pages}{1--4}.
\newblock


\bibitem[\protect\citeauthoryear{OConnor and Reeves}{OConnor and
  Reeves}{2008}]%
        {oconnor2008bluetooth}
\bibfield{author}{\bibinfo{person}{Terrence OConnor} {and}
  \bibinfo{person}{Douglas Reeves}.} \bibinfo{year}{2008}\natexlab{}.
\newblock \showarticletitle{Bluetooth network-based misuse detection}. In
  \bibinfo{booktitle}{\emph{2008 Annual Computer Security Applications
  Conference (ACSAC)}}. IEEE, \bibinfo{pages}{377--391}.
\newblock


\bibitem[\protect\citeauthoryear{Oh, Kim, and Ro}{Oh et~al\mbox{.}}{2014}]%
        {oh2014malicious}
\bibfield{author}{\bibinfo{person}{Doohwan Oh}, \bibinfo{person}{Deokho Kim},
  {and} \bibinfo{person}{Won Ro}.} \bibinfo{year}{2014}\natexlab{}.
\newblock \showarticletitle{A malicious pattern detection engine for embedded
  security systems in the Internet of Things}.
\newblock \bibinfo{journal}{\emph{Sensors}} \bibinfo{volume}{14},
  \bibinfo{number}{12} (\bibinfo{year}{2014}), \bibinfo{pages}{24188--24211}.
\newblock


\bibitem[\protect\citeauthoryear{Patcha and Park}{Patcha and Park}{2007}]%
        {patcha2007overview}
\bibfield{author}{\bibinfo{person}{Animesh Patcha} {and}
  \bibinfo{person}{Jung-Min Park}.} \bibinfo{year}{2007}\natexlab{}.
\newblock \showarticletitle{An overview of anomaly detection techniques:
  Existing solutions and latest technological trends}.
\newblock \bibinfo{journal}{\emph{Computer networks}} \bibinfo{volume}{51},
  \bibinfo{number}{12} (\bibinfo{year}{2007}), \bibinfo{pages}{3448--3470}.
\newblock


\bibitem[\protect\citeauthoryear{Patel}{Patel}{2014}]%
        {IOTPrivacySurv2014}
\bibfield{author}{\bibinfo{person}{J.~Sathish Kumar Dhiren~R. Patel}.}
  \bibinfo{year}{2014}\natexlab{}.
\newblock \showarticletitle{A Survey on Internet of Things: Security and
  Privacy Issues}.
\newblock \bibinfo{journal}{\emph{International Journal of Computer
  Applications}} (\bibinfo{year}{2014}).
\newblock


\bibitem[\protect\citeauthoryear{Pongle and Chavan}{Pongle and Chavan}{2015}]%
        {pongle2015real}
\bibfield{author}{\bibinfo{person}{Pavan Pongle} {and}
  \bibinfo{person}{Gurunath Chavan}.} \bibinfo{year}{2015}\natexlab{}.
\newblock \showarticletitle{Real time intrusion and wormhole attack detection
  in internet of things}.
\newblock \bibinfo{journal}{\emph{International Journal of Computer
  Applications}} \bibinfo{volume}{121}, \bibinfo{number}{9}
  (\bibinfo{year}{2015}).
\newblock


\bibitem[\protect\citeauthoryear{Raza, Wallgren, and Voigt}{Raza
  et~al\mbox{.}}{2013}]%
        {raza2013svelte}
\bibfield{author}{\bibinfo{person}{Shahid Raza}, \bibinfo{person}{Linus
  Wallgren}, {and} \bibinfo{person}{Thiemo Voigt}.}
  \bibinfo{year}{2013}\natexlab{}.
\newblock \showarticletitle{SVELTE: Real-time intrusion detection in the
  Internet of Things}.
\newblock \bibinfo{journal}{\emph{Ad hoc networks}} \bibinfo{volume}{11},
  \bibinfo{number}{8} (\bibinfo{year}{2013}), \bibinfo{pages}{2661--2674}.
\newblock


\bibitem[\protect\citeauthoryear{Sabhnani and Serpen}{Sabhnani and
  Serpen}{2004}]%
        {sabhnani2004machine}
\bibfield{author}{\bibinfo{person}{Maheshkumar Sabhnani} {and}
  \bibinfo{person}{Gursel Serpen}.} \bibinfo{year}{2004}\natexlab{}.
\newblock \showarticletitle{Why machine learning algorithms fail in misuse
  detection on KDD intrusion detection data set}.
\newblock \bibinfo{journal}{\emph{Intelligent data analysis}}
  \bibinfo{volume}{8}, \bibinfo{number}{4} (\bibinfo{year}{2004}),
  \bibinfo{pages}{403--415}.
\newblock


\bibitem[\protect\citeauthoryear{Scarfone and Mell}{Scarfone and Mell}{2007}]%
        {scarfone2007guide}
\bibfield{author}{\bibinfo{person}{Karen Scarfone} {and} \bibinfo{person}{Peter
  Mell}.} \bibinfo{year}{2007}\natexlab{}.
\newblock \showarticletitle{Guide to intrusion detection and prevention systems
  (idps)}.
\newblock \bibinfo{journal}{\emph{NIST special publication}}
  \bibinfo{volume}{800}, \bibinfo{number}{2007} (\bibinfo{year}{2007}),
  \bibinfo{pages}{94}.
\newblock


\bibitem[\protect\citeauthoryear{Sedjelmaci and Senouci}{Sedjelmaci and
  Senouci}{2013}]%
        {sedjelmaci2013efficient}
\bibfield{author}{\bibinfo{person}{Hichem Sedjelmaci} {and}
  \bibinfo{person}{Sidi~Mohammed Senouci}.} \bibinfo{year}{2013}\natexlab{}.
\newblock \showarticletitle{Efficient and lightweight intrusion detection based
  on nodes' behaviors in wireless sensor networks}. In
  \bibinfo{booktitle}{\emph{Global Information Infrastructure Symposium-GIIS
  2013}}. IEEE, \bibinfo{pages}{1--6}.
\newblock


\bibitem[\protect\citeauthoryear{Sedjelmaci, Senouci, and Al-Bahri}{Sedjelmaci
  et~al\mbox{.}}{2016}]%
        {sedjelmaci2016lightweight}
\bibfield{author}{\bibinfo{person}{Hichem Sedjelmaci},
  \bibinfo{person}{Sidi~Mohammed Senouci}, {and} \bibinfo{person}{Mohamad
  Al-Bahri}.} \bibinfo{year}{2016}\natexlab{}.
\newblock \showarticletitle{A lightweight anomaly detection technique for
  low-resource IoT devices: A game-theoretic methodology}. In
  \bibinfo{booktitle}{\emph{2016 IEEE International Conference on
  Communications (ICC)}}. IEEE, \bibinfo{pages}{1--6}.
\newblock


\bibitem[\protect\citeauthoryear{Shreenivas, Raza, and Voigt}{Shreenivas
  et~al\mbox{.}}{2017}]%
        {shreenivas2017intrusion}
\bibfield{author}{\bibinfo{person}{Dharmini Shreenivas},
  \bibinfo{person}{Shahid Raza}, {and} \bibinfo{person}{Thiemo Voigt}.}
  \bibinfo{year}{2017}\natexlab{}.
\newblock \showarticletitle{Intrusion Detection in the RPL-connected 6LoWPAN
  Networks}. In \bibinfo{booktitle}{\emph{Proceedings of the 3rd ACM
  International Workshop on IoT Privacy, Trust, and Security}}. ACM,
  \bibinfo{pages}{31--38}.
\newblock


\bibitem[\protect\citeauthoryear{Sommer and Paxson}{Sommer and Paxson}{2010}]%
        {sommer2010outside}
\bibfield{author}{\bibinfo{person}{Robin Sommer} {and} \bibinfo{person}{Vern
  Paxson}.} \bibinfo{year}{2010}\natexlab{}.
\newblock \showarticletitle{Outside the closed world: On using machine learning
  for network intrusion detection}. In \bibinfo{booktitle}{\emph{2010 IEEE
  symposium on security and privacy}}. IEEE, \bibinfo{pages}{305--316}.
\newblock


\bibitem[\protect\citeauthoryear{Song, Chen, and Li}{Song
  et~al\mbox{.}}{2010}]%
        {song2010weak}
\bibfield{author}{\bibinfo{person}{Xianfeng Song}, \bibinfo{person}{Guangxi
  Chen}, {and} \bibinfo{person}{Xiaolong Li}.} \bibinfo{year}{2010}\natexlab{}.
\newblock \showarticletitle{A Weak Hidden Markov Model based intrusion
  detection method for wireless sensor networks}. In
  \bibinfo{booktitle}{\emph{2010 International Conference on Intelligent
  Computing and Integrated Systems}}. IEEE, \bibinfo{pages}{887--889}.
\newblock


\bibitem[\protect\citeauthoryear{Stelte and Rodosek}{Stelte and
  Rodosek}{2013}]%
        {stelte2013thwarting}
\bibfield{author}{\bibinfo{person}{Bj{\"o}rn Stelte} {and}
  \bibinfo{person}{Gabi~Dreo Rodosek}.} \bibinfo{year}{2013}\natexlab{}.
\newblock \showarticletitle{Thwarting attacks on zigbee-removal of the
  killerbee stinger}. In \bibinfo{booktitle}{\emph{Proceedings of the 9th
  International Conference on Network and Service Management (CNSM 2013)}}.
  IEEE, \bibinfo{pages}{219--226}.
\newblock


\bibitem[\protect\citeauthoryear{Summerville, Zach, and Chen}{Summerville
  et~al\mbox{.}}{2015}]%
        {summerville2015ultra}
\bibfield{author}{\bibinfo{person}{Douglas~H Summerville},
  \bibinfo{person}{Kenneth~M Zach}, {and} \bibinfo{person}{Yu Chen}.}
  \bibinfo{year}{2015}\natexlab{}.
\newblock \showarticletitle{Ultra-lightweight deep packet anomaly detection for
  internet of things devices}. In \bibinfo{booktitle}{\emph{2015 IEEE 34th
  International Performance Computing and Communications Conference (IPCCC)}}.
  IEEE, \bibinfo{pages}{1--8}.
\newblock


\bibitem[\protect\citeauthoryear{Tan, Killourhy, and Maxion}{Tan
  et~al\mbox{.}}{2002}]%
        {tan2002undermining}
\bibfield{author}{\bibinfo{person}{Kymie~MC Tan}, \bibinfo{person}{Kevin~S
  Killourhy}, {and} \bibinfo{person}{Roy~A Maxion}.}
  \bibinfo{year}{2002}\natexlab{}.
\newblock \showarticletitle{Undermining an anomaly-based intrusion detection
  system using common exploits}. In \bibinfo{booktitle}{\emph{International
  Workshop on Recent Advances in Intrusion Detection}}. Springer,
  \bibinfo{pages}{54--73}.
\newblock


\bibitem[\protect\citeauthoryear{Thanigaivelan, Nigussie, Kanth, Virtanen, and
  Isoaho}{Thanigaivelan et~al\mbox{.}}{2016}]%
        {thanigaivelan2016distributed}
\bibfield{author}{\bibinfo{person}{Nanda~Kumar Thanigaivelan},
  \bibinfo{person}{Ethiopia Nigussie}, \bibinfo{person}{Rajeev~Kumar Kanth},
  \bibinfo{person}{Seppo Virtanen}, {and} \bibinfo{person}{Jouni Isoaho}.}
  \bibinfo{year}{2016}\natexlab{}.
\newblock \showarticletitle{Distributed internal anomaly detection system for
  internet-of-things}. In \bibinfo{booktitle}{\emph{2016 13th IEEE Annual
  Consumer Communications \& Networking Conference (CCNC)}}. IEEE,
  \bibinfo{pages}{319--320}.
\newblock


\bibitem[\protect\citeauthoryear{Vora, Oza, et~al\mbox{.}}{Vora
  et~al\mbox{.}}{2013}]%
        {vora2013survey}
\bibfield{author}{\bibinfo{person}{Pritesh Vora}, \bibinfo{person}{Bhavesh
  Oza}, {et~al\mbox{.}}} \bibinfo{year}{2013}\natexlab{}.
\newblock \showarticletitle{A survey on k-mean clustering and particle swarm
  optimization}.
\newblock \bibinfo{journal}{\emph{International Journal of Science and Modern
  Engineering}} \bibinfo{volume}{1}, \bibinfo{number}{3}
  (\bibinfo{year}{2013}), \bibinfo{pages}{1--14}.
\newblock


\bibitem[\protect\citeauthoryear{Wallgren, Raza, and Voigt}{Wallgren
  et~al\mbox{.}}{2013}]%
        {wallgren2013routing}
\bibfield{author}{\bibinfo{person}{Linus Wallgren}, \bibinfo{person}{Shahid
  Raza}, {and} \bibinfo{person}{Thiemo Voigt}.}
  \bibinfo{year}{2013}\natexlab{}.
\newblock \showarticletitle{Routing Attacks and Countermeasures in the
  RPL-based Internet of Things}.
\newblock \bibinfo{journal}{\emph{International Journal of Distributed Sensor
  Networks}} \bibinfo{volume}{9}, \bibinfo{number}{8} (\bibinfo{year}{2013}),
  \bibinfo{pages}{794326}.
\newblock


\bibitem[\protect\citeauthoryear{Wang and Megalooikonomou}{Wang and
  Megalooikonomou}{2005}]%
        {wang2005clustering}
\bibfield{author}{\bibinfo{person}{Qiang Wang} {and} \bibinfo{person}{Vasileios
  Megalooikonomou}.} \bibinfo{year}{2005}\natexlab{}.
\newblock \showarticletitle{A clustering algorithm for intrusion detection}. In
  \bibinfo{booktitle}{\emph{Data Mining, Intrusion Detection, Information
  Assurance, and Data Networks Security 2005}}, Vol.~\bibinfo{volume}{5812}.
  International Society for Optics and Photonics, \bibinfo{pages}{31--39}.
\newblock


\bibitem[\protect\citeauthoryear{Watkins and Timmis}{Watkins and
  Timmis}{2004}]%
        {watkins2004exploiting}
\bibfield{author}{\bibinfo{person}{Andrew Watkins} {and} \bibinfo{person}{Jon
  Timmis}.} \bibinfo{year}{2004}\natexlab{}.
\newblock \showarticletitle{Exploiting parallelism inherent in AIRS, an
  artificial immune classifier}. In \bibinfo{booktitle}{\emph{International
  Conference on Artificial Immune Systems}}. Springer,
  \bibinfo{pages}{427--438}.
\newblock


\bibitem[\protect\citeauthoryear{Weaver, Weaver, and Farwood}{Weaver
  et~al\mbox{.}}{2013}]%
        {weaver2013guide}
\bibfield{author}{\bibinfo{person}{Randy Weaver}, \bibinfo{person}{Dawn
  Weaver}, {and} \bibinfo{person}{Dean Farwood}.}
  \bibinfo{year}{2013}\natexlab{}.
\newblock \bibinfo{booktitle}{\emph{Guide to network defense and
  countermeasures}}.
\newblock \bibinfo{publisher}{Cengage Learning}.
\newblock


\bibitem[\protect\citeauthoryear{Wendt and Potkonjak}{Wendt and
  Potkonjak}{2015}]%
        {IOTSecChallenges2015}
\bibfield{author}{\bibinfo{person}{Teng Xu James~B Wendt} {and}
  \bibinfo{person}{Miodrag Potkonjak}.} \bibinfo{year}{2015}\natexlab{}.
\newblock \showarticletitle{Security of IoT Systems: Design Challenges and
  Opportunities}.
\newblock \bibinfo{journal}{\emph{Informational System Frontiers}}
  (\bibinfo{year}{2015}).
\newblock


\bibitem[\protect\citeauthoryear{Wu and Banzhaf}{Wu and Banzhaf}{2010}]%
        {wu2010use}
\bibfield{author}{\bibinfo{person}{Shelly~Xiaonan Wu} {and}
  \bibinfo{person}{Wolfgang Banzhaf}.} \bibinfo{year}{2010}\natexlab{}.
\newblock \showarticletitle{The use of computational intelligence in intrusion
  detection systems: A review}.
\newblock \bibinfo{journal}{\emph{Applied soft computing}}
  \bibinfo{volume}{10}, \bibinfo{number}{1} (\bibinfo{year}{2010}),
  \bibinfo{pages}{1--35}.
\newblock


\bibitem[\protect\citeauthoryear{Yadav and Srinivasan}{Yadav and
  Srinivasan}{2010}]%
        {yadav2010itrust}
\bibfield{author}{\bibinfo{person}{Kuldeep Yadav} {and}
  \bibinfo{person}{Avinash Srinivasan}.} \bibinfo{year}{2010}\natexlab{}.
\newblock \showarticletitle{iTrust: an integrated trust framework for wireless
  sensor networks}. In \bibinfo{booktitle}{\emph{Proceedings of the 2010 ACM
  Symposium on Applied Computing}}. ACM, \bibinfo{pages}{1466--1471}.
\newblock


\bibitem[\protect\citeauthoryear{Yu and Tsai}{Yu and Tsai}{2008}]%
        {yu2008framework}
\bibfield{author}{\bibinfo{person}{Zhenwei Yu} {and}
  \bibinfo{person}{Jeffrey~JP Tsai}.} \bibinfo{year}{2008}\natexlab{}.
\newblock \showarticletitle{A framework of machine learning based intrusion
  detection for wireless sensor networks}. In \bibinfo{booktitle}{\emph{2008
  IEEE International Conference on Sensor Networks, Ubiquitous, and Trustworthy
  Computing (sutc 2008)}}. IEEE, \bibinfo{pages}{272--279}.
\newblock


\bibitem[\protect\citeauthoryear{Zarpelao, Miani, Kawakani, and
  de~Alvarenga}{Zarpelao et~al\mbox{.}}{2017}]%
        {zarpelao2017survey}
\bibfield{author}{\bibinfo{person}{Bruno~Bogaz Zarpelao},
  \bibinfo{person}{Rodrigo~Sanches Miani}, \bibinfo{person}{Cl{\'a}udio~Toshio
  Kawakani}, {and} \bibinfo{person}{Sean~Carlisto de Alvarenga}.}
  \bibinfo{year}{2017}\natexlab{}.
\newblock \showarticletitle{A survey of intrusion detection in Internet of
  Things}.
\newblock \bibinfo{journal}{\emph{Journal of Network and Computer
  Applications}}  \bibinfo{volume}{84} (\bibinfo{year}{2017}),
  \bibinfo{pages}{25--37}.
\newblock


\bibitem[\protect\citeauthoryear{Zhang, Wang, Sun, Green, and Alam}{Zhang
  et~al\mbox{.}}{2011}]%
        {zhang2011artificial}
\bibfield{author}{\bibinfo{person}{Yichi Zhang}, \bibinfo{person}{Lingfeng
  Wang}, \bibinfo{person}{Weiqing Sun}, \bibinfo{person}{Robert~C Green}, {and}
  \bibinfo{person}{Mansoor Alam}.} \bibinfo{year}{2011}\natexlab{}.
\newblock \showarticletitle{Artificial immune system based intrusion detection
  in a distributed hierarchical network architecture of smart grid}. In
  \bibinfo{booktitle}{\emph{2011 IEEE Power and Energy Society General
  Meeting}}. IEEE, \bibinfo{pages}{1--8}.
\newblock


\bibitem[\protect\citeauthoryear{Zurada, Marks, and Robinson}{Zurada
  et~al\mbox{.}}{1995}]%
        {zurada1995review}
\bibfield{author}{\bibinfo{person}{JM Zurada}, \bibinfo{person}{RJ Marks},
  {and} \bibinfo{person}{J Robinson}.} \bibinfo{year}{1995}\natexlab{}.
\newblock \bibinfo{title}{Review of computational intelligence: imitating
  life}.
\newblock
\newblock


\end{thebibliography}

\appendix
\newpage
\section{Survey IDS Request Template Letter}

Dear <INSERT AUTHOR LIST>,\\

\noindent I am a PhD Student working on IoT Security at Newcastle University \url{(My Info)} and I am currently working on developing an IoT security testbed. 
My work hopes to test out various security solutions in a range of test IoT environments in the form of embedded virtual machines hosted in various network configurations. 
I am contacting you in regards to your work <PAPER TITLE HERE>. 
<INSERT UNIQUE COMMENT ON PAPER>. 
I was hoping you could provide me with an implementation of your work to use as I was unable to find a publicly available repository?\\  

\noindent Please let me know how you would prefer I cite your work in the case of publication and what license restrictions might be applicable.\\

\noindent If you have any questions about my work or how your implementation will be used please don't hesitate to ask.\\

\noindent Yours sincerely,\\
Luca Arnaboldi

\section{Survey IDS Request Example Letter}

Dear \sout{\textit{REDACTED}},\\

\noindent I am a PhD Student working on IoT Security at Newcastle University \url{(My Info)} and I am currently working on developing an IoT security testbed. 
My work hopes to test out various security solutions in a range of test IoT environments in the form of embedded virtual machines hosted in various network configurations. 
I am contacting you in regards to your work \sout{\textit{REDACTED}}. 
\textit{Your presented detection methodology is very powerful showing some excellent accuracy with low false positives, I would be interested in seeing how well it would scale in larger systems and different configurations.} 
I was hoping you could provide me with an implementation of your work to use as I was unable to find a publicly available repository?\\  

\noindent Please let me know how you would prefer I cite your work in the case of publication and what license restrictions might be applicable.\\

\noindent If you have any questions about my work or how your implementation will be used please don't hesitate to ask.\\

\noindent Yours sincerely,\\
Luca Arnaboldi
\end{document}